\documentclass{pthesis}
\usepackage{color}
\usepackage{graphics}
\usepackage{graphicx}
\usepackage{epsfig}
\usepackage{amssymb}
\usepackage{verbatim}
\usepackage{graphics,amsmath,amsfonts,amssymb} 
\usepackage{epsfig,verbatim}
 \newtheorem{theorem}{Theorem}[section]
 \newtheorem{corollary}[theorem]{Corollary}
 \newtheorem{lemma}[theorem]{Lemma}
 
 \newtheorem{proposition}[theorem]{Proposition}
 \theoremstyle{definition}
 \newtheorem{definition}[theorem]{Definition}
 \theoremstyle{remark}
 \newtheorem{remark}[theorem]{Remark}
 \newtheorem*{example}{Example}
 \newtheorem*{examples}{Examples}
 \newtheorem*{algorithm}{Algorithm}
  
\input xy
\xyoption{all}
\def\sympow{{\setbox0\hbox{$\bigcirc$}\setbox1\hbox to\wd0{\hss$s$\hss}%
\wd1 0pt\box1\box0}}
\begin{document}

\pagenumbering{roman}
\title{PhD. Dissertation \bigskip \\ \textbf{GALOISIAN APPROACH TO SUPERSYMMETRIC QUANTUM
MECHANICS}}
\author{Primitivo Bel\'en Acosta-Hum\'anez\\ \medskip \\ \medskip\textbf{ADVISORS}\\ \medskip Juan J. Morales-Ruiz -- Universidad Polit\'ecnica de Madrid \smallskip \\ Jacques-Arthur Weil -- Universit\'e de Limoges}
\bigskip
\date{\bigskip Departament de Matem\'atica Aplicada I\\ \medskip
Universitat Polit\`ecnica de Catalunya\medskip \\ May 26, 2009}
 \maketitle
\newpage
\
\thispagestyle{empty}
\newpage
\chapter*{ }
Doctoral Program in Applied Mathematics\\

\noindent Dissertation presented as requirement to obtain the
degree of Doctor in Applied Mathematics by Universitat
Polit\`ecnica de Catalunya.\\

\bigskip

\bigskip

\bigskip

\bigskip

\bigskip

\bigskip

\bigskip

\bigskip

\bigskip

\bigskip

\bigskip

\bigskip

\bigskip

\bigskip

\bigskip

\begin{center}
We certify that this dissertation has been\\ done by Primitivo
Bel\'en Acosta-Hum\'anez\\ with our supervision and co-direction.
\end{center}
\bigskip
\centerline{Barcelona, May 26, 2009}
\bigskip

\bigskip

\bigskip

\bigskip

\bigskip

\bigskip

\bigskip

\bigskip

\bigskip

\bigskip

\bigskip

\bigskip

\begin{displaymath}
\textrm{Juan J. Morales-Ruiz}\qquad \qquad \qquad \qquad
\textrm{Jacques-Arthur Weil}
\end{displaymath}
\chapter*{ }
\begin{raggedleft}

\hspace{1.cm} \emph{Todo lo puedo en Cristo} \\
\hspace{1.cm} \emph{que me fortalece.} \\

\medskip

\hspace{1.cm } \emph{Filipenses 4, 13.}

\end{raggedleft}

\bigskip

\noindent To my inspirators: Primitivo Bel\'en Hum\'anez Vergara,
Jairo
Charris Casta\~neda and Jerry Kovacic, in memoriam.\\

\noindent To my professors: Juan Jos\'e Morales-Ruiz,
Jacques-Arthur Weil,  H\'ector Roger Maya Taboada, Joaqu\'\i n
Luna and Jes\'us Hernando P\'erez (Pelusa).
\\

\noindent To my family:\begin{itemize} \item My daughter Angie
Marcela, princess and tenderness that maintains my life in calm.
\item My son Sergio Gabriel, friend and colleague that give me
happiness every day.
\item My wife Mar\'{\i}a Cristina. My mother Carmen Olinda. My father Manuel de los \'Angeles. My brothers German, Sandro
Manuel, Jes\'us David (with occasion of his 30th birthday) and
Manuel Fernando. My sisters Nordith, Elinor and Carmen Roc\'{\i}o.
My nephews German Javier, Pedro, German Manuel, Jos\'e David and
Richard Antonio. My nieces Nerys, Yeiny, Shirley Vanessa, Uldis
Shamara and Elis Paola. My godchildrens, Yessica, Rodrigo Antonio
and Juan Pablo. My Hum\'anez parents, descendants of the brothers
Eulalio and Lucas Hum\'anez Ruiz. My Acosta parents, descendants
of the union between Francisco Vidal and Melchora Acosta.
\end{itemize}

\newpage













































\chapter*{Acknowledgements}
I want to thank to my advisors Juan Jos\'e Morales Ruiz and
Jacques-Arthur Weil by their orientation and by encourage me in
the development of this work. Juanjo, thank you very much because
you believed in me and you gave me the opportunity negated by
others, please keep in mind that every thing that I have learned
with you is the angular stone in which my research and scientific
life will be holden. Jacques-Arthur, thank you for give me the
opportunity to work with you and to accept be my co-advisor, this
time in Limoges working with you and Moulay Barkatou has been
algorithmically fruitful and allowed to improve this work. Moulay,
thank you a lot for your help, advices, financial support and for
consider me as another member of your team.

\medskip

I also acknowledge my academic-doctoral brothers David
Bl\'azquez-Sanz and Sergi Sim\'on i Estrada, for their help and
support during these years.
\medskip

I also want to thank my doctoral professors at \emph{Universitat
Polit\`ecnica de Catalunya -- Universitat de Barcelona}: Amadeu
Delshams, Pere Guti\'errez, Tere Seara, Josep Masdemont, Carles
Sim\'o, Nuria Fagella, Pere Mumbr\'u and Juan Morales-Ruiz for
encouraged me to study very hard and to understand every subject
in a serious way. Specially I am very thanked with Amadeu Delshams
because he received me cordially when I moved to Barcelona five
years ago to work in his dynamical system project with the grant
FPI Spanish Government, project BFM2003-09504-C02-02.

\medskip

I also want to thank my professors at \emph{Universidad Sergio
Arboleda}: Jairo Charris Casta\~neda, Joaqu\'\i n Luna and Jes\'us
Hernando P\'erez Alc\'azar (Pelusa), for their continuous support,
both in academical and personal affairs. In a much more concrete
way, I have to thank Pelusa for embedding me in the world of
Differential Galois theory, Joaqu\'in for embedding me in the
world of dynamical systems and Jairo for encouraging me to
continue developing my research and personal life from algebra.
Nowadays, Universidad Sergio Arboleda continues being an important
support for me, due to the excellent management with me on part of
Reinaldo N\'u\~nez and Germ\'an Quintero. Reinaldo and German,
thank you a lot, for consider my work, really I feel me
\emph{Sergista}.

\medskip

  I also want to mention my Mexican colleagues from
\emph{Universidad Aut\'onoma Metropolitana - Iztapalapa},
Joaqu\'\i n Delgado and Martha \'Alvarez-Ram\'\i rez, who hosted
me in M\'exico in the winter of 2007 and again in the summer of
2008. My stay there was fruitful because we can write one paper
and set one un project for further research collaborations.
\medskip

There are a lot of people who helped me throughout these studies,
but there is not enough space here to list their names. Thus, I
will just mention some of them.

\medskip

H\'ector Roger Maya Taboada, my godfather, who encouraged me to
research when I was a scholar student in \emph{Colegio la Salle de
Monter\'\i a}. He also help me to born into the mathematical world
because he altruistically worked with me preparing me for my first
mathematical talk and publication that was known as \emph{Ley
Costeana}. Armando Potes Guti\'errez and Abelardo Arrieta Vergara,
my teachers in Colegio la Salle de Monter\'\i a and after in
\emph{Universidad de C\'ordoba} who believed in me from the start.
A special mention for Arturo Pelaez and the rest of the people in
Colegio la Salle de Monter\'\i a who shared with me the most
beautiful memories leading to my research life. \medskip

My professors and friends in Universidad de C\'ordoba,
\emph{Universidad Nacional de Colombia}, Universidad Sergio
Arboleda, \emph{Universidad Pedag\'ogica Nacional} and
\emph{Universidad Ideas} with whom I learned that the research
life is very hard but also very pretty. Llu\'\i s Alsed\'a, Gemma
Huguet, Ariadna Farr\'es, Alejandro Luque, Marina Gonchenko, Chema
Benita, Oswaldo Larreal, Chara Pantazi, Carme Oliv\'e, Tom\'as
L\'azaro (my tutor), Arturo Vieiro and the rest of the people of
\emph{Dance net} who shared with me very important success in the
winter schools. Mayte Alita, Ainhoa Aparicio, Salva Rodr\'\i guez,
Manuel Marcote and Mar\'\i a Fernanda Gonz\'alez who shared with
me movies and enjoyable moments. Anna Demier, Montse Manubens,
Mar\'\i a Saumell and
Clemens Huemer, who shared with me the pleasure of the dance.\\

Francisco Marcell\'an, Jes\'us Palaci\'an and Patricia Yanguas for
our mathematical and fruitful discussions. Jerry Kovacic, William
Sit and Li Guo for allow me to be as speaker in \emph{AMS special
session in differential algebra} and \emph{DART II}. I clarified
some things about Kovacic algorithm with Jerry, was a good
experience. The professors Michael Singer, Rick Churchill, Marius
Van der Put, Mark Van Hoej, Andy Magid, Guy Casale, B. Heinrich
Matzat, Julia Hartman, Lourdes Juan, Jean-Pierre Ramis, Bernard
Malgrange, Jorge Mozo, Jos\'e Cano, Teresa Crespo, Zbigniew Hajto
and Phillys Cassidy, who eventually discussed with me in some
conference, seminar or meeting; their vision about the
differential algebra reached my own point of view.
\bigskip

Finally, I would like to thanks to the most important
\textit{being} in my life, who lives into my heart. Thank you
because every thing I do and every thing I am is inspired by you.

\newpage
 \tableofcontents \cleardoublepage

\chapter*{Introduction}\label{introduction}\thispagestyle{empty}
\markboth{Introduction }{Introduction}
\addcontentsline{toc}{chapter}{Introduction}
\pagenumbering{arabic}\setcounter{page}{1}
\section*{Structure of the Thesis}
The main object studied in this thesis, in the differential Galois
theory framework, is the one-dimensional stationary
non-relativistic Schr\"odinger equation
$$\partial_x^2\Psi=(V-\lambda)\Psi,\quad V\in K,$$ where $K$ is a
differential field containing $x$, closed algebraically and of
characteristic zero. In particular, $K$ is considered as the
smallest differential field containing the potential $V$.\\

This thesis is divided in two parts.

\subsection*{Chapter 1. Theoretical Background} In this part there are
not original results. Summaries of Picard-Vessiot theory and
supersymmetric quantum mechanics, necessaries to understand the
next chapter, are presented here.

\subsection*{Chapter 2. Differential Galois Theory Approach to
Supersymmetric Quantum Mechanics} This part contain the original
results of this thesis, which were developed using the previous
chapter. Up to specific cases; theorems, propositions, corollaries
and lemmas given in this chapter are considered as original
results of
this thesis.\\

Two different Galoisian approaches are studied in this chapter,
which depends on the differential field: the first one is
$\mathbb{C}(x)$ and the second one is
$K=\mathbb{C}(z(x),\partial_z(x))$, where $z=z(x)$ is a
\textit{Hamiltonian change of variable}. This concept allow us to
introduce an useful derivation $\widehat{\partial}_z$, important
tool to transforms differential equations with non rational
coefficients into differential equations with rational
coefficients, to apply the results given
in the case of $\mathbb{C}(x)$.\\

This chapter is divided in three parts or sections.\\

\subsubsection*{Section 2.1} Here is introduced the
set $\Lambda$ as the set of values of $\lambda$ in which the
Schr\"odinger equation is integrable in the sense of
Picard-Vessiot theory. With this new set $\Lambda$, we define in
an easy way the concepts of \textit{algebraically solvable},
\textit{algebraically quasi-solvable} and \textit{algebraically
non-solvable} potentials.\\

Also we define in this section the concepts of
\textit{iso-galoisian}, \textit{virtually iso-galoisian} and
\textit{strong iso-galoisian} transformations. So, we give in
proposition 2.1.5 some conditions in which the transformation of
the second order linear differential equation into a reduced form
(the term in $\partial_x y$ is absent) is strong iso-galoisian or
virtually strong iso-galoisian. In particular case, corollary
2.1.7 shows that the reduction of Sturm-Liouville problem is a
virtually strong iso-galoisian transformation.

\subsubsection*{Section 2.2} This section is devoted to the
analysis of supersymmetric quantum mechanics from a differential
Galois theory point of view when $V$ is a rational function. We
start considering the polynomial case proving that the only one
possibility to get algebraically solvable potential is when $V$ is
a polynomial of degree 2. In case of algebraically quasi-solvable
potentials are presented as examples the well known quartic and
sextic anharmonic oscillators. Some results such as lemma 2.2.1
and theorem 2.2.2 were previously
published in \cite{acbl} and used in \cite{acbl2}.\\

Kovacic's algorithm is applied to solve the some Schr\"odinger
equations with rational potential (shape invariant potentials)
obtaining also their differential differential Galois groups and
eigenrings. In proposition 2.2.6 and corollary 2.2.7 is shown that
the Schr\"odinger equation with rational potential does not fall
in case 3 of Kovacic's algorithm.\\

Darboux transformation and Crum iteration are written in a
Galoisian sense (theorem 2.2.8, proposition 2.2.11 and proposition
2.2.12), so that by proposition 2.2.9 and proposition 2.2.10, such
transformations can be seen as isogaloisian and for instance they
preserves the Eigenrings. We prove in proposition 2.2.13 that the
supersymmetric partner potentials and the superpotential are
rational functions and in particular, corollary 2.2.14 says that
the Darboux transformation is strong iso-galoisian when the
superpotential is a rational function.\\

Finally, we define, in the context of differential Galois theory,
the concept of shape invariant potentials giving an algorithm to
check whether a potential satisfy the shape invariance condition
(remark 2.2.16) and providing an important result (theorem 2.2.17)
in where the Schr\"odinger equations obtained with the procedure
of shape invariant potentials preserves the differential Galois
groups and the Eigenrings.

\subsubsection*{Section 2.3} This section is devoted to the
supersymmetric quantum mechanics with non-rational potentials. We
start considering the proper pullback between two second order
linear differential equations in which one of them has rational
coefficients (\textit{algebrization process} when the first
differential equation with non-rational coefficients is
transformed into a differential equation with rational
coefficients). Proposition 2.3.2 shows that the change of
independent variable preserves the connected identity component of
the differential Galois group and give one formula of the
transformed equation (see also \cite{acbl}).\\

In proposition 2.3.3, are presented the relationships between the
differential fields, Picard-Vessiot extensions and differential
Galois groups corresponding to this change of independent
variable. Using these propositions, in remark 2.3.4 we introduce
the concept of \textit{Hard algebrization} giving an algorithm to
obtain it whether is possible. Proposition 2.3.5, which also
appears in \cite{acbl}, shows the algebrization of some families
of second order differential equations in where the differential
field is given by $\mathbb{C}(x,e^{\int f})$, being
$f\in\mathbb{C}(x)$.\\

The definition of \textit{Hamiltonian change of variable} is
introduced, leading to the concept of \textit{Hamiltonian
algebrization}, which allows to arrive to one systematic
algebrization of second order linear differential equations
presented in proposition 2.3.9, remark 2.3.10 and proposition
2.3.11 (these results also can be found in \cite{acbl} and were
applied in \cite{ac,acalde}). We show through examples that the
algebrization process can be used as transformation between
differential equations without expecting to obtain differential
equations with rational coefficients.\\

 We introduce a new
derivation $\widehat{\partial}_z=\sqrt{\alpha}\partial_z$, in
where $z=z(x)$ is a Hamiltonian change of variable and
$\partial_xz=\sqrt{\alpha}$. The most important theoretical result
here is that the differential field
$K=\mathbb{C}(z(x),\partial_xz(x))$ is isomorphic to the
differential field $\widehat{K}=\mathbb{C}(z,\sqrt{\alpha})\supset
\mathbb{C}(z)$, which allows to preserve the differential Galois
groups and Eigenrings in the algebrization process (theorem 2.3.13
and proposition 2.3.16). So, we transform a lot of differential
equations with non-rational coefficients into differential
equations with rational coefficients (Riccati, systems, etc...)
and in the case of second order linear differential equations,
Kovacic's algorithm can be used successfully.\\

We recover the results of section 2.2 rewriting everything with
$\widehat{}$ (hat over the symbols used in section 2.2), in this
way the \textit{algebrized supersymmetric quantum mechanics} is
presented with its elements: \textit{algebrized Schr\"odinger
operator} $\widehat{H}$, \textit{algebrized superpotential}
$\widehat{W}$, \textit{algebrized supersymmetric partner
potentials} $\widehat{V}_\pm$, \textit{algebrized shape invariant
potentials}, \textit{algebrized Darboux transformation}
$\widehat{\mathrm{DT}}$, \textit{algebrized Crum iteration}
$\widehat{\mathrm{CI}}_n$, \textit{algebrized ladder operators}
$\widehat{A}, \widehat{A}^\dagger$, \textit{algebrized wave
functions} $\widehat{\Psi}$, etc.. An important fact is given in
theorem 2.3.20, i.e.,
$\mathrm{DT}\varphi=\varphi\widehat{\mathrm{DT}}$, where $\varphi$
is the Hamiltonian algebrization.\\

Using the Hamiltonian algebrization and Kovacic's algorithm we
solve some Schr\"odinger equations with non-rational potentials
(shape invariant potentials), obtaining the differential Galois
groups and the Eigenrings of these differential equations.\\

Finally, we give a mechanism to search new exactly solvable
potentials through the inverse process of the Hamiltonian
algebrization, using known parameterized differential equations.

\section*{Historical Outline}

This historical outline begins with two mathematicians: Gast\'on
Darboux and Emile Picard. Darboux published in 1882 the paper
\cite{da1} in where he presents a proposition in a general way,
which in particular case the history proved to be a notable
theorem today known as \emph{Darboux transformation}. Darboux had
shown that whenever one knows to integrate the equation
$$\partial_x^2y = (f(x) + m)y$$ for all the values of the
constant $m$, one can obtain an infinite set of equations,
displaying the variable parameter in the same way, which are
integrable for any value of the parameter. This proposition also
can be found in his book \cite[p. 210]{da2}.\\

One year after, in 1883, Picard published the paper \cite{pi3} in
which he gave the starting point to a \emph{Galois theory for
linear differential equations}. Although the analogies between the
linear differential equations and the algebraic equations for a
long time were announced and continued in different directions,
Picard developed an analogue theory to the Galois theory for
algebraic equations, arriving to a proposition which seems to
correspond to the fundamental Galois theorem, in where he
introduces the concept of \emph{group of linear transformations
corresponding to the linear differential equation}, which today is
known as \emph{Differential Galois Group} (the group of
differential automorphism leaving fixed the elements of the field
base). Another contribution of Picard to this Galois theory was
the paper \cite{pi2} in 1887.\\

Five years after, in 1892, Ernest Vessiot, doctoral former student
of Picard, published his thesis \cite{ve} giving consolidation to
the new Galois theory for linear differential equations, the
so-called \emph{Picard-Vessiot theory}. Two years later, in 1894,
Picard published the paper \cite{pi4}, summarizing the results
presented in \cite{pi2,pi3,ve} which also can be found in his book
\cite[\S
7 ]{pi}.\\

Curiously, Picard-Vessiot theory and Darboux transformation were
forgotten during decades. The Picard-Vessiot theory was recovered
by Joseph Fels Ritt (in 1950, see \cite{ri}), Irving Kaplansky (in
1957, see \cite{ka}), and fundamentally by Ellis Kolchin (in 1948,
see \cite{kol} and references therein). Kolchin wrote the
\emph{Differential Galois Theory} in a modern language
(algebraic group theory).\\

Darboux transformation was presented as an exercise in 1926 by
Ince (see exercises 5, 6 and 7 \cite[p. 132]{in}), which follows
closely
the formulation of Darboux given in \cite{da1,da2}.\\

In 1930, P. Dirac publishes \textit{The Principles of Quantum Mechanics}, in where he gave a mathematically rigorous formulation of quantum mechanics.\\

 In 1938, J. Delsarte wrote the
paper \cite{de}, in which he introduced the notion of
transformation (transmutation) operator, today know as
\emph{intertwining operator} which is closely
related with Darboux transformation and ladder operators.\\

In 1941, E. Schr\"odinger published the paper \cite{schr} in which
he factorized in several ways the hypergeometric equation. This
was a byproduct of his \textit{factorization method} originating
an approach that can be traced back to Dirac's raising and
lowering operators for the harmonic oscillator.\\

Ten years later, in 1951, another factorization method was
presented. L. Infeld and T. E. Hull published the paper
\cite{inhu} in where they gave the classification of their
factorizations of linear second order differential equations for
eigenvalue problems of wave mechanics.\\

In 1955, M.M. Crum inspired in the Liouville's work about
Sturm-Liouville systems (see \cite{lio,lio2}), published the paper
\cite{cr} giving one kind of iterative generalization of Darboux
transformation. Crum surprisingly did not mention Darboux.\\

In 1971, G.A. Natanzon published the paper  \cite{nat}, in which
he studied a general form of the transformation that converts the
hypergeometric equation to the Schr\"odinger equation writing down
the most general \emph{solvable potential}, potential for which
the Schr\"odinger equation can be reduced to hypergeometric or
confluent hypergeometric form, a concept introduced
by himself.\\

Almost one hundred years later than Darboux's proposition, in
1981, Edward Witten in his renowned paper \cite{wi} gave birth to
the \emph{Supersymmetric Quantum Mechanics}, discussing general
conditions for dynamical supersymmetry breaking.\\

Since the work of Witten, thousands of papers, about
supersymmetric quantum mechanics, has been written. We mention here some relevant papers.\\

In 1983, L. \'E. Gendenshtein published the paper \cite{ge} in
where the \emph{Shape invariance} condition, i.e. preserving the
shape under Darboux transformation, was presented and used to find
the complete spectra for a broad class of problems including all
known exactly solvable problems of quantum mechanics (bound state
and reflectionless potentials). Today this kind of exactly
solvable potentials satisfying the shape invariance condition are
called \emph{Shape invariant
potentials}.\\

In 1986, A. Turbiner in \cite{tu} introduces the concept of
\textit{quasi-exactly solvable potentials}, giving an example that
is well known as \textit{Turbiner's potential.}\\

 In 1991, V.B. Matveev and M. Salle published the book
\cite{masa} in where they focused on Darboux transformations and
their relation with solitons. Matveev and Salle interpreted the
Darboux transformation as Darboux covariance of a Sturm-Liouville
problem and also proved that Witten's supersymetric quantum
mechanics is equivalent to a single
Darboux transformation.\\

In 1996, C. Bender and G. Dunne studied the \textit{sextic
anharmonic oscillator} in \cite{bedu}, which is a quasi-exactly
solvable model derived from the Turbiner's potentials. They found
that a portion of the spectrum correspond to the roots of
polynomials in the energy. These polynomials are orthogonal and
are called \textit{Bender-Dunne polynomials}.\\

Relationships between the spectral theory and differential Galois
theory have been studied by V. Spiridonov \cite{sp}, F. Beukers
\cite{be2} and Braverman et. al. \cite{bretga}. As far as we know,
Spiridonov was the first author that considered the usefulness of
the Picard-Vessiot theory in the context of the quantum mechanics.
This thesis agrees with his point of view. \\

\chapter{Theoretical Background}
\label{background}

 In this chapter we set the main theoretical background
 needed to understand the results of this thesis. We start setting conventions and notations that will be used along this work.
\begin{itemize}
\item The sets $\mathbb{Z}_+$, $\mathbb{Z}_-$, $\mathbb{Z}^*_+$ and $\mathbb{Z}^*_-$ are defined as $$\mathbb{Z}_+=\{n\in\mathbb{Z}:\quad n\geq 0\},\quad
\mathbb{Z}_-=\{n\in\mathbb{Z}:\quad n\leq 0\},\quad
\mathbb{Z}_+^*=\mathbb{Z}^+,\quad \mathbb{Z}^{*}_{-}
=\mathbb{Z}^-.$$

\item The cardinality of the set $A$ will be denoted by
$\mathrm{Card}(A)$.

\item The determinant of the matrix $A$ will be denoted by
$\det A$.

\item The set of matrices $n\times n$ with entries in $\mathbb{C}$ and determinant
non-null, the general linear group over $\mathbb{C}$, will be
denoted by $\mathrm{GL}(n,\mathbb{C})$.

\item The derivation $d/d\xi$ will be denoted by $\partial_\xi$.
For example, the derivations $'=d/dx$ and $\dot{} = d/dt$ are
denoted by $\partial_x$ and $\partial_t$ respectively.

\end{itemize}

\section{Picard-Vessiot theory}

Picard-Vessiot theory is the Galois theory of linear differential
equations. In the classical Galois theory, the main object is a
group of permutations of the roots, while in the Picard-Vessiot
theory it is a linear algebraic group. For polynomial equations we
want a solution in terms of radicals, which from classical Galois
theory it is well if the Galois group is a solvable group.

An analogous situation holds for linear homogeneous differential
equations (see \cite{be,crhamo,mo,vasi}). The following definition
is true in general dimension, but for simplicity we are
restricting to matrices $2\times 2.$
\subsection{Definitions and Known Results}
\begin{definition} An
algebraic group of matrices $2\times 2$ is a subgroup $G\subset
\mathrm{GL}(2,\mathbb{C})$, defined by algebraic equations in its
matrix elements and in the inverse of its determinant. That is,
for $A\in\mathrm{GL}(2,\mathbb{C})$ given by
$$A=\left(\begin{array}{cc} x_{11} & x_{12} \\ x_{21} & x_{22}
\end{array}\right),\quad \det A=x_{11}x_{22}-x_{21}x_{22}$$
 there exists a set of polynomials
$$\{P_i(x_{11},x_{12},x_{21},x_{22},1/\det A)\}_{i\in I},$$
such that
$$A\in G \quad\Leftrightarrow\quad \forall i\in I,
P_i(x_{11},x_{12},x_{21},x_{22},1/\det A) = 0.$$
\end{definition}
In this case we say that $G$ is an algebraic manifold endowed with
a group structure.
\begin{examples}[Known algebraic groups] The following algebraic
groups should be kept in mind throughout this work.
\begin{itemize}
\item Special linear group group:\\
 $\mathrm{SL}(2,\mathbb{C})=\left\{\begin{pmatrix}a&b\\c&d\end{pmatrix},\quad ad-bc=1, \quad a,b,c,d\in\mathbb{C}\right\}$
\item Borel group: $\mathbb{B}=\mathbb{C}^*\ltimes\mathbb{C}=\left\{\begin{pmatrix}c&d\\0&c^{-1}\end{pmatrix},\quad c\in\mathbb{C}^*, \quad d\in\mathbb{C}\right\}$
\item Multiplicative group: $\mathbb{G}_m=\left\{\begin{pmatrix}c&0\\0&c^{-1}\end{pmatrix},\quad c\in\mathbb{C}^*\right\}$
\item Additive group: $\mathbb{G}_a=\left\{\begin{pmatrix}1&d\\0&1\end{pmatrix},\quad d\in\mathbb{C}\right\}$
\item Infinite dihedral group (also called meta-abelian group):\\
 $\mathbb{D}_{\infty}=\left\{\begin{pmatrix}c&0\\0&c^{-1}\end{pmatrix},\quad c\in\mathbb{C}^*\right\}\cup\left\{\begin{pmatrix}0&d\\-d^{-1}&0\end{pmatrix},\quad d\in\mathbb{C}^*\right\}$
\item $n-$quasi-roots: $\mathbb{G}^{\{n\}}=\left\{\begin{pmatrix}c&d\\0&c^{-1}\end{pmatrix},\quad c^n=1, \quad d\in\mathbb{C}\right\}$
\item $n-$roots: $\mathbb{G}^{[n]}=\left\{\begin{pmatrix}c&0\\0&c^{-1}\end{pmatrix},\quad c^n=1\right\}$
\item Identity group: $e=\left\{\begin{pmatrix}1&0\\0&1\end{pmatrix}\right\}$
\item The tetrahedral group $A_4^{\mathrm{SL}_2}$ of order $24$
is generated by matrices
$$M_1=\begin{pmatrix}\xi&0\\0&\xi^{-1}\end{pmatrix}\quad \textrm{and}\quad
M_2=\frac13(2\xi-1)\begin{pmatrix}1&1\\2&-1\end{pmatrix},$$ where
$\xi$ denotes a primitive sixth root of unity, that is,
$\xi^2-\xi+1=0$.
\item The octahedral group $S_4^{\mathrm{SL}_2}$ of order $48$
is generated by matrices
$$M_1=\begin{pmatrix}\xi&0\\0&\xi^{-1}\end{pmatrix}\quad \textrm{and}\quad
M_2=\frac12\xi(\xi^2+1)\begin{pmatrix}1&1\\1&-1\end{pmatrix},$$
where $\xi$ denotes a primitive eighth root of unity, that is,
$\xi^4+1=0$.
\item The icosahedral group $A_5^{\mathrm{SL}_2}$ of order $120$
is generated by matrices
$$M_1=\begin{pmatrix}\xi&0\\0&\xi^{-1}\end{pmatrix}\quad \textrm{and}\quad
M_2=\frac15\begin{pmatrix}\phi&\psi\\\psi&-\phi\end{pmatrix},$$
where $\xi$ denotes a primitive tenth root of unity, that is,
$\xi^4-\xi^3+\xi^2-\xi+1=0$, $\phi=\xi^3-\xi^2+4\xi-2$ and
$\psi=\xi^3+3\xi^2-2\xi+1$.
\end{itemize}
\end{examples}

Recall that a group $G$ is called solvable if and only if there
exists a chain of normal subgroups
$$e=G_0\triangleleft G_1 \triangleleft \ldots \triangleleft G_n=G$$ such that the
quotient $G_i/G_j$ is abelian for all $n\geq i\geq j\geq 0$. Also
recall that an algebraic group $G$ has a unique connected normal
algebraic subgroup $G^0$ of finite index. This means that the
identity connected component $G^0$ is the largest connected
algebraic subgroup of $G$ containing the identity. For instance,
if $G=G^0$, we say that $G$ is a \textit{connected group}.

Furthermore, if $G^0$ satisfy some property, then we say that $G$
virtually satisfy such property. In this way, virtually
solvability of $G$ means solvability of $G^0$ and virtually
abelianity of $G$ means abelianity of $G^0$ (see \cite{we2}).

\begin{theorem}[Lie-Kolchin]\label{LiKo}
Let $G\subseteq \mathrm{GL}(2,\mathbb{C})$ be a virtually solvable
group. Then $G^0$ is triangularizable, that is, conjugate to a
subgroup of upper triangular matrices.
\end{theorem}

\begin{definition}
Let $G\subseteq \mathrm{GL}(2,\mathbb{C})$ be a group acting on a
vector space $V$. We say that (the action of) $G$ is either:
\begin{enumerate}
\item \textit{Reducible}, if there exists a non-trivial subspace $W\subset V$
such that $G(W)\subset W$. We say that $G$ is \textit{irreducible}
if $G$ is not reducible.
\item \textit{Imprimitive}, if $G$ is irreducible and there exists
subspaces $V_i$ such that $V=V_1\otimes\cdots\otimes V_m$, where
$G$ permutes transitively the $V_i$, i.e $\forall i=1,\ldots,m$,
$\forall g\in G$, $\exists j\in\{1,\dots,m\}$ such that
$g(V_i)=V_j$. We say that $V_1,\ldots, V_m$ form a system of
\textit{imprimitivity} for $G$.

\item \textit{Primitive}, if $G$ is irreducible and not imprimitive.

\end{enumerate}
\end{definition}
\begin{examples}
Any subgroup of the Borel group is reducible, the infinite
dihedral group is imprimitive and the groups $A^{\mathrm{SL}_2}$,
$S^{\mathrm{SL}_2}_4$, $A_5^{\mathrm{SL}_2}$,
$\mathrm{SL}(2,\mathbb{C})$ are primitives (see \cite{vasi,we2}).
\end{examples}
\begin{definition}[Differential Fields]\label{defdiff}
Let $K$ (depending on $x$) be a commutative field of
characteristic zero, $\partial_x$ a derivation, that is, a map
$\partial_x : K\rightarrow K$ satisfying $\partial_x (a + b) =
\partial_x a + \partial_x b$ and $\partial_x(ab) = \partial_x a \cdot b + a
\cdot\partial_x b$ for all $a,b\in K$. By $\mathcal C$ we denote
the field of constants of $K$ $$\mathcal C = \{c\in K |
\partial_x c = 0\}$$ which is also of characteristic zero and will be assumed algebraically closed. In this terms,
we say that $K$ is a {\it{differential field}} with the derivation
$\partial_x$.
\end{definition}

Along this work, up to some specifications, \textit{we consider as
differential field the smallest differential containing the
coefficients}. Furthermore, up to special considerations, we
analyze second order linear homogeneous differential equations,
that is, equations in the form
\begin{equation}\label{soldeq}
\mathcal L:= \partial^2_xy+a\partial_x y+by=0,\quad a,b\in K.
\end{equation}

\begin{definition}[Picard-Vessiot Extension] Suppose that $y_1, y_2$ is a basis of solutions of $\mathcal L$ given in equation \eqref{soldeq}, i.e., $y_1, y_2$ are linearly
independent over $K$ and every solution is a linear combination
over $\mathcal{C}$ of these two. Let $L= K\langle y_1, y_2 \rangle
=K(y_1, y_2, \partial_xy_1, \partial_xy_2)$ the differential
extension of $K$ such that $\mathcal C$ is the field of constants
for $K$ and $L$. In this terms, we say that $L$, the smallest
differential field containing $K$ and $\{y_{1},y_{2}\}$, is the
\textit{Picard-Vessiot extension} of $K$ for $\mathcal L$.
\end{definition}

\begin{definition}[Differential Galois Groups]
Assume $K$, $L$ and $\mathcal L$ as in previous definition. The
group of all differential automorphisms (automorphisms that
commutes with derivation) of $L$ over $K$ is called the
{\it{differential Galois group}} of $L$ over $K$ and is denoted by
${\rm DGal}(L/K)$. This means that for $\sigma\in
\mathrm{DGal}(L/K)$, $\sigma(\partial_xa)=\partial_x(\sigma(a))$
for all $a\in L$ and $\forall a\in K,$ $\sigma(a)=a$.
\end{definition}
Assume that $\{y_1,y_2\}$ is a fundamental system of solutions
(basis of solutions) of $\mathcal L$. If $\sigma \in
\mathrm{DGal}(L/K)$ then $\{\sigma y_1, \sigma y_2\}$ is another
fundamental system of $\mathcal L$. Hence there exists a matrix

$$A_\sigma=
\begin{pmatrix}
a & b\\
c & d
\end{pmatrix}
\in \mathrm{GL}(2,\mathbb{C}),$$ such that
$$\sigma
\begin{pmatrix}
y_{1}\\
y_{2}
\end{pmatrix}
=
\begin{pmatrix}
\sigma (y_{1})\\
\sigma (y_{2})
\end{pmatrix}
=\begin{pmatrix} y_{1}& y_{2}
\end{pmatrix}A_\sigma,$$ in a natural way, we can extend to
systems:
$$\sigma
\begin{pmatrix}
y_{1}&y_2\\
\partial_xy_1&\partial_xy_{2}
\end{pmatrix}
=
\begin{pmatrix}
\sigma (y_{1})&\sigma (y_2)\\
\sigma (\partial_xy_1)&\sigma (\partial_xy_{2})
\end{pmatrix}
=\begin{pmatrix} y_{1}& y_{2}\\\partial_xy_1&\partial_xy_2
\end{pmatrix}A_\sigma.$$

This defines a faithful representation $\mathrm{DGal}(L/K)\to
\mathrm{GL}(2,\mathbb{C})$ and it is possible to consider
$\mathrm{DGal}(L/K)$ as a subgroup of $\mathrm{GL}(2,\mathbb{C})$.
It depends on the choice of the fundamental system $\{y_1,y_2\}$,
but only up to conjugacy.

One of the fundamental results of the Picard-Vessiot theory is the
following theorem (see \cite{ka,kol}).

\begin{theorem}  The differential Galois group $\mathrm{DGal}(L/K)$ is an
algebraic subgroup of $\mathrm{GL}(2,\mathbb{C})$.
\end{theorem}
\begin{examples}
 Consider the following differential equations:

\begin{itemize}
\item $\mathcal L:=\partial^2_xy=0$,
the basis of solutions is given by $y_{1}=1$, $y_2=x$. If we set
as differential field $K=\mathbb{C}(x)$, we can see that
$\sigma(1)=1$, $\sigma(x)=x$, then the Picard-Vessiot extension
$L=K$ and for instance $\mathrm{DGal}(L/K)=e$:
$$\sigma\begin{pmatrix}
  y_1 \\
  y_2
\end{pmatrix}=\begin{pmatrix}
  y_1 &  y_2
\end{pmatrix}\begin{pmatrix}
  1 & 0 \\
 0 & 1
\end{pmatrix}=\begin{pmatrix}
  y_1 \\
  y_2
\end{pmatrix}.$$
 Now, if we set $K=\mathbb{C}$, then
$L=K\langle x \rangle$, $\partial_xx\in\mathbb{C},$
$\partial_x(\sigma(x))=\sigma(\partial_xx)=\sigma(1)=1=\partial_xx$,
so $\sigma(x)=x+d$, $d\in\mathbb{C}$ and for instance
$\mathrm{DGal}(L/K)=\mathbb{G}_a$:
$$\sigma\begin{pmatrix}
  y_1 \\
  y_2
\end{pmatrix}=\begin{pmatrix}
  y_1 &
  y_2
\end{pmatrix}\begin{pmatrix}
  1 & d \\
 0 & 1
\end{pmatrix}=\begin{pmatrix}
  y_1 \\
 dy_1+ y_2
\end{pmatrix}.$$
\item $\mathcal L:=\partial^2_xy=\kappa y$, $\kappa\in\mathbb{C}^*$,
the basis of solutions is given by $y_{1}=e^{\sqrt \kappa x}$,
$y_2=e^{-\sqrt \kappa x}$, with $\kappa\neq 0$. If we set as
differential field $K=\mathbb{C}(x)$, we can see that $L=K\langle
e^{\sqrt k x}\rangle=K(e^{\sqrt k x})$,
 $$\sigma\left({\partial_xy_1\over y_1}\right)={\partial_x(\sigma(y_1))\over \sigma(y_1)}=
 {\partial_xy_1\over y_1},\quad \sigma\left({\partial_xy_2\over y_2}\right)={\partial_x(\sigma(y_2))\over \sigma(y_2)}={\partial_xy_2\over y_2},$$
 $\sigma(y_1y_2)=\sigma(y_1)\sigma(y_2)=y_1y_2=1,$
$\sigma(y_1)=cy_1$, $\sigma(y_2)=dy_2$, $c,d\in\mathbb{C}$, but
$cd=1$ and for instance $\mathrm{DGal}(L/K)=\mathbb{G}_m$:
$$\sigma\begin{pmatrix}
  y_1 \\
  y_2
\end{pmatrix}=\begin{pmatrix}y_1&y_2\end{pmatrix}\begin{pmatrix}
  c & 0 \\
  0 & c^{-1}
\end{pmatrix}=\begin{pmatrix}
  cy_1 \\
  c^{-1}y_2
\end{pmatrix},$$
Now, if we set $K=\mathbb{C}$, we obtain the same result.

\item $\mathcal L:=\partial_x^2y+\frac{n-1}{nx} \partial_xy=0,$ the basis of solutions is given by $y_{1}= z$, where $z^n=x$, $y_2=1$.
 If we set $K=\mathbb{C}(x)$, then
$L=K\left\langle x^{\frac1n} \right\rangle$, $y_1^n=
x\in\mathbb{C}(x),$ $\sigma^n(y_1)=\sigma(y_1^n)= x$,
$\sigma(y_1)=cy_1$, so that $c^n=1$ and for instance
$\mathrm{DGal}(L/K)$ is given by:
$$\sigma\begin{pmatrix}
  y_1 \\
  y_2
\end{pmatrix}=\begin{pmatrix}y_1&y_2\end{pmatrix}\begin{pmatrix}
  c & 0 \\
 0 & 1
\end{pmatrix}=\begin{pmatrix}
  cy_1 \\
  y_2
\end{pmatrix},\quad c^n=1.$$

\item $\mathcal L:=\partial_x^2y+\frac{n^2-1}{4n^2x^2} y=0,$ $n\in\mathbb{Z}$, the basis of solutions is given by
 $y_{1}= x^{n+1\over 2n}$, $y_2=x^{n-1\over 2n}$.
If we set $K=\mathbb{C}(x)$ and $n$ even, then $L=K\left\langle
x^{\frac1{2n}} \right\rangle,$ $$ \sigma(y_1)=cy_1, \quad
\sigma^{2n}(y_1)=c^{2n}y_1^{2n}=\sigma(y_1^{2n})= y_1^{2n},\quad
c^{2n}=1,$$
$$\sigma(y_2)=dy_2, \quad \sigma^{2n}(y_2)=d^{2n}y_2^{2n}=\sigma(y_2^{2n})=
y_2^{2n},\quad d^{2n}=1 ,$$
$\sigma(y_1y_2)=y_1y_2=\sigma(y_1)\sigma(y_2)=cdy_1y_2$ so that
$cd=1$ and for instance $\mathrm{DGal}(L/K)=\mathbb{G}^{[2n]}$:
$$\sigma\begin{pmatrix}
  y_1 \\
  y_2
\end{pmatrix}=\begin{pmatrix}y_1&y_2\end{pmatrix}\begin{pmatrix}
  c & 0 \\
 0 & c^{-1}
\end{pmatrix}=\begin{pmatrix}
  cy_1 \\
  c^{-1}y_2
\end{pmatrix},\quad c^{2n}=1,\quad n>1.$$
Now, if we consider $n$ odd, then $L=K\left\langle x^{\frac1{n}}
\right\rangle,$ and $\mathrm{DGal}(L/K)=\mathbb{G}^{[2n]}$.
\item Cauchy-Euler equation
$$\mathcal L:=\partial_x^2y={m(m+1)\over x^2}y,\quad m\in\mathbb{C},$$
the basis of solutions is $y_1=x^{m+1}$, $y_2=x^{-m}$. Setting
$K=\mathbb{C}(x)$, we have the following possible cases:

\begin{itemize}
\item for $m\in\mathbb{Z}$, $L=K$ and $\mathrm{DGal}(L/K)=e$,
\item for $m\in\mathbb{Q}\setminus\mathbb{Z}$,
$L=K(x^m)$ and $\mathrm{DGal}(L/K)=\mathbb{G}^{[d]}$, where
$m=n/d$,
\item for $m\in\mathbb{C}\setminus\mathbb{Q}$,
$L=K(x^m)$ and $\mathrm{DGal}(L/K)=\mathbb{G}_m$.
\end{itemize}
\end{itemize}
\end{examples}

\begin{definition}[Integrability]\label{integrability} Consider the linear differential equation
$\mathcal L$ such as in equation \eqref{soldeq}. We say that
$\mathcal L$ is \textit{integrable} if the Picard-Vessiot
extension $L\supset K$ is obtained as a tower of differential
fields $K=L_0\subset L_1\subset\cdots\subset L_m=L$ such that
$L_i=L_{i-1}(\eta)$ for $i=1,\ldots,m$, where either
\begin{enumerate}
\item $\eta$ is {\emph{algebraic}} over $L_{i-1}$, that is $\eta$ satisfies a
polynomial equation with coefficients in $L_{i-1}$.
\item $\eta$ is {\emph{primitive}} over $L_{i-1}$, that is $\partial_x\eta \in L_{i-1}$.
\item $\eta$ is {\emph{exponential}} over $L_{i-1}$, that is $\partial_x\eta /\eta \in L_{i-1}$.
\end{enumerate}
\end{definition}
We recall that the differential field of coefficients has been
fixed before, i.e., the smallest differential field containing the
coefficients.\\

We remark that the usual terminology in differential algebra for
integrable equations is that the corresponding Picard-Vessiot
extensions are called \textit{Liouvillian}.\\

\begin{theorem}[Kolchin]\label{LK}
The equation $\mathcal L$ given in \eqref{soldeq} is integrable if
and only if $\mathrm{DGal}(L/K)$ is virtually solvable.
\end{theorem}

Consider the differential equation
\begin{equation}\label{LDE}
\mathcal L:=\partial_x^2\zeta=r\zeta,\quad r\in K.
\end{equation}

We recall that equation \eqref{LDE} can be obtained from equation
\eqref{soldeq} through the change of variable
\begin{equation}\label{redsec}
y=e^{-{1\over 2}\int_{}^{}a}\zeta,\quad r={a^2\over
4}+{\partial_xa\over 2}-b
\end{equation} and equation \eqref{LDE} is called the \textit{reduced
form} of equation \eqref{soldeq}.

On the other hand, introducing the change of variable
$v=\partial_x\zeta/\zeta$ we get the associated Riccati equation
to equation (\ref{LDE})
\begin{equation}\label{Riccatti}
\partial_xv=r-v^2,\quad v={\partial_x\zeta\over \zeta},
\end{equation}
where $r$ is obtained by equation (\ref{redsec}).

\begin{theorem}[Singer 1981, \cite{sil}]\label{singer}
The Riccatti equation \eqref{Riccatti} has one algebraic solution
over the differential field $K$ if and only if the differential
equation \eqref{LDE} is integrable.
\end{theorem}

For $\mathcal L$ given by equation
 (\ref{LDE}), it is very well
known (see \cite{ka,kol,vasi}) that ${\rm DGal}_K(\mathcal L)$ is
an algebraic subgroup of ${\rm SL}(2,\mathbb{C})$. The well known
classification of subgroups of $\mathrm{SL}(2,\mathbb{C})$ (see
\cite[p.31]{ka}, \cite[p.7,27]{ko}) is the following.

\begin{theorem}\label{subgroups} Let $G$ be an algebraic subgroup of ${\rm SL}(2,\mathbb{C})$.
Then, up to conjugation, one of the following cases occurs.
\begin{enumerate}
\item $G\subseteq \mathbb{B}$ and then $G$ is reducible and
triangularizable.
\item $G\nsubseteq\mathbb{B}$, $G\subseteq \mathbb{D}_\infty$ and then $G$ is imprimitive.
\item $G\in\{A_4^{\mathrm{SL}_2},S_4^{\mathrm{SL}_2},A_5^{\mathrm{SL}_2}\}$ and then $G$ is primitive (finite)
\item $G = {\rm SL}(2,\mathbb{C})$ and then $G$ is primitive (infinite).
\end{enumerate}
\end{theorem}

\begin{definition}  Consider the differential equation $\mathcal L$ given by equation \eqref{LDE}.
 Let $\{\zeta_1, \zeta_2\}$ be a fundamental system of $\mathcal L$.  Let
$f=f(Y_1,Y_2)\in\mathcal{C}[Y_1,Y_2]$ be a homogeneous polynomial,
we say that:
\begin{enumerate}
\item The polynomial $f$ is an {\em invariant} with respect to $\mathcal L$ if its evaluation on a $\mathcal{C}$-basis
$\{\zeta_1,\zeta_2\}$ of solutions is invariant under the action
of $\mathrm{DGal}(L/K)$, that is, for every $\sigma \in
\mathrm{DGal}(L/K)$, $\sigma h(x) = h(x)$, where
$h(x)=f(\zeta_1(x),\zeta_2(x))\in K$. The function $h(x)$ is
called the \emph{value} of the invariant polynomial $f$.
\item The polynomial $f$ is a {\em semi-invariant} with respect to
$\mathcal L$ if the logarithmic derivative ${\partial_xh\over h}$
of its evaluation $h(x)=f(\zeta_1(x),\zeta_2(x))$ on any
$\mathcal{C}$-basis $\{\zeta_1,\zeta_2\}$ is an element of $K$,
that is, for every $\sigma \in \mathrm{DGal}(L/K)$, $\sigma \theta
= \theta$, where $\theta=\partial_xh(x)/h(x)\in K$.
\end{enumerate}
\end{definition}

\begin{theorem}[Kovacic, \cite{ko}]\label{invariants} Let $\{\zeta_1,\zeta_2\}$ be a fundamental system of solutions
of $\mathcal L$ given by the differential equation \eqref{LDE}.
Then, for some $i\in\{1,2\}$ and for every $\sigma \in
\mathrm{DGal}(L/K)$, exclusively one of the following cases holds.
\begin{enumerate}
\item $\mathrm{DGal}(L/K)$ is reducible and then $f=Y_i$ is
semi-invariant with respect to $\mathcal L$, i.e $\partial_x(\ln
\zeta_i)\in K$.

\item $\mathrm{DGal}(L/K)$ is imprimitive and then $f_1=\zeta_1\zeta_2$ is semi-invariant with respect to $\mathcal L$,
 $f_2=(Y_1Y_2)^2$ is invariant with respect to $\mathcal L$, i.e $[K\langle\partial_x(\ln \zeta_i)\rangle:K]=2$.

\item $\mathrm{DGal}(L/K)$ is finite primitive and then the invariants with respect to $\mathcal L$ is either $f_1=(Y_1^4 +
8Y_1Y_2^3)^3$, or $f_2=(Y_1^5Y_2 - Y_1Y_2^5)^2$ or
$f_3=Y_1^{11}Y_2 - 11Y_1^6Y_2^6 - Y_1Y_2^{11}$, i.e
$[K\langle\partial_x(\ln \zeta_i)\rangle:K]=4,6,12$ .

\item  $\mathrm{DGal}(L/K)$ is infinite primitive, i.e there are no non-trivial semi-invariants.
\end{enumerate}
\end{theorem}
Statements and proofs of theorems \ref{LK}, \ref{singer} and
\ref{subgroups} can be found in \cite{vasi}.

\subsection{Kovacic's Algorithm} Considering $K=\mathbb{C}(x)$, $\mathcal C=\mathbb{C}$ in theorems \ref{LK}, \ref{subgroups}
 and \ref{invariants}, Kovacic in 1986 (\cite{ko}) introduced an
algorithm to solve the differential equation (\ref{LDE}) showing
that (\ref{LDE}) is integrable if and only if the solution of the
Riccati equation (\ref{Riccatti}) is a rational function (case 1),
is a root of polynomial of degree two (case 2) or is a root of
polynomial of degree 4, 6, or 12 (case 3). For more details see
reference \cite{ko}. Improvements for this algorithm are given in
references \cite{dulo,fa,ulwe}. Here, we follow the original
version given by Kovacic in reference \cite{ko} with an adapted
version presented in reference \cite{acbl}.
\\

Each case in Kovacic's algorithm is related with each one of the
algebraic subgroups of ${\rm SL}(2,\mathbb{C})$ and the associated
Riccatti equation
$$\partial_xv=r-v^{2}=\left( \sqrt{r}-v\right)
\left(  \sqrt{r}+v\right),\quad v={\partial_x\zeta\over \zeta}.$$

According to Theorem \ref{subgroups}, there are four cases in
Kovacic's algorithm. Only for cases 1, 2 and 3 we can solve the
differential equation, but for the case 4 the differential
equation is not integrable. It is possible that Kovacic's
algorithm can provide us only one solution ($\zeta_1$), so that we
can obtain the second solution ($\zeta_2$) through
\begin{equation}\label{second}
\zeta_2=\zeta_1\int\frac{dx}{\zeta_1^2}.
\end{equation}

{\bf\large Notations.} For the differential equation given by
$$\partial_x^2\zeta=r\zeta,\qquad r={s\over t},\quad s,t\in \mathbb{C}[x],$$
we use the following notations.
\begin{enumerate}
\item Denote by $\Gamma'$ be
the
set of (finite) poles of $r$, $\Gamma^{\prime}=\left\{  c\in\mathbb{C}%
:t(c)=0\right\}$.

\item Denote by
$\Gamma=\Gamma^{\prime}\cup\{\infty\}$.
\item By the order of $r$ at
$c\in \Gamma'$, $\circ(r_c)$, we mean the multiplicity of $c$ as a
pole of $r$.

\item By the order of $r$ at $\infty$, $\circ\left(
r_{\infty}\right) ,$ we mean the order of $\infty$ as a zero of
$r$. That is $\circ\left( r_{\infty }\right)
=\mathrm{deg}(t)-\mathrm{deg}(s)$.

\end{enumerate}
\textbf{The four cases}
\\

{\bf\large Case 1.} In this case $\left[ \sqrt{r}\right] _{c}$ and
$\left[ \sqrt{r}\right] _{\infty}$ means the Laurent series of
$\sqrt{r}$ at $c$ and the Laurent series of $\sqrt{r}$ at $\infty$
respectively. Furthermore, we define $\varepsilon(p)$ as follows:
if $p\in\Gamma,$ then $\varepsilon\left( p\right) \in\{+,-\}.$
Finally, the complex numbers $\alpha_{c}^{+},\alpha_{c}^{-},\alpha_{\infty}%
^{+},\alpha_{\infty}^{-}$ will be defined in the first step. If
the differential equation has no poles it only can fall in this
case.
\medskip

{\bf Step 1.} Search for each $c \in \Gamma'$ and for $\infty$ the
corresponding situation as follows:

\medskip

\begin{description}

\item[$(c_{0})$] If $\circ\left(  r_{c}\right)  =0$, then
$$\left[ \sqrt {r}\right] _{c}=0,\quad\alpha_{c}^{\pm}=0.$$

\item[$(c_{1})$] If $\circ\left(  r_{c}\right)  =1$, then
$$\left[ \sqrt {r}\right] _{c}=0,\quad\alpha_{c}^{\pm}=1.$$

\item[$(c_{2})$] If $\circ\left(  r_{c}\right)  =2,$ and $$r= \cdots
+ b(x-c)^{-2}+\cdots,\quad \textrm{then}$$
$$\left[ \sqrt {r}\right]_{c}=0,\quad \alpha_{c}^{\pm}=\frac{1\pm\sqrt{1+4b}}{2}.$$

\item[$(c_{3})$] If $\circ\left(  r_{c}\right)  =2v\geq4$, and $$r=
(a\left( x-c\right)  ^{-v}+...+d\left( x-c\right)
^{-2})^{2}+b(x-c)^{-(v+1)}+\cdots,\quad \textrm{then}$$ $$\left[
\sqrt {r}\right] _{c}=a\left( x-c\right) ^{-v}+...+d\left(
x-c\right) ^{-2},\quad\alpha_{c}^{\pm}=\frac{1}{2}\left(
\pm\frac{b}{a}+v\right).$$

\item[$(\infty_{1})$] If $\circ\left(  r_{\infty}\right)  >2$, then
$$\left[\sqrt{r}\right]  _{\infty}=0,\quad\alpha_{\infty}^{+}=0,\quad\alpha_{\infty}^{-}=1.$$

\item[$(\infty_{2})$] If $\circ\left(  r_{\infty}\right)  =2,$ and
$r= \cdots + bx^{2}+\cdots$, then $$\left[
\sqrt{r}\right]  _{\infty}=0,\quad\alpha_{\infty}^{\pm}=\frac{1\pm\sqrt{1+4b}%
}{2}.$$

\item[$(\infty_{3})$] If $\circ\left(  r_{\infty}\right) =-2v\leq0$,
and
$$r=\left( ax^{v}+...+d\right)  ^{2}+ bx^{v-1}+\cdots,\quad \textrm{then}$$
$$\left[  \sqrt{r}\right]  _{\infty}=ax^{v}+...+d,\quad
and\quad \alpha_{\infty}^{\pm }=\frac{1}{2}\left(
\pm\frac{b}{a}-v\right).$$
\end{description}
\medskip

{\bf Step 2.} Find $D\neq\emptyset$ defined by
$$D=\left\{
n\in\mathbb{Z}_{+}:n=\alpha_{\infty}^{\varepsilon
(\infty)}-%
{\displaystyle\sum\limits_{c\in\Gamma^{\prime}}}
\alpha_{c}^{\varepsilon(c)},\forall\left(  \varepsilon\left(
p\right) \right)  _{p\in\Gamma}\right\}  .$$ If $D=\emptyset$,
then we should start with the case 2. Now, if
$\mathrm{Card}(D)>0$, then for each $n\in D$ we search $\omega$
$\in\mathbb{C}(x)$ such that
$$\omega=\varepsilon\left(
\infty\right)  \left[  \sqrt{r}\right]  _{\infty}+%
{\displaystyle\sum\limits_{c\in\Gamma^{\prime}}}
\left(  \varepsilon\left(  c\right)  \left[  \sqrt{r}\right]  _{c}%
+{\alpha_{c}^{\varepsilon(c)}}{(x-c)^{-1}}\right).$$
\medskip

{\bf Step 3}. For each $n\in D$, search for a monic polynomial
$P_n$ of degree $n$ with
\begin{equation}\label{recu1}
\partial_x^2P_n + 2\omega \partial_xP_n + (\partial_x\omega + \omega^2 - r) P_n = 0.
\end{equation}
If success is achieved then $\zeta_1=P_n e^{\int\omega}$ is a
solution of the differential equation.  Else, case 1 cannot hold.
\bigskip

{\bf\large Case 2.}  Search for each $c \in \Gamma'$ and for
$\infty$ the corresponding situation as follows:
\medskip

{\bf Step 1.} Search for each $c\in\Gamma^{\prime}$ and $\infty$
the sets $E_{c}\neq\emptyset$ and $E_{\infty}\neq\emptyset.$ For
each $c\in\Gamma^{\prime}$ and for $\infty$ we define
$E_{c}\subset\mathbb{Z}$ and $E_{\infty}\subset\mathbb{Z}$ as
follows:
\medskip

\begin{description}
\item[($c_1$)] If $\circ\left(  r_{c}\right)=1$, then $E_{c}=\{4\}$

\item[($c_2$)] If $\circ\left(  r_{c}\right)  =2,$ and $r= \cdots +
b(x-c)^{-2}+\cdots ,\ $ then $$E_{c}=\left\{
2+k\sqrt{1+4b}:k=0,\pm2\right\}.$$

\item[($c_3$)] If $\circ\left(  r_{c}\right)  =v>2$, then $E_{c}=\{v\}$

\item[$(\infty_{1})$] If $\circ\left(  r_{\infty}\right)  >2$, then
$E_{\infty }=\{0,2,4\}$

\item[$(\infty_{2})$] If $\circ\left(  r_{\infty}\right)  =2,$ and
$r= \cdots + bx^{2}+\cdots$, then $$E_{\infty }=\left\{
2+k\sqrt{1+4b}:k=0,\pm2\right\}.$$

\item[$(\infty_{3})$] If $\circ\left(  r_{\infty}\right)  =v<2$,
then $E_{\infty }=\{v\}$
\medskip
\end{description}

{\bf Step 2.} Find $D\neq\emptyset$ defined by
$$D=\left\{
n\in\mathbb{Z}_{+}:\quad n=\frac{1}{2}\left(  e_{\infty}-
{\displaystyle\sum\limits_{c\in\Gamma^{\prime}}} e_{c}\right)
,\forall e_{p}\in E_{p},\quad p\in\Gamma\right\}.$$ If
$D=\emptyset,$ then we should start the case 3. Now, if
$\mathrm{Card}(D)>0,$ then for each $n\in D$ we search a rational
function $\theta$ defined by
$$\theta=\frac{1}{2}
{\displaystyle\sum\limits_{c\in\Gamma^{\prime}}}
\frac{e_{c}}{x-c}.$$
\medskip

{\bf Step 3.} For each $n\in D,$ search a monic polynomial $P_n$
of degree $n$, such that
\begin{equation}\label{recu2}
\partial_x^3P_n+3\theta
\partial_x^2P_n+(3\partial_x\theta+3\theta
^{2}-4r)\partial_xP_n+\left(
\partial_x2\theta+3\theta\partial_x\theta
+\theta^{3}-4r\theta-2\partial_xr\right)P_n=0.
\end{equation}
 If $P_n$ does not
exist, then case 2 cannot hold. If such a polynomial is found, set
$\phi = \theta + \partial_xP_n/P_n$ and let $\omega$ be a solution
of
$$\omega^2 + \phi \omega + {1\over2}\left(\partial_x\phi + \phi^2 -2r\right)=
0.$$

Then $\zeta_1 = e^{\int\omega}$ is a solution of the differential
equation.
\bigskip

{\bf\large Case 3.} Search for each $c \in \Gamma'$ and for
$\infty$ the corresponding situation as follows:
\medskip

{\bf Step 1.} Search for each $c\in\Gamma^{\prime}$ and $\infty$
the sets $E_{c}\neq\emptyset$ and $E_{\infty}\neq\emptyset.$ For
each $c\in\Gamma^{\prime}$ and for $\infty$ we define
$E_{c}\subset\mathbb{Z}$ and $E_{\infty}\subset\mathbb{Z}$ as
follows:
\medskip

\begin{description}

\item[$(c_{1})$] If $\circ\left(  r_{c}\right)  =1$, then
$E_{c}=\{12\}$

\item[$(c_{2})$] If $\circ\left(  r_{c}\right)  =2,$ and $r= \cdots +
b(x-c)^{-2}+\cdots$, then
\begin{displaymath}
E_{c}=\left\{ 6+k\sqrt{1+4b}:\quad
k=0,\pm1,\pm2,\pm3,\pm4,\pm5,\pm6\right\}.
\end{displaymath}

\item[$(\infty)$] If $\circ\left(  r_{\infty}\right)  =v\geq2,$ and $r=
\cdots + bx^{2}+\cdots$, then {\small $$E_{\infty }=\left\{
6+{12k\over m}\sqrt{1+4b}:\textrm{ }
k=0,\pm1,\pm2,\pm3,\pm4,\pm5,\pm6\right\},\textrm{ }
m\in\{4,6,12\}.$$}
\medskip
\end{description}

{\bf Step 2.} Find $D\neq\emptyset$ defined by
$$D=\left\{
n\in\mathbb{Z}_{+}:\quad n=\frac{m}{12}\left(
e_{\infty}-{\displaystyle\sum\limits_{c\in\Gamma^{\prime}}}
e_{c}\right)  ,\forall e_{p}\in E_{p},\quad p\in\Gamma\right\}.$$
In this case we start with $m=4$ to obtain the solution,
afterwards $m=6$ and finally $m=12$. If $D=\emptyset$, then the
differential equation is not integrable because it falls in the
case 4. Now, if $\mathrm{Card}(D)>0,$ then for each $n\in D$ with
its respective $m$, search a rational function
$$\theta={m\over 12}{\displaystyle\sum\limits_{c\in\Gamma^{\prime}}}
\frac{e_{c}}{x-c}$$ and a polynomial $S$ defined as $$S=
{\displaystyle\prod\limits_{c\in\Gamma^{\prime}}} (x-c).$$

{\bf Step 3}. Search for each $n\in D$, with its respective $m$, a
monic polynomial $P_n=P$ of degree $n,$ such that its coefficients
can be determined recursively by
$$\bigskip P_{-1}=0,\quad P_{m}=-P,$$
$$P_{i-1}=-S\partial_xP_{i}-\left( \left( m-i\right)
\partial_xS-S\theta\right)  P_{i}-\left( m-i\right)  \left(
i+1\right)  S^{2}rP_{i+1},$$ where $i\in\{0,1\ldots,m-1,m\}.$ If
$P$ does not exist, then the differential equation is not
integrable because it falls in Case 4. Now, if $P$ exists search
$\omega$ such that $$ {\displaystyle\sum\limits_{i=0}^{m}}
\frac{S^{i}P}{\left( m-i\right)  !}\omega^{i}=0,$$ then a solution
of the differential equation is given by $$\zeta=e^{\int
\omega},$$ where $\omega$ is solution of the previous polynomial
of degree $m$.
\bigskip

\begin{remark}[\cite{acbl}]\label{rkov2}
If the differential equation falls only in the case 1 of Kovacic's
algorithm, then its differential Galois group is given by one of
the following groups:

\begin{description}

\item[I1] $e$ when the algorithm provides two
rational solutions.

\item[I2] $\mathbb{G}^{[n]}$ when the algorithm provides two algebraic solutions $\zeta_1,\zeta_2$ such that $\zeta_1^n,\zeta_2^n\in\mathbb{C}(x)$
and $\zeta_1^{n-1},\zeta_2^{n-1}\notin\mathbb{C}(x)$.

\item[I3] $\mathbb{G}^{\{n\}}$ when the algorithm provides only one
algebraic solution $\zeta$ such that $\zeta^n\in\mathbb{C}(x)$
with $n$ minimal.

\item[I4] $\mathbb{G}_m$ when the algorithm provides two
non-algebraic solutions.

\item[I5] $\mathbb{G}_a$ when the algorithm provides one
rational solution and the second solution is not algebraic.

\item[I6] $\mathbb{B}$ when the algorithm only provides one
solution $\zeta$ such that $\zeta$ and its square are not rational
functions.

\end{description}
\end{remark}

\subsubsection{Kovacic's Algorithm in Maple}
In order to analyze second order linear differential equations
with rational coefficients, generally without parameters, a
standard procedure is using \textsc{Maple}, and especially
commands \texttt{dsolve} and \texttt{kovacicsols}. Whenever the
command \texttt{kovacicsols} yields an output ``\texttt{[ ]}", it
means that the second order linear differential equation being
considered is not integrable, and thus its Galois group is non-virtually solvable.\\

In some cases, moreover, \texttt{dsolve} makes it possible to
obtain the solutions in terms of special functions such as
\emph{Airy functions}, \emph{Bessel functions} and
\textbf{hypergeometric functions}, among others
(see \cite{abst}).\\

 There is a number of second
order linear equations whose coefficients are not rational, and
whose solutions \textsc{Maple} can find with the command
\texttt{dsolve} but the presentation of the solutions is very
complicated, furthermore the command \texttt{kovacicsols} does not
work with such coefficients. These problems, in some cases, can be
solved by our \emph{algebrization method} (see section \ref{sec21}
and see also \cite{acbl}).

\subsubsection{Beyond Kovacic's Algorithm}
According to the works of Michael Singer and Felix Ulmer
\cite{sil,siul1,siul3,ul1,ul2}, we can have another perspective of
the Kovacic's algorithm by means of the $m$-th \textit{symmetric
power} of a linear differential equation $\mathcal L$, which is
denoted as $\mathcal L^{\sympow m}$.

\begin{theorem}  Let $\mathcal L$ be a linear homogeneous differential equation
of arbitrary order $n$.  For any $m \geq 1$ there is another
linear homogeneous differential equation, denoted by $\mathcal
L^{\sympow m}$, with the following property. If
$\zeta_1,\dots,\zeta_n$ are any solutions of $\mathcal L$ then any
homogeneous polynomial in $\zeta_1,\dots,\zeta_n$ of degree $m$ is
a solution of $\mathcal L^{\sympow m}$.
\end{theorem}

In this way, Kovacic's algorithm can be stated as follows (see
also \cite{dulo,fa,ulwe}).

\begin{algorithm} Consider $\mathcal L = \partial_x^2\zeta -
r\zeta$.\\

\begin{description}

\item[{\bf Step 1.}]  Check if $\mathcal L$ has an exponential solution $\zeta=e^{\int u}$ (with
$u\in\mathbb{C}(x)$). If so then $\zeta=e^{\int u}$.

\item[{\bf Step 2.}]  Check if $\mathcal L^{\sympow 2}$ has an exponential solution $\zeta=e^{\int u}$ (with $u\in\mathbb{C}(x)$). If so, let $v$ be a root of
\[
    v^2 + uv + \left(\frac12\partial_xu + \frac12u^2 - u\right) = 0.
\]
Then $\zeta=e^{\int v}$.

\item[{\bf Step 3.}]  Check if $\mathcal L^{\sympow 4}$, $\mathcal L^{\sympow 6}$,
or $\mathcal L^{\sympow 12}$ has an exponential solution
$\zeta=e^{\int u}$ (with $u\in\mathbb{C}(x)$).  If so, then there
is a polynomial of degree $4$, $6$ or $12$ (respectively) such
that if $v$ is a solution of it then $\zeta=e^{\int v}$.
\end{description}
\end{algorithm}

This algorithm can be generalized using the results of Michael
Singer in \cite{sil}. The trick is to find the correct numbers
(like $2,4,6,12$ of the Kovacic's algorithm).

\begin{theorem}  Suppose a linear homogeneous differential
equation of order $n$ is integrable.  Then it has a solution of
the form
\[
    \zeta = e^{\int v}
\]
where $v$ is algebraic over $\mathbb{C}(x)$.  The degree of $v$ is
bounded by $I(n)$, which is defined inductively by
\begin{align*}
    I(0) &= 1   \\
    I(n) &= \max\{J(n),\  n!I(n-1)\}    \\
    J(n) &= \left(\sqrt{8n} + 1\right)^{2n^2} -
        \left(\sqrt{8n} - 1\right)^{2n^2}.
\end{align*}
\end{theorem}
The following theorem is due to Singer and Ulmer (see
\cite{siul3}).

\begin{theorem}[Singer \& Ulmer, \cite{siul3}] Let $\mathcal L$ be
a homogeneous $n$-th order linear differential equation over $K$
with differential Galois group $\mathrm{DGal}(L/K)\subseteq
\mathrm{GL}(n,\mathcal{C})$.

\begin{enumerate}
\item If $\mathcal L$ has a Liouvillian solution whose logarithmic
derivative is algebraic of degree $m$, then there is a
$\mathrm{DGal}(L/K)$-semi-invariant of degree $m$ in $\mathcal
C[Y_1,\ldots,Y_n]$ which factors into linear forms.

\item If there is a
$\mathrm{DGal}(L/K)$-semi-invariant of degree $m$ in $\mathcal
C[Y_1,\ldots,Y_n]$ which factors into linear forms, then $\mathcal
L$ has a Liouvillian solution whose logarithmic derivative is
algebraic of degree $\leq m$
\end{enumerate}

\end{theorem}

This algorithm is not considered implementable; the numbers $I(n)$
are simply much too large.  For example $I(2) = 384064$, so this
algorithm would require checking if $\mathcal L^{\sympow m}$ has a
solution $\zeta$ with $\partial_x\zeta/\zeta \in \mathbb{C}(x)$
for $m=1,2,\dots,384064$. However, specific algorithms exist for
order three and higher, see
\cite{hes,siul3,siul2,ul1,ul2,horaulwe}.

\subsection{Eigenrings}
We consider two different formalisms for Eigenrings, the matrix
and operators formalism. We start with the Matrix formalism of
Eigenrings following M. Barkatou in \cite{ba}, but restricting
again to $2\times 2$ matrices.

Let $K$ be a differential field and let $A$ be a matrix in ${\rm
GL}(2,K)$ such that,
\begin{equation}\label{eigen1}\partial_x\mathbf{X}=-A\mathbf{X}.\end{equation}
Consider a matrix equation \eqref{eigen1} and let $P\in {\rm
GL}(2,K)$. The substitution $\mathbf{X} = P\mathbf{Y}$ leads to
the matrix equation
\begin{equation}\label{eigen2}\partial_x\mathbf{Y}=-B\mathbf{Y},\quad B=P^{-1}(\partial_xP+AP).\end{equation}

\begin{definition}
The matrices $A$ and $B$ are \textit{equivalents} over $K$,
denoted by $A\sim B$, when there exists a matrix $P\in{\rm
GL}(2,K)$ satisfying equation \eqref{eigen2}. The systems
\eqref{eigen1} and \eqref{eigen2} are equivalent, denoted by
$[A]\sim [B]$, when $A$ and $B$ are equivalent.
\end{definition}

By equation \eqref{eigen2}, we have $PB=\partial_xP+AP$. In
general, assuming $PB=PA$, where $P$ is a $2\times 2$ matrix, i.e,
$P$ is not necessarily in ${\rm GL}(2,K)$, we obtain
$PA=\partial_xP+AP$, which leads us to the following definition.

\begin{definition}
 The Eigenring of the system $[A]$, denoted by $\mathcal E(A)$, is
the set of $2\times 2$ matrices $P$ in $K$ satisfying
\begin{equation}\label{eigen4} \partial_xP = PA - AP.
\end{equation}
\end{definition}
Equation \eqref{eigen4} can be viewed as a system of $4$
first-order linear differential equations over $K$. Thus,
$\mathcal E(A)$ is a $\mathcal C-$vector space of finite dimension
$\leq 4$. Owing to the product of two elements of $\mathcal E(A)$
is also an element of $\mathcal E(A)$ and the identity matrix
$I_2$ belongs to $\mathcal E(A)$, we have that $\mathcal E(A)$ is
an algebra over $\mathcal C$, i.e.,  $\mathcal E(A)$ is a
$\mathcal C-$algebra. As a consequence, we have the following
results that can be found in \cite{ba}.
\begin{proposition}[\cite{ba}]\label{bark1} Any element $P$ of $\mathcal E(A)$ has:

\begin{itemize}
\item a minimal polynomial with coefficients in $\mathcal C$ and
\item all its eigenvalues are constant.
\end{itemize}
\end{proposition}

\begin{proposition}[\cite{ba}]\label{bark2}
If two systems $[A]$ and $[B]$ are equivalent, their eigenrings
$\mathcal E(A)$ and $\mathcal E(B)$ are isomorphic as $\mathcal
C-$algebras. In particular, one has $\dim_{\mathcal C}\mathcal
E(A) = \dim_{\mathcal C}\mathcal E(B).$
\end{proposition}

\begin{definition} The system $[A]$ is
called \textit{reducible} when $A\sim B$, being $B$ given by
$$B=\begin{pmatrix}b_{11}&0\\b_{21}&b_{22}\end{pmatrix}.$$
When $[A]$ is reducible and $b_{21}=0$, the system $[A]$ is called
\textit{decomposable} or \textit{completely reducible}. The system
$[A]$ is called \textit{irreducible} or \textit{indecomposable}
when $[A]$ is not reducible.
\end{definition}

Assume that the eigenring $\mathcal E(A)$ is known.

\begin{theorem}[\cite{ba}]\label{bark3} If $\mathcal E(A)$ is not a division ring then $[A]$ is reducible and
the reduction can be carried out by a matrix $P\in
\mathrm{GL}(2,K)$ that can be computed explicitly.
\end{theorem}

The condition \textit{$\mathcal E(A)$ is not a division ring}
implies $\dim_{\mathcal C} \mathcal E(A) > 1$. Indeed, if $P\in
\mathcal E(A) \setminus \{0\}$ is not invertible, then the family
$\{I, P\}$ is linearly independent (over $\mathcal C$) and hence
$\dim_{\mathcal C}\mathcal E(A) > 1$. In our case the converse is
true, due to the field of constants $\mathcal C$ is algebraically
closed. Indeed, suppose that $\dim_{\mathcal C}\mathcal E(A)> 1$
then there exists $P\in \mathcal E(A)$ such that the family $\{I,
P\}$ be linearly independent. Since $\mathcal C$ is algebraically
closed, there exists $\lambda\in\mathcal C$ such that $\det (P -
\lambda I) = 0$. Hence $\mathcal E(A)$ contains an element, namely
$P - \lambda I$, which
is non-zero and non invertible.\\

 The computation of eigenrings of the
system $[A]$ is implemented in \textit{ISOLDE} (Integration of
Systems of Ordinary Linear Differential Equations). The function
is \texttt{eigenring}, the calling sequence is
\texttt{eigenring(A, x)} being the parameters: $A$ - a square
rational function matrix with coefficients in an algebraic
extension of the rational numbers and $x$ - the independent
variable (a name). ISOLDE was written in Maple V and it is
available at \texttt{http://isolde.sourceforge.net/}.\\

In operators formalism, we restrict ourselves to second order
differential operators and we follow the works of Singer, Barkatou
and Van Hoeij (see \cite{si,ba,ho,ho1,ho2}). A differential
equation $\mathcal L:=\partial_x^2y+a\partial_xy+by=0$ with
$a,b\in K$ corresponds to a differential operator
$f=\partial_x^2+a\partial_x+b$ acting on $y$. The differential
operator $f$ is an element of the
non-commutative ring $K[\partial_x]$.\\

The factorization of operators is very important to solve
differential equations, that is, a factorization $f=\mathfrak{LR}$
where $\mathfrak L,\mathfrak R\in K[\partial_x]$ is useful for
computing solutions of $f$ because solutions of the right-hand
factor $\mathfrak R$ are solutions of $f$ as well.

\begin{definition} Let $\mathfrak{L}$ be a second order differential operator,
i.e $\mathcal{L}:=\mathfrak{L}(y)=0$. Denote $V(\mathfrak{L})$ as
the solution space of $\mathcal{L}$. The \textit{Eigenring} of
$\mathfrak{L}$, denoted by $\mathcal{E}(\mathfrak{L})$, is the set
of all operators $\mathfrak R$ for which $\mathfrak R(V(\mathfrak
L))$ is a subset of $V(\mathfrak L)$, that is
$\mathfrak{LR}=\mathfrak{SL}$, where $\mathfrak S$ is also an
operator.
\end{definition}

As consequence of the previous definition, $\mathfrak R$ is an
endomorphism of the solution space $V(\mathfrak {L})$. This means
that we can think of $\mathfrak R$ as a linear map $V\rightarrow
V$ and choosing one local basis of $V$ we obtain, by linear
algebra, that $\mathfrak R$ has a matrix $M_{\mathfrak R}$. The
characteristic polynomial of this map can be computed with the
classical methods of linear algebra. For endomorphisms $\mathfrak
R$, the product of $\mathfrak L$ and $\mathfrak R$ is divisible on
the right by $\mathfrak L$. This means that if $\mathfrak L(y)=0$,
then $\mathfrak L(\mathfrak R(y))=0$, so that $\mathfrak R$ map
$V\rightarrow V$.\\

For the general case of operators
$$\mathfrak L=\sum_{k=0}^nb_k\partial_x^k, \quad
\mathfrak R=\sum_{i=0}^ma_i\partial_x^i,\quad a_i,b_i\in K,\quad
n>m,\quad \mathcal L:= \mathfrak L(y)=0$$ and
$G=\mathrm{DGal}(L/K)$, we can see $\mathfrak R$ as a $G$-map.
Now, denoting by $P$ the characteristic polynomial of $\mathfrak
R$, assume that there exists polynomials $P_1$, $P_2$ with
$\gcd(P_1,P_2)=1$ such that $P=P_1P_2$.  By Cayley-Hamilton
theorem we have that $P(M_{\mathfrak R})=0$ and by kernel theorem
we have that $V=\ker(P_1(M_{\mathfrak
R}))\oplus\ker(P_2(M_{\mathfrak R}))$, in where
$\ker(P_1(M_{\mathfrak R}))$
and $\ker(P_2(M_{\mathfrak R}))$ are invariants under $G$.\\

Let $\lambda$ denote an eigenvalue of $\mathfrak R$; there exists
a non-trivial eigenspace $V_{\lambda}\subseteq V$, which means
that $\mathfrak L$ and $\mathfrak R-\lambda$ has common solutions
and therefore $\mathrm{right}-\gcd(\mathfrak L,\mathfrak
R-\lambda)$ is non-trivial, this means that
$\mathrm{right}-\gcd(\mathfrak L,\mathfrak R-\lambda)$ it is a
factor of $\mathfrak L$.

Returning to the second order operators, we establish the
relationship between the Eigenring of the system $[A]$ and the
Eigenring of the operator $\mathfrak L$. We start recalling that
$\mathcal L$ given by $\partial_x^2y+a\partial_xy+by=0$, $a,b\in
K$, can be
written as the system $$\partial_x\left(\begin{array}{c} y\\
 \partial_x y
\end{array}\right) =\left(\begin{array}{cc} 0 & 1  \\ -b & -a
  \end{array}\right)\left(\begin{array}{c} y \\
 \partial_x y
\end{array}\right),$$ and the system of linear differential
equations

$$ \partial_x\left(\begin{array}{c} y \\
 z
\end{array}\right) =\left(\begin{array}{cc} a & b \\ c & d
  \end{array}\right)\left(\begin{array}{c} y \\
 z
\end{array}\right),\quad a,b,c,d\in K,
$$ by means of an elimination process, is equivalent
to the second-order equation
\begin{equation}\label{sisor2}
\partial^2_xy-\left(a+d+\frac {\partial_xb}{b}\right)\partial_xy
-\left(\partial_xa+bc-ad-a\frac{\partial_xb}{b}\right)y=0,\quad
b\neq 0.
\end{equation}
In this way, we can go from operators to systems and reciprocally
computing the Eigenrings in both formalism. In particular, we
emphasize in the operator $\mathfrak
L=\partial_x^2+p\partial_x+q$, which is equivalent to the system
$[A]$, where $A$ is given by
$$A=\begin{pmatrix}0&-1\\q&p\end{pmatrix}\quad  p,q\in K.$$ As
immediate consequence we have the following lemmas.

\begin{lemma}
Consider $\mathfrak L$, $A$ and $P$ as follows:
$$\mathfrak
L=\partial_x^2+p\partial_x+q,\quad
A=\begin{pmatrix}0&-1\\q&p\end{pmatrix},\quad
P=\begin{pmatrix}a&b\\c&d\end{pmatrix},\quad a,b,c,d,p,q\in K.$$
The following statements holds
\begin{enumerate}
\item If $P\in\mathcal E(A)$, then $\mathfrak
R=a+b\partial_x\in\mathcal{E}(\mathfrak L)$.
\item If $\mathfrak R=a+b\partial_x\in\mathcal E (\mathfrak
L)$, then $P\in\mathcal E(A)$, where $P$ is given by
$$P=\begin{pmatrix}a&b\\\partial_xa-bq&a+\partial_xb-bp\end{pmatrix}.$$
\item $1\leq \dim_{\mathcal C }\mathcal{E}(\mathfrak L)\leq 4$.
\item $P\in
\mathrm{GL}(2,K)\Leftrightarrow
\frac{\partial_xa}{a}-\frac{a}{b}+p\neq\frac{\partial_xb}{b}-\frac{b}{a}q$.
\end{enumerate}
\end{lemma}

\begin{proof} Suppose that $V_{\mathfrak L}$ and $V_A$ are solution spaces of $\mathcal L$ and $[A]$ respectively.

\begin{enumerate}
\item If $P\in\mathcal E(A)$, then for a solution
$Y=\begin{pmatrix}y_1\\y_2\end{pmatrix}$ of $\partial_xY=-AY$, if
we set $Z=PY$ then $\partial_xZ=-AZ$. Setting
$$Z=\begin{pmatrix}\zeta_1\\\zeta_2\end{pmatrix}=P\begin{pmatrix}y_1\\y_2\end{pmatrix}$$
we have
$$\begin{array}{l}\partial_x\zeta_1=\zeta_2,\\ \\
\zeta_1=ay+b\partial_xy.\end{array}$$ Owing to $\mathfrak
L(\zeta_1)=0$ for all $y\in V_{\mathfrak L}$, $a+b\partial_x$ maps
$V_{\mathfrak L}$ to  $V_{\mathfrak L}$ and so $\mathfrak
R:=a+b\partial_x\in\mathcal E(\mathfrak L)$.
\item If $\mathfrak R=a+b\partial_x\in\mathcal E(\mathfrak L)$, then
let $\zeta:=\mathfrak R(y)=ay+b\partial_xy.$ Then $\mathfrak
L(\mathfrak R(y))=\mathfrak L(\zeta)=0$ and
$$\partial_x\begin{pmatrix}\zeta\\ \partial_x\zeta\end{pmatrix}=-A\begin{pmatrix}\zeta\\\partial_x\zeta\end{pmatrix},
\quad \forall
\begin{pmatrix}y\\\partial_xy\end{pmatrix}\in\mathcal E(A).$$
Thus, there exists $P$ satisfying
$$\begin{pmatrix}\zeta\\\partial_x\zeta\end{pmatrix}=P\begin{pmatrix}y\\\partial_xy\end{pmatrix},$$
which lead us to the expression
$$\begin{array}{l}\zeta=ay+b\partial_xy,\\ \\
\partial_x\zeta=(\partial_xa-bq)y+(a+b\partial_x-bp)\partial_xy.\end{array}$$
\end{enumerate}
Statements 3 and 4 can be obtained immediately.
\end{proof}
\medskip

\begin{remark}
Using this $P$, one can obtain the characteristic polynomial of
$P$ which is the characteristic polynomial of $\mathfrak R$ and so
obtain eigenvalues of $\mathfrak R$ needed.
\end{remark}
\medskip

\begin{remark}
Let $\mathfrak L$ be the differential operator $\partial_x^2+b$,
where $b\in K$, $\mathcal L:= \mathfrak L(y)=0$. The dimension of
the eigenring of $\mathfrak L$ is related with: \begin{itemize}
\item the number of solutions over $K$ of the differential equation
$\mathcal L$ and its second symmetric power $\mathcal L^{\sympow
2}$ and \item the type of differential Galois group (see
\cite{ba,si,ho,ho1,ho2}).
\end{itemize}
\end{remark}
\noindent The previous remark is detailed in the following lemma.
\begin{lemma}\label{lemmaeige}
 Assume $\mathfrak L=\partial_x^2+b$, where $b\in
K$, $\mathcal L:= \mathfrak L(y)=0$. The following statements
holds.

\begin{enumerate}
\item If $\dim_{\mathcal C}\mathcal{E}(\mathfrak L)=1$, then either
differential Galois group is irreducible ($\mathbb{D}_{\infty}$ or
primitive), or indecomposable ($G\subseteq \mathbb{B}$, $G\notin
\{e,\mathbb{G}_m,\mathbb{G}_a, \mathbb{G}^{\{n\}},\mathbb{G}^{[n]}
\}$).

\item If $\dim_{\mathcal C}\mathcal{E}(\mathfrak L)=2$, then either, the differential Galois group is the additive group or is contained in the
multiplicative group, but never is the identity group. Thus, we
can have two solutions but not over the differential field $K$.

\item If $\dim_{\mathcal C}\mathcal{E}(\mathfrak L)=4$, then the differential Galois group is the
identity group. In this case we have 2 independent solutions
$\zeta_1$ and $\zeta_2$ in which $\zeta_1^2$, $\zeta_2^2$ and
$\zeta_1\zeta_2$ are elements of the differential field $K$, i.e.
the solutions of $\mathcal L ^{\sympow 2}$ belongs to the
differential field $K$.
\end{enumerate}
\end{lemma}

\begin{proof}
Assume that $G=\mathrm{DGal}(L/K)$, then $G$ commutes with each
element $P\in\mathcal E(\mathfrak L$), i.e., $GP=PG$. Computing
the dimension of the set of matrices satisfying $GP=PG$ we obtain
$\dim_{\mathcal C}\mathcal{E}(\mathfrak L)$ for each $G\subseteq
\mathrm{SL}(2,\mathbb{C})$.
\end{proof}

The eigenring for a differential operator $\mathfrak L$ has been
implemented in Maple. The function is \texttt{eigenring}, the
calling sequences are \texttt{eigenring(L,domain)} and
\texttt{endomorphism-charpoly(L, R, domain)}, where $\mathfrak L$
is a differential operator, $\mathfrak R$ is the differential
operator in the output of eigenring. The argument \texttt{domain}
describes the differential algebra. If this argument is the list
\texttt{[Dx,x]} then the differential operators are notated with
the symbols \texttt{Dx} and \texttt{x}, where \texttt{Dx} is the
operator $\partial_x$.

\begin{example} Consider $K=\mathbb{C}(x)$, $\mathfrak L=\partial_x^2-\frac6{x^2}$ and $[A]$ where $A$ is given by
$$A=\begin{pmatrix}0&-1\\-6x^{-2}&0\end{pmatrix}.$$ The Eigenring
of $[A]$ and the Eigenring of $\mathfrak L$ are given by
$$\mathcal E
(A)=\mathrm{Vect}\left(\begin{pmatrix}1&0\\0&1\end{pmatrix},\begin{pmatrix}-1&x\\6x^{-1}&0\end{pmatrix},
\begin{pmatrix}-3x^5&x^6\\-9x^4&3x^5\end{pmatrix}, \begin{pmatrix}2x^{-5}&x^{-4}\\-4x^{-6}&-2x^{-5}\end{pmatrix} \right),$$
$$\mathcal E(\mathfrak L)=\mathrm{Vect}\left(1,x\partial_x-1,x^6\partial_x-3x^5,{\partial_x\over x^4}+\frac2{x^5}\right).$$
\end{example}

\subsection{Riemann's Equation}\label{riemansection}
The Riemann's equation is an important differential equation which
has been studied for a long time, since Gauss, Riemann, Schwartz,
etc., see for example \cite{iksy,po}. We are interested in the
relationship with the Picard-Vessiot theory. Thus, we follow the
works of Kimura \cite{ki}, Martinet \& Ramis \cite{marram} and
Duval \& Loday-Richaud \cite{dulo}.

\begin{definition} The \textit{Riemann's equation} is an
homogeneous ordinary linear differential equation of the second
order over the Riemann's sphere with at most three singularities
which are of the regular type. Assuming $a$, $b$ and $c$ as
regular singularities, the Riemann's equation may be written in
the form
\begin{eqnarray}\label{riemmaneq1}
&& \partial_x^2y+ \left(\frac{1-\rho-\rho'}{x-a} + \frac{1-\sigma-\sigma'}{x-b}+\frac{1-\tau-\tau'}{x-c}\right)\partial_xy\\
&&  + \left(\frac{\rho\rho'(a-b)(a-c)}{(x-a)^2(x-b)(x-c)} +
\frac{\sigma\sigma'(b-a)(b-c)}{(x-b)^2(x-a)(x-c)} +
\frac{\tau\tau'(c-a)(c-b)}{(x-c)^2(x-a)(x-b)} \right) y = 0,
\nonumber
\end{eqnarray}
where $(\rho , \rho')$, $(\sigma , \sigma')$ and $(\tau, \tau')$
are the exponents at the singular points $a,b,c$ respectively and
must satisfy the Fuchs relation $\rho + \rho' + \sigma +
\sigma'+\tau + \tau'= 1$. The quantities $\rho'-\rho$,
$\sigma'-\sigma$ and $\tau'-\tau$ are called the \textit{exponent
differences} of the Riemann's equation \eqref{riemmaneq1} at $a$,
$b$ and $c$ respectively and are denoted by $\widetilde{\lambda}$,
$\widetilde{\mu}$ and $\widetilde{\nu}$ as follows:
$$\widetilde{\lambda}=\rho'-\rho,\quad \widetilde{\mu}=\sigma'-\sigma,\quad \widetilde{\nu}=\tau'-\tau.$$ The complete set of solutions of the
Riemann's equation \eqref{riemmaneq1} is denoted by the symbol
$$y=P\left\{\begin{array}{cccc}a&b&c&\\\rho&\sigma&\tau&x\\\rho'&\sigma'&\tau'&\end{array}\right\}$$
and is called \textit{Riemann's $P$-function}.
\end{definition}

Now, we will briefly describe here the Kimura's theorem that gives
necessary and sufficient conditions for the integrability of the
Riemann's differential equation.

\begin{theorem}[Kimura, \cite{ki}]\label{kimurath}
The Riemann's differential equation \eqref{riemmaneq1} is
integrable if and only if, either
\begin{enumerate}
\item[(i)] At least one of the four numbers $\widetilde{\lambda}+\widetilde{\mu}+\widetilde{\nu}$,
$-\widetilde{\lambda}+\widetilde{\mu}+\widetilde{\nu}$,
$\widetilde{\lambda}-\widetilde{\mu}+\widetilde{\nu}$,
$\widetilde{\lambda}+\widetilde{\mu}-\widetilde{\nu}$ is an odd
integer, or
\item[(ii)] The numbers $\widetilde{\lambda}$ or $-\widetilde{\lambda}$, $\widetilde{\mu}$
or $-\widetilde{\mu}$ and $\widetilde{\nu}$ or $-\widetilde{\nu}$
belong (in an arbitrary order) to some of the following fifteen
families
\end{enumerate}
$$
\begin{array}{|c|c|c|c|c|}\hline
1 & 1/2+l & 1/2+m & \mbox{arbitrary complex number} &\\ \hline 2 &
1/2+l & 1/3+m & 1/3+q &\\ \hline 3 & 2/3+l & 1/3+m & 1/3+q &
l+m+q\mbox{ even}\\ \hline 4 & 1/2+l & 1/3+ m & 1/4+q & \\ \hline
5 & 2/3+l & 1/4+m & 1/4+q & l+m+q\mbox{ even}\\ \hline 6 & 1/2+l &
1/3+m & 1/5+q & \\ \hline 7 & 2/5+l & 1/3+m & 1/3+q & l+m+q\mbox{
even}\\ \hline 8 & 2/3+l & 1/5+m & 1/5+q & l+m+q\mbox{ even}\\
\hline 9 & 1/2+l & 2/5+m & 1/5+q & l+m+q\mbox{ even}\\ \hline 10 &
3/5+l & 1/3+m & 1/5+q & l+m+q\mbox{ even}\\ \hline 11 &2/5+l &
2/5+m & 2/5+q & l+m+q\mbox{ even}\\ \hline 12 &2/3+l & 1/3+m &
1/5+q & l+m+q\mbox{ even}\\ \hline 13 & 4/5+l & 1/5+m & 1/5+q &
l+m+q\mbox{ even}\\ \hline 14 & 1/2+l &2/5+m & 1/3+q & l+m+q\mbox{
even}\\ \hline 15 & 3/5+l & 2/5+m & 1/3+q & l+m+q\mbox{ even}\\
\hline
\end{array}
$$
Here $n,m,q$ are integers.
\end{theorem}

Using the M\"oebius transformation \cite{hi}, also known as
homographic substitution, in the Riemann's equation
\eqref{riemmaneq1}, we can map $x=a,b,c$ to $x'=a',b',c'$,
respectively:
$$x'={px+q\over rx+s}.$$ In particular, we can place the
singularities at $x = 0, 1, \infty$ to obtain the following
Riemann's equation:
\begin{eqnarray}\label{hypergeometric1}
&& \partial_x^2y+ \left(\frac{1-\rho-\rho'}{x} + \frac{1-\sigma-\sigma'}{x-1}\right)\partial_xy\\
&& \qquad\qquad \qquad  + \left(\frac{\rho\rho'}{x^2} +
\frac{\sigma\sigma'}{(x-1)^2} +
\frac{\tau\tau'-\rho\rho'\sigma\sigma'}{x(x-1)}\right)y = 0,
\nonumber
\end{eqnarray}

where the set of solutions is
$$y=P\left\{\begin{array}{cccc}0&1&\infty&\\\rho&\sigma&\tau&x\\\rho'&\sigma'&\tau'&\end{array}\right\}.$$
Sometimes it is very useful to map $x=0,1,\infty$ to
$x'=-1,1,\infty$ in the Riemann's equation
\eqref{hypergeometric1}, for example, setting $\rho=0$, we can
state the substitution:
$$P\left\{\begin{array}{cccc}0&1&\infty&\\0&\sigma&\tau&x\\\frac12&\sigma'&\tau'&\end{array}\right\}=
P\left\{\begin{array}{cccc}-1&1&\infty&\\\sigma&\sigma&2\tau&\sqrt{x}\\\sigma'&\sigma'&2\tau'&\end{array}\right\}.$$

We can transforms equation \eqref{hypergeometric1} to the
\textit{Gauss Hypergeometric equation} as follows:

$$P\left\{\begin{array}{cccc}0&1&\infty&\\\rho&\sigma&\tau&x\\\rho'&\sigma'&\tau'&\end{array}\right\}=x^\rho(x-1)^\sigma
P\left\{\begin{array}{cccc}0&1&\infty&\\0&0&\kappa&x\\1-\gamma&\gamma-\kappa-\beta&\beta&\end{array}\right\},$$
where $\kappa=\rho+\sigma+\tau$, $\beta=\rho+\sigma+\tau'$ and
$\gamma=1+\rho-\rho'$. Then
$$y=P\left\{\begin{array}{cccc}0&1&\infty&\\0&0&\kappa&x\\1-\gamma&\gamma-\kappa-\beta&\beta&\end{array}\right\},$$
is the set of solutions of the Gauss Hypergeometric differential
equation\footnote{In general, for the Hypergeometric differential
equation, is used $\alpha$ instead of $\kappa$, but we want to
avoid further confusions.}
\begin{equation}\label{hypergeometric}
\partial_x^2y+{\left(\gamma-(\kappa+\beta+1)x\right)\over
x(1-x)}\partial_xy-{\kappa\beta\over x(1-x)} y=0,
\end{equation}
where the Fuchs relation is trivially satisfied and the exponent
differences are given by
$$\widetilde{\lambda}=1-\gamma,\quad \widetilde{\mu}=1-\gamma-\beta, \quad\widetilde{\nu}=\beta-\kappa.$$
We remark that the Galoisian structure of the Riemann's equation
do not change with the M\"oebius transformation.\\

The \textit{confluent Hypergeometric equation} is a degenerate
form of the Hypergeometric differential equation where two of the
three regular singularities merge into an irregular singularity.
For example, making ``$1$ tend to $\infty$" in a suitable way, the
Hypergeometric equation \eqref{hypergeometric} has two classical
forms:
\begin{itemize}
\item \textit{Kummer's} form \begin{equation}\label{kummer} \partial_x^2y+{c-x\over x}\partial_xy-{a\over x}y=0\end{equation}

\item \textit{Whittaker's} form \begin{equation}\label{whittaker} \partial_x^2y=\left(\frac14-{\kappa\over x}+{4\mu^2-1\over 4x^2}\right)y\end{equation}
\end{itemize}
where the parameters of the two equations are linked by
$\kappa=\frac{c}2-a$ and $\mu=\frac{c}2-\frac12$. Furthermore,
using the expression \eqref{redsec}, we can see that the
Whittaker's equation is the reduced form of the Kummer's equation.
The Galoisian structure of these equations has been deeply studied
in \cite{marram,dulo}.

\begin{theorem}[Martinet \& Ramis, \cite{marram}]\label{thmarram}
The Whittaker's differential equation \eqref{whittaker} is
integrable if and only if either,
$\kappa+\mu\in\frac12+\mathbb{N}$, or
$\kappa-\mu\in\frac12+\mathbb{N}$, or
$-\kappa+\mu\in\frac12+\mathbb{N}$, or
$-\kappa-\mu\in\frac12+\mathbb{N}$.

\end{theorem}

The \textit{Bessel's equation} is a particular case of the
confluent Hypergeometric equation and is given by
\begin{equation}\label{bessel} \partial_x^2y+{1\over x}\partial_xy+{x^2-n^2\over
x^2}y=0.\end{equation} Under a suitable transformation, the
reduced form of the Bessel's equation is a particular case of the
Whittaker's equation. Thus, we can obtain the following well known
result, see \cite[p. 417]{kol} and see also \cite{ko,mo}.

\begin{corollary}\label{corbessel}
The Bessel's differential equation \eqref{bessel} is integrable if
and only if $n\in \frac12+\mathbb{Z}$.
\end{corollary}

We point out that the integrability of Bessel's equation for half
integer of the parameter was known by Daniel Bernoulli \cite{wa}.
By double confluence of the Hypergeometric equation
\eqref{hypergeometric}, that is making ``$0$ and $1$ tend to
$\infty$" in a suitable way, one gets the \textit{parabolic
cylinder equation} (also known as \textit{Weber's} equation):
\begin{equation}\label{weber} \partial_x^2y=\left(\frac14x^2-{1\over
2}-n\right)y,\end{equation} which is integrable if and only if
$n\in\mathbb{Z}$, see \cite{ko, dulo}. Setting
$n=\frac{b^2-c}{2a}-\frac12$ and making the change
$x\mapsto\sqrt{\frac2a}(ax+b)$, one can gets the Rehm's form of
the Weber's equation:
\begin{equation}\label{rehm}
\partial_x^2y=\left(ax^2+2bx+c\right)y,\quad a\neq
0,\end{equation} so that ${b^2-c\over a}$ is an odd integer.

The Hypergeometric equation, including confluences, is a
particular case of the differential equation
\begin{equation}\label{eqorpol}
\partial_x^2y+\frac{L}{Q}\partial_xy+\frac{\lambda}{Q} y,\quad
\lambda\in\mathbb{C},\quad L=a_0+a_1x,\quad Q=b_0+b_1x+b_2x^2.
\end{equation}
We recall that the \textit{classical orthogonal polynomials} and
\textit{Bessel polynomials} are solutions of equation
\eqref{eqorpol}, see \cite{ch,is,niuv}:
\begin{itemize}
\item Hermite, denoted by $H_n$,
\item Chebyshev of first kind, denoted by $T_n$,\item Chebyshev of second
kind, denoted by $U_n$, \item Legendre, denoted by $P_n$, \item
Laguerre, denoted by $L_n$,
\item associated Laguerre, denoted by $L_n^{(m)}$, \item Gegenbauer, denoted by $C_n^{(m)}$
\item Jacobi polynomials, denoted by $\mathcal{P}_n^{(m,\nu)}$ and
\item Bessel polynomials, denoted by $B_n$.
\end{itemize}
\medskip

In the following table we give $Q$, $L$ and $\lambda$
corresponding to equation \eqref{eqorpol} for classical orthogonal
polynomials and Bessel polynomials.
\medskip

\begin{tabular}[t]{|l|l|l|l|l|}

\hline

\textbf{Polynomial}& $\boldsymbol{Q}$&$\boldsymbol{L}$&$\boldsymbol{\lambda}$\\

\hline

$H_n$&$1$&$-2x$&$2n$\\

\hline

$T_n$&$1-x^2$&$-x$&$n^2$\\

\hline

$U_n$&$1-x^2$&$-3x$&$n(n+2)$\\

\hline

$P_n$&$1-x^2$&$-2x$&$n(n+1)$\\

\hline

$L_n$&$x$&$1-x$&$n$\\

\hline

$L_n^{(m)}$&$x$&$m+1-x$&$n$\\

\hline

$C_n^{(m)}$&$1-x^2$&$-(2m+1)x$&$n(n+2m)$\\

\hline

$\mathcal P_n^{(m,\nu)}$&$1-x^2$&$\nu-m-(m+\nu+2)x$&$n(n+1+m+\nu)$\\

\hline
$B_n$&$x^2$&$2(x+1)$&$-n(n+1)$\\

\hline
\end{tabular}%
\bigskip

The \textit{associated Legendre polynomials}, denoted by
$P_n^{(m)}$, does not appear in the previous table. They are
solutions of the differential equation
\begin{equation}\label{assleg} \partial_x^2y -{2x\over 1-x^2}\partial_xy +
\left({n(n+1) - \frac{m^2}{1-x^2}\over 1-x^2}\right)y = 0.
\end{equation} This equation can be transformed into a Riemann's
differential equation through the change $x\mapsto\frac1{1-x^2}$.
Thus, the complete set of solutions of equation \eqref{assleg} is
given by
$$P\left\{\begin{array}{cccc}0&1&\infty&\\-\frac12 n&0&\frac12m&\frac1{1-x^2}\\\frac12+\frac12n&\frac12&-\frac12m&\end{array}\right\},$$
the exponent differences are $\widetilde{\lambda}=\frac12$,
$\widetilde{\mu}=\frac12$ and $\widetilde{\nu}=0$. By Kimura's
theorem this equation is integrable.\\

Finally, we remark that integrability conditions and solutions of
differential equations with solutions orthogonal polynomials,
including Bessel polynomials, can be obtained applying Kovacic's
algorithm. In the same way, we can apply Kovacic's algorithm to
obtain the same results given by Kimura \cite{ki} and Martinet \&
Ramis \cite{marram}. Also we recall that Duval \& Loday-Richaud
applied Kovacic's algorithm to some families of special functions
\cite{dulo}.

\section{Supersymmetric Quantum Mechanics}\label{susysection}

In this section we establish the basic information on
Supersymmetric Quantum Mechanics. We only consider the case of
non-relativistic quantum mechanics.

\subsection{The Schr\"odinger Equation}
In classical mechanics the Hamiltonian corresponding to the energy
(kinetic plus potential) is given by
$$H={\|\overrightarrow{p}\|^2\over 2m}+U(\overrightarrow{x}),\quad \overrightarrow{p}=(p_1,\ldots,p_n),\quad \overrightarrow{x}=(x_1,\ldots,x_n),$$
while in quantum mechanics the momentum $\overrightarrow{p}$ is
given by $\overrightarrow{p}=-\imath\hbar\nabla$, the Hamiltonian
operator is the Schr\"odinger (non-relativistic, stationary)
operator which is given by
$$H=-{\hbar^2\over 2m}\nabla^2+U(\overrightarrow{x})$$
and the Schr\"odinger equation is $H\Psi=E\Psi$, where
 $\overrightarrow{x}$ is the {\it{coordinate}}, the eigenfunction $\Psi$
is the {\it{wave function}}, the eigenvalue $E$ is the {\it{energy
level}}, $U(\overrightarrow{x})$ is the {\it{potential or
potential energy}} and the solutions of the Schr\"odinger equation
are the {\it{states}} of the particle. Furthermore, it is known
that $H^\dagger=H$, i.e., the Schr\"odinger operator is a
\textit{self-adjoint operator} in a suitable \textit{complex and
separable Hilbert space}. Thus, $H$ has a purely real
\textit{spectrum} $\mathrm{spec}(H)$ and its spectrum
$\mathrm{spec}(H)$ is the disjoint union of the \textit{point
spectrum} $\mathrm{spec}_p(H)$ and the continuous spectrum
$\mathrm{spec}_c(H)$, i.e.,
$\mathrm{spec}(H)=\mathrm{spec}_p(H)\cup \mathrm{spec}_c(H)$ with
$\mathrm{spec}_p(H)\cap \mathrm{spec}_c(H)=\emptyset$. See for example \cite{bana,pru,te}.\\

 Along this thesis we only
consider the one-dimensional Schr\"odinger equation written as
follows:

\begin{equation}\label{hams}
H\Psi=E\Psi, \quad H=-\partial_z^2+V(z),
\end{equation}
 where $z=x$ (cartesian coordinate) or $z=r$ (radial coordinate) and $\hbar=2m=1$.
  We denote by $\Psi_n$ the wave function for $E=E_n$. The potentials should satisfy some conditions depending of
 the physic situation such as barrier, scattering, etc., see
 \cite{cokasu,gapa,lali,masa,sc}.
\begin{definition}[Bound States]\label{defbs} The solution $\Psi_n$ is called a \textit{bound state}
  when $E$ belongs to the point spectrum of $H$ and its norm is finite, i.e.,
\begin{equation}\label{condqm}
E_n\in \mathrm{spec}_p(H),\quad \int|\Psi_n(x)|^2dx<\infty, \quad
n\in\mathbb{Z}_+.
\end{equation}
\end{definition}

An interesting property of bound states is given by the Sturm's
theorem, see \cite{bana,te}.

\begin{theorem}[Sturm's Theorem]
If $\Psi_0,\Psi_1,\ldots,\Psi_n,\ldots$ are the wave functions of
the bound states with energies $E_0<E_1<\cdots<E_n<\ldots$, then
$\Psi_n$ has $n$ nodes (zeros). Furthermore, between two
consecutive nodes of $\Psi_n$, there is a node of $\Psi_{n-1}$,
and moreover $\Psi_{n+r}$ has at least one zero for all $r\geq 1$.
\end{theorem}

\begin{definition}[Ground State and Excited States]  Assume $\Psi_0,\Psi_1,\ldots,\Psi_n,\ldots$ as in the Sturm's theorem. The state $\Psi_0$,
 which is state with minimum energy
 is called the \textit{ground state} and the states $\Psi_1,\ldots,\Psi_n,\ldots$ are
 called the \textit{excited states}.
  \end{definition}

\begin{definition}[Scattering States]  The solution $\Psi$ corresponding to the level energy $E$ is called a \textit{scattering state}
  when $E$ belongs to the continuous spectrum of $H$ and its norm is
  infinite.
\end{definition}
\medskip

The wave function belonging to the continuous spectrum have two
typical boundary conditions: the first ones, barrier potentials
and the second one periodic boundary conditions. The
\textit{transmission and reflection coefficients} are related with
the barrier potentials, \cite{gapa}.\\

 When the particle moves in one dimension, we
 use the classical one dimensional Schr\"odinger equation with
 cartesian coordinate $x$.\\

\begin{example}[The Harmonic Oscillator] We consider the Hamiltonian operator $H$ given by \eqref{hams}. Normalizing the angular velocity ($\omega=1$),
 the one-dimensional
harmonic oscillator potential is $V=\frac14x^2$. The
\textit{creator} (raising) and \textit{annihilator} (lowering)
operators, denoted respectively by $a^\dagger$ and $a$, given by
$$a^\dagger=-\partial_x+\frac12x,\quad a=\partial_x+\frac12x$$ lead
us to the relations \begin{equation}\label{harmosci}H=a^\dagger
a+\frac12,\quad [a,a^\dagger]=1,\quad
[a^\dagger,H]=-a^\dagger,\quad [a,H]=a.
\end{equation}

We want to solve $H\Psi=E\Psi$, in particular we are interested in
the case $H\Psi_n=E_n\Psi_n$, where $\Psi_0$ is the ground state
and $\Psi_1,\ldots,\Psi_n$ are the excited states. Considering
$H\Psi_0=E_0\Psi_0$ and assuming that $a^\dagger a\Psi_0=0$, we
obtain the energy $E_0$ and the ground state $\Psi_0$:
$$E_0=\frac12,\quad \Psi_0=e^{-\frac14x^2}.$$

Using the relation $\Psi_n=(a^\dagger)^n\Psi_0,$ we obtain the
rest of wave functions
$$\begin{array}{lll}\Psi_0=e^{-\frac14x^2},&\Psi_1=a^\dagger\Psi_0=xe^{-\frac14x^2},&\Psi_2=a^\dagger\Psi_1=(x^2-1)e^{-\frac14x^2},\\
& &\\
 \Psi_3=(x^3-3x)\Psi_0,&
 \Psi_4=(x^4-6x^2+3)\Psi_0,\ldots&\Psi_n=a^\dagger\Psi_{n-1}=H_n\Psi_0,\end{array}$$
 where $H_n$ are the Hermite's polynomials presented in the previous
 section. Now, to obtain the complete energy spectrum we use the relations
\eqref{harmosci}, thus $$E_0=\frac12,\quad E_1=1+E_0=\frac32,\quad
E_2=1+E_1=\frac52,\quad \ldots,\quad E_n=\frac{2n+1}2.$$ Another
way to obtain $E_n$ is given by the formula
$$E_n=\frac14x^2-\frac{\partial_x^2\Psi_n}{\Psi_n}.$$
\end{example}
\medskip

 Also we can consider the particle
 moving in three dimensions, this means that $\overrightarrow{p}=(p_x,p_y,p_z)$,
 where $p_x=-\imath\partial_x$, $p_y=-\imath\partial_y$,
 $p_z=-\imath\partial_z$ and $p=\|\overrightarrow{p}\|$. The \textit{angular momentum operator} is given
 by $\overrightarrow{L}=(L_x,L_y,L_z)$ where $L_x=yp_z-zp_y$,
 $L_y=zp_x-xp_z$, $L_z=xp_y-yp_x$ and $L=\|\overrightarrow{L}\|$. The square of the angular
 momentum operator $\|\overrightarrow{L}\|^2=L^2=L_x^2+L_y^2+L_z^2$
 commutes with all components of the angular momentum operator.

In spherical coordinates $x=r\sin\nu\cos\varphi$,
$y=r\sin\nu\sin\varphi$, $z=r\cos\nu$, $L^2$ is given by
$$L^2=-\triangle_{\nu,\varphi},\quad \triangle_{\nu,\varphi}=\frac1{\sin\nu}\partial_\nu\left(\sin\nu\partial_\nu\right)+\frac1{\sin^2\nu}\partial_{\varphi}^2,$$
where we denote by $\triangle_{\nu,\varphi}$ that part of the
Laplacian acting on the variables $\nu$ and $\varphi$ only. The
kinetic energy given by
$p^2=\partial_x^2+\partial_y^2+\partial_z^2$ reads in polar
coordinates as
$$p^2=p_r^2+{1\over r^2}L^2,\quad p_r^2=-\frac1{r^2}\partial_r\left({r^2}\partial_r\right)=-\left(\partial_r^2+\frac2r\partial_r\right).$$
Now, for \textit{central potentials}, where the potential
$U(\overrightarrow{r})$ is spherically symmetric, i.e.,
$U(\overrightarrow{r})=U(r)$, we can reduce the Schr\"odinger
equation to an one dimensional problem, the so-called
\textit{radial equation}.

We start writing the eigenfunctions and eigenvalues of the
operator $L^2$:
$$L^2Y_{\ell,m}(\nu,\varphi)=\ell(\ell+1)Y_{\ell,m}(\nu,\varphi),$$
the eigenfunctions $Y(\nu,\varphi)$ are the \textit{spherics
harmonics} which are related with the associated Legendre
Polynomials
$$Y_{\ell,m}(\nu,\varphi)=P_\ell^m(\cos\nu)e^{\imath m\varphi}.$$

Assuming $\Phi$ as eigenfunctions of $p^2+U(r)$ satisfying
$\Phi=R_\ell(r)Y_{\ell,m}(\nu,\varphi)$, i.e., the \textit{partial
wave function decomposition} see (\cite{gapa,lali, sc}), we have

$$\left(p_r^2+{1\over r^2}L^2+U(r)-E\right)R_\ell(r)Y_{\ell,m}(\nu,\varphi)=0,$$ so that we obtain
$$\left(p_r^2+U(r)-E\right)R_\ell(r) +{R_\ell(r)\over r^2Y_{\ell,m}(\nu,\varphi)}L^2Y_{\ell,m}(\nu,\varphi)=0$$
and owing to
$L^2Y_{\ell,m}(\nu,\varphi)=\ell(\ell+1)Y_{\ell,m}(\nu,\varphi)$
we have the radial equation
$$\left(p_r^2+{\ell(\ell+1)\over
r^2}+U(r)\right)R_\ell(r)=ER_\ell(r).$$ Applying the expression
\eqref{redsec}, the radial equation can be reduced to the
Schr\"odinger equation \eqref{hams} as follows:
$$H\Psi=E\Psi,\quad H=-\partial_r^2+V(r),\quad V(r)={\ell(\ell+1)\over r^2}+U(r),\quad \Psi=rR_\ell(r).$$

The equation for the angular part is always solved through
spherics harmonics, while for the radial part, the analysis
depends on the spherically symmetric potential $U(r)$. One example
of the radial equation is the Coulomb potential. The complete set
of physical and mathematical conditions for the potentials,
spectrum and wave functions in one or three dimensions can be
found in any book of quantum mechanics, including bound states and scattering cases, see for example \cite{gapa,lali,sc}.\\

\subsection{Darboux Transformation}\label{darbouxsection}
The following theorem is the most general case for Darboux
transformation in the case of second order linear differential
equations, which is taken faithfully from \cite{da1}.

\begin{theorem}[Darboux, \cite{da1}]\label{darbouxgenth} Suppose that we know how to integrate, for any
value of the constant $m$, the following equation
\begin{equation}\label{orda7}\partial_x^2y+ P\partial_xy+(Q -mR)y = 0.\end{equation} If
$\theta\neq 0$ is an integral of the equation
\begin{displaymath}\partial_x^2\theta+P\partial_x\theta+Q\theta=0,\end{displaymath} then the function
\begin{equation}\label{orda9} u={\partial_xy-{\partial_x\theta\over \theta}y\over
\sqrt{R}},\end{equation} will be an integral of the equation
\begin{equation}\label{orda10}
\partial_x^2u+P\partial_xu+\left(\theta\sqrt{R}\partial_x\left({P\over
\theta\sqrt{R}}\right)-\theta\sqrt{R}\partial_x^2\left({1\over
\theta\sqrt{R}}\right)-mR\right)u=0,\end{equation} for $m\neq
0$.\end{theorem} Darboux in \cite{da1,da2} presented the
particular case for $R=1$ and $P=0$, which today is known as
\textit{Darboux transformation}, but really is a corollary of the
\textit{general Darboux transformation} given in theorem
\ref{darbouxgenth}.

\begin{corollary}[Darboux, \cite{da1,da2}]
Suppose that we know how to integrate
\begin{equation}\label{orda11}
\partial_x^2y= (f(x) + m)y \end{equation} for any value of $m$. If $\theta$
satisfies the equation $\partial_x^2\theta= (f(x)+m_1)\theta$, the
function
$$u = \partial_xy-{\partial_x\theta\over\theta}y$$ will be an integral of the
equation
\begin{equation}\label{orda12} \partial_x^2u= \left(\theta\partial_x^2\left(1\over \theta\right)-m_1+m \right)u, \end{equation}
for $m\neq m_1$. Furthermore, $$\theta\partial_x^2\left(1\over
\theta\right)-m_1=f(x)-2\partial_x\left({\partial_x\theta\over
\theta}\right)=2\left({\partial_x\theta\over
\theta}\right)^2-f(x)-2m_1.$$

\end{corollary}

\begin{remark}
In practice, we need two values of $m$ to apply the Darboux's
results.
\end{remark}
\medskip

\begin{example} Consider the equation $\partial_x^2y= my$ . Employing the
solution $\theta = x,$ we shall get
$$\partial_x^2y=\left({1\cdot 2\over x^2} + m\right)y.$$
Applying the same method to the latter equation, but taking now
$\theta = x^2$, we shall have
$$\partial_x^2y=\left({2\cdot 3\over x^2} + m\right)y$$ and so
on. The cases $m_1=0$ and $m_1=-1$ also can be found as exercises
in the Ince's book  \cite[p. 132]{in}.
\end{example}
\medskip

We can see that equation \eqref{orda11} coincides with the
Schr\"odinger equation \eqref{hams}. Thus, we can apply the
Darboux transformation in where $m=-E$ and $m_1=-E_0$.\\

 The following definition corresponds with Delsarte's
transformation operators (isomorphisms of transmutations), which
today are called \textit{intertwiner operators}, see \cite{de}.

\begin{definition} Two operators $\mathfrak L_0$ and $\mathfrak L_1$ are said to be
\textit{intertwined} by an operator $T$ if
\begin{equation}\label{del4}\mathfrak L_1T = T\mathfrak L_0.\end{equation}
\end{definition}

We can relate the intertwiner operators with the Darboux
transformation of equation \eqref{hams}, where $\mathfrak L_1$ and
$\mathfrak L_0$ are Schr\"odinger operators and $T$ can be
either $\mp\partial_x+{\partial_x\Psi_0\over \Psi_0}$.\\

Crum, inspired by the works of Liouville \cite{lio,lio2} obtained
one kind of iterative generalization of Darboux's result giving
emphasis in the Sturm-Liouville systems, i.e., he proved that the
Sturm-Liouville conditions are preserved under Darboux
transformations, see \cite{cr}. The Crum's result is presented in
the following theorem, defining the Wronskian determinant $W$ of
$k$ functions $f_1, f_2, \ldots, f_k$ by
$$W(f_1,\ldots,f_k)=\det A,\quad A_{ij}=\partial_x^{i-1}f_j,\quad i,j=1,2,\ldots, k.$$

\begin{theorem}[Crum, \cite{cr}]
Let  $\Psi_1,  \Psi_2, \ldots \Psi_n$ be solutions of the
Schr\"odinger equation \eqref{hams} for fixed, arbitrary energy
levels $E = E_1, E_2, \ldots,E_n$, respectively. Then, we obtain
the Schr\"odinger equation
$$H^{[n]}\Psi[n]=E\Psi[n],\quad E\neq E_i, 1\leq i\leq n, \quad H^{[n]}=-\partial_x^2 +
V[n],$$ where $$\Psi[n] = {W(\Psi_1,\ldots, \Psi_n,\Psi)\over
W(\Psi_1, \ldots, \Psi_n)},\quad V[n] = V - 2\partial_x^2 \ln
W(\Psi_1, \ldots, \Psi_n).$$
\end{theorem}

Darboux transformation coincides with Crum's result in the case
$n=1$ and the iterations of Darboux transformation coincides with
Crum iteration, see \cite{ornoro}. Both formalisms allow us to
obtain new families of Schr\"odinger equations preserving the
spectrum and the Sturm-Liouville conditions, see
\cite{masa,ornoro}. Furthermore, there are extensions of Crum's
iteration connecting the
Sturm-Liouville theory with orthogonal polynomial theory \cite{kr}.\\

Schr\"odinger in \cite{schr} factorized the Hypergeometric
equation \eqref{hypergeometric}. He started making the change of
variable $2x-1=\cos\theta$, after, using the expression
\eqref{redsec}, he reduced the Hypergeometric equation to obtain
conditions of factorization. In this way, setting $\kappa\beta=E$,
we obtain families of Schr\"odinger equations \eqref{hams}. This
result was used by Natanzon in \cite{nat} to obtain the well known
\textit{Natanzon's potentials}, i.e, potentials which can be
obtained by transformations of the Hypergeometric equation and its
confluences, see \cite{cokasu,cokasu2}. In particular, the
\textit{Ginocchio potentials} are obtained through the Gegenbauer
polynomials.\\

Witten in \cite[\S 6]{wi} presented some models in where
\textit{dynamical breaking of supersymmetry} is plausible. The
first model is a model in potential theory-supersymmetric quantum
mechanics, although is not a  model in the field theory.

\begin{definition}
 A \textit{supersymmetric quantum
mechanical system} is one in which there are operators $Q_i$, that
commute with the hamiltonian $\mathcal H$,
\begin{equation}\label{wit1} [Q_i,\mathcal H]=0,\quad i=1,\ldots,n
\end{equation} and satisfy the algebra \begin{equation}\label{wit2} \{Q_i,Q_j\}=\delta_{ij}\mathcal H,\quad \{Q_i,Q_j\}=Q_iQ_j+Q_jQ_i. \end{equation}
\end{definition}
The simplest example of a supersymmetric quantum mechanical system
corresponds to the case $n=2$, which in physical sense involves a
\textit{spin one half particle} moving on the line. This case is
the main object of this thesis. The wave function of $\mathcal
H\Phi=E\Phi$ is therefore a two-component \textit{Pauli spinor},
$$\Phi(x)=\begin{pmatrix}\Psi_+(x)\\\Psi_-(x)\end{pmatrix}.$$

The \textit{supercharges} $Q_i$ are defined as
\begin{equation}\label{wit3}
Q_\pm=\frac{\sigma_1p\pm\sigma_2W(x)}2,\quad Q_+=Q_1,\, Q_-=
Q_2,\quad p = -i\partial_x,
\end{equation} where the \textit{superpotential} $W$ is an arbitrary
function of $x$ and $\sigma_i$ are the usual \textit{Pauli spin
matrices}
$$ \sigma_1 =
\begin{pmatrix} 0&1\\ 1&0
\end{pmatrix},\,
    \sigma_2 = \begin{pmatrix} 0&-i\\ i&0
    \end{pmatrix},\,
    \sigma_3 = \begin{pmatrix} 1&0\\ 0&-1 \end{pmatrix}. $$
Using the expressions \eqref{wit1}, \eqref{wit2} and \eqref{wit3}
we obtain $\mathcal H$:
\begin{equation}\label{wit4} \mathcal H=2Q_-^2=2Q_+^2=\frac{I_2p^2+I_2W^2(x)+\sigma_3\partial_xW(x)}2,\quad I_2=\begin{pmatrix} 1&0\\ 0&1
    \end{pmatrix}.
\end{equation}
The \textit{supersymmetric partner Hamiltonians} $H_\pm$ are given
by
$$H_\pm=-{1\over 2}\partial_x^2+V_\pm,\quad V_\pm=\left({W\over \sqrt{2}}\right)^2\pm{1\over \sqrt{2}}\partial_x\left({W\over \sqrt{2}}\right).$$

The potentials $V_\pm$ are called \textit{supersymmetric partner
potentials} and are linked with the superpotential $W$ through a
Riccati equation. So that equation \eqref{wit4} can be written as
$$\mathcal H=\begin{pmatrix} H_+&0\\ 0&H_- \end{pmatrix},$$ which
lead us to the Schr\"odinger equations $H_+\Psi_+=E\Psi_+$ and
$H_-\Psi_-=E\Psi_-$, and for instance, to solve $\mathcal
H\Phi=E\Phi$ is equivalent to solve simultaneously
$H_+\Psi_+=E\Psi_+$ and $H_-\Psi_-=E\Psi_-$.\\

We analyze equation $Q_i\Phi = 0$, which must be satisfied by a
\textit{supersymmetric state}. This is due to $Q_1^2 = Q_2^2
=\frac12\mathcal H$ and for instance $Q_2 = -i\sigma_3Q_1$, which
implies that $Q_1\Phi = 0$, or $\sigma_1p\Phi = -\sigma_2W\Phi$.
Now, multiplying by $\sigma_1$ and using the facts that $p =
-i\partial_x$, $\sigma_1\sigma_2=i\sigma_3$, this equation becomes
\begin{equation}\label{wit5} \partial_x\Phi=
W(x)\sigma_3\Phi(x),
\end{equation} and the solution is \begin{equation}\label{wit55} \Phi(x)=e^{\int W(x)\sigma_3dx}.
\end{equation}

In agreement with \cite{wi}, one important generalization of the
previous model could be its extension to four dimensions, i.e., the case $n=4$.\\

V.B. Matveev and M. Salle in \cite{masa} interpret the Darboux
Theorem as Darboux covariance of a Sturm-Liouville problem and
proved the following result, see also \cite[\S 5-6]{ro}.\\

 \begin{theorem}[Matveev \& Salle, \cite{masa}]
The case $n=2$ in Supersymetric Quantum Mechanics is equivalent to
a single Darboux transformation.
\end{theorem}

According to Natanzon \cite{nat}, a \emph{solvable potential},
also known as \emph{exactly solvable potentials}, is a potential
in which the Schr\"odinger equation can be reduced to
hypergeometric or confluent hypergeometric form. The following are
examples of solvable potentials.

\begin{equation}\label{eqexp}
\begin{array}{ll}
\textbf{Potential} &\textbf{Name}\\&\\
 V(x)=\Big\{\begin{array}{ll} 0,& x\in[0,L] \\ \infty, & x \notin [0,L]\end{array}&\textit{Infinite square
well}
\\
V(r)=\Big\{\begin{array}{ll}{\ell(\ell+1)\over r^2},& r\in[0,L]\\
\infty, & r
\notin [0,L]\end{array}&\textit{Radial infinite square well}\\
V(r)= -{\mu\over e^{\kappa r}-1},&\textit{Hulth\'en}\\
V(r)= -{\mu\over e^{\kappa r}-1}+{2\kappa^2e^{\kappa r}\over
(e^{\kappa r}-1)^2},&\textit{Generalized Hulth\'en}
\\ V(x) =\frac14\omega^2x^2+g_a{x^2-a^2\over (x^2+a^2)^2},\quad g_a>0& -- -- -- \cite{caperasa}
\end{array}
\end{equation}

We remark that our definition of integrability, definition
\ref{integrability},  is different of the concept of solvability
given by Natanzon. In the next chapter we come back on this
problem.\\

According to Dutt et al. \cite{dukava}, a \emph{conditionally
solvable potential} is a potential in which the entire bound state
spectrum can be analytically obtained, where the parameters in the
potential satisfies a specific relation. The following potentials
are two examples of conditionally solvable potentials
\begin{equation}\label{cspdukava}
V(x)={A\over 1+e^{-2x}} - {B\over \sqrt{1+e^{-2x}}} - {3\over
4(1+e^{-2x})^2}\quad \text{and}$$ $$V(x)={A\over 1+e^{-2x}} -{B
e^{-x}\over \sqrt{e^{-2x}+1}} - {3\over
4(1+e^{-2x})^2}.\end{equation}

As a generalization of the method to solve the harmonic oscillator
\cite{ge,duka}, the ladder (raising and lowering) operators are
defined as
$$A^+=-\partial_x-{\partial_x\Psi_0\over \Psi_0},\quad
A=\partial_x-{\partial_x\Psi_0\over \Psi_0},$$ which are very
closed with the supercharges $Q_\pm$ in the Witten's formalism.
Thus,
$$A\Psi_0=0,\quad A^+A=H_-,\quad AA^+=H_+=-\partial_x^2+V_+(x),\quad \textrm{where}$$
$$V_+(x)=V_-(x)-2\partial_x\left({\partial_x\Psi_0\over\Psi_0}\right)=-V_-(x)+2\left({\partial_x\Psi_0\over\Psi_0}\right)^2.$$
The supersymmetric partner potentials $V_+$ and $V_-$ have the
same energy levels, except for $E_0^{(-)}=0$. In terms of the
superpotential $W(x)$, the operators $A$ and $A^+$ are given by
$$A^+=-\partial_x+W(x),\quad A=\partial_x+W(x).$$

In the same way, the supersymmetric partner potentials
$V_{\pm}(x)$ and the superpotential $W(x)$ satisfies:
$${V_+(x) + V_-(x)\over 2}=W^2(x),\quad [A,A^+]=2\partial_x W(x).$$
Let $\Psi_n^{(-)}$ and $\Psi_n^{(+)}$ denote the eigenfunctions of
the supersymmetric Hamiltonians $H_-$ and $H_+$ respectively, with
eigenvalues $E_n^{(-)}$ and $E_n^{(+)}$.  The integer
$n=0,1,2,\ldots,$ denotes the number of nodes in the wave
function.

\begin{theorem}[Dutt et al., \cite{duka}]\label{susypo}
If $\Psi_n^{(-)}$ is any eigenfunction of $H_-$ with eigenvalue
$E_n^{(-)}$, then $A\Psi_n^{(-)}$ is an eigenfunction of $H_+$
with the same eigenvalue. Furthermore
$$E_n^{(+)}=E_{n+1}^{(-)},\quad \Psi_n^{(+)}={A\over\sqrt{E_{n+1}^{(-)}}}\Psi_{n+1}^{(-)}.$$
\end{theorem}

Considering $E_0=0$ and $n>0$ in theorem \ref{susypo}, we can see
that the supersymmetric partner potentials $V_+$ and $V_-$ have
the same spectrum. The ground state energy $E_0=0$ of $V_-$ has no
corresponding level for $V_+$. Furthermore, if the eigenfunction
$\Psi_{n+1}^{(-)}$ of $H_-$ is normalized, then also
the eigenfunction $\Psi_{n}^{(+)}$ of $H_+$ is normalized.\\

The operator $A$ converts an eigenfunction of $H_-$ into an
eigenfunction of $H_+$ with the same energy, whilst the operator
$A^+$ converts an eigenfunction of $H_+$ into an eigenfunction of
$H_-$ with the same energy. Furthermore, the operator $A$ destroys
a node ($\Psi_{n+1}^{(-)}$ has $n+1$ nodes, whilst
$\Psi_{n}^{(+)}$ has $n$ nodes) and the operator $A^+$ creates a
node. In summary, the \textit{annihilator and creator operators}
($A$ and $A^+$ respectively) connect states of the same energy for
two different supersymmetric partner potentials.\\

Gendenshte\"{\i}n, in his remarkable paper \cite{ge}, introduced
the concept of \textit{shape invariance}, which is a property or
condition of some classes of potentials with respect to their
parameters. Assuming $a$ as a family of parameters, the shape
invariance condition can be seen such as follows:
$$V_{n+1}(x; a_n) = V_n(x; a_{n+1}) + R(a_n),\quad V_-=V_0,\quad V_+=V_1,$$ where $R$ is a
remainder, which does not depends on $x$. In this way we say that
$V=V_-=V_0$ is a \textit{shape invariant potential}.\\

The potentials, corresponding to a Schr\"odinger operator,
satisfying the shape invariance property allow us to obtain a
fully algebraic scheme for the spectrum and wave functions. This
is illustrated in the following theorem obtained by
Gendenshte\"{\i}n.

\begin{theorem}[Gendenshte\"{\i}n, \cite{ge}]\label{geth} Consider the
Schr\"odinger equation $H\Psi=E_n\Psi$, where $V=V_-=V_0$ is a
shape invariant potential. If we fix the first level of energy
$E_0 = 0$, then the excited spectrum and the wave functions are
given respectively by
\begin{equation}\label{dukage} E_n = \sum_{k=2}^{n+1}R(a_k),\quad \Psi_n^{(-)}(x;
a_1) = \prod_{k=1}^nA^+(x; a_k)\Psi^{(-)}_0(x;
a_{n+1}).\end{equation}
\end{theorem}
\medskip

The complete statement and proof of theorem \ref{geth} can be
found in \cite{ge,duka}. As consequences we have the following
facts (see \cite{ge,duka}):

\begin{itemize}
\item $$\Psi_n^{(+)}(x;a_0)=\Psi_{n}^{(-)}(x;a_1),\quad \Psi_{n+1}^{(-)}(x;a_0)=A^\dagger(x;a_0)\Psi_n^{(-)}(x;a_1).$$
\item $$H^{(n)}=-\partial_x^2+V_-(x;a_n)+\sum_{k=1}^nR(a_k)$$ has the same spectrum
of $H^{(n+1)}$ for $n>0$, where $$H^{(0)}=H_-,\quad
H^{(1)}=H_+,\quad a_n=f^n(a_0).$$
\item $$H^{(n+1)}=-\partial_x^2+V_+(x;a_n)+\sum_{k=1}^{n}R(a_k).$$
\end{itemize}
Following \cite{cokasu,duka} we present the list of shape
invariant potentials given in expression (\ref{eqshin}).

\begin{equation}\label{eqshin}
\begin{array}{ll}
\textbf{Potential} \quad $V$&\textbf{Name}\\&\\
{1\over 4}\omega^2x^2-{\omega\over 2}&\textit{Harmonic Oscillator}\\&\\
{1\over 4}\omega^2r^2+{\ell(\ell+1)\over r^2}-\left(\ell+{3\over
2}\right)\omega&\textit{3D Harmonic Oscillator}\\&\\ -{e^2\over
r}+{\ell(\ell+1)\over
r^2}+{e^4\over8(\ell+1)^2}&\textit{Coulomb}\\&\\
A^2+B^2e^{-2ax}-2B\left(A+{a\over2}\right)e^{-ax}&\textit{Morse}\\&\\
A^2+{B^2\over A^2}-2B\coth ar+A{A-a\over
\sinh^2 ar}&\textit{Eckart}\\&\\
A^2+{B^2\over A^2}+2B\tanh ax-A{A+a\over
\cosh^2 ax}&\textit{Rosen-Morse Hyp.}\\&\\
-A^2+{B^2\over A^2}+2B\cot ax+A{A+a\over
\sin^2 ar}&\textit{Rosen-Morse Trig.}\\&\\
A^2+{B^2-A^2-Aa\over \cosh^2ax}+{B\left(2A+a\right)\sinh ax \over
\cosh^2 ax}&\textit{Scarf Hyp. I}\\&\\
A^2+{B^2+A^2+Aa\over \sinh^2ar}-{B\left(2A+a\right)\cosh ar \over
\sinh^2 ar}&\textit{Scarf Hyp. II}\\&\\
-A^2+{B^2+A^2-Aa\over \cos^2ax}-{B\left(2A-a\right)\sin ax \over
\cos^2 ax}&\textit{Scarf Trig. I}\\&\\
-A^2+{B^2+A^2-Aa\over \sin^2ax}-{B\left(2A-a\right)\cos ax \over
\sin^2 ax}&\textit{Scarf Trig. II}\\&\\
-(A+B)^2+{A\left(A-a\right)\over\cos^2
ax}+{B\left(B-a\right)\over\sin^2
ax}&\textit{P\"oschl-Teller 1}\\&\\
(A-B)^2-{A\left(A+a\right)\over\cosh^2
ar}+{B\left(B-a\right)\over\sinh^2 ar}&\textit{P\"oschl-Teller
2}\end{array}
\end{equation}
\\

We recall that a good short survey about Darboux transformations
can be found in \cite{ro}.

\chapter[Differential Galois Theory Approach to SUSY QM]{Differential Galois Theory Approach to Supersymmetric Quantum Mechanics}
\label{darboux}

 In this chapter we present our original results of this thesis,
 which corresponds to the Galoisian approach to Supersymmetric quantum mechanics. We start rewriting in a Galoisian context some
points of the section \ref{susysection}, chapter \ref{background}.
The results presented here are also true for any differential
field $K$ with field of constants $\mathcal C$ in agreement with
definition \ref{defdiff}. We emphasize in the differential fields
$K_0=\mathbb{C}(x)$ and $K_1=\mathbb{C}(z(x),\partial_xz(x))$,
where in both cases $\mathcal C=\mathbb{C}$.
\section{Preliminaries}\label{algebrizationsec}
The main object of our Galoisian analysis is the Schr\"odinger
equation \eqref{hams}, which now is written as
\begin{equation}\label{equ1}
\mathcal L_\lambda:=H\Psi=\lambda\Psi, \quad
H=-\partial_x^2+V(x),\quad V\in K,
\end{equation}
where $K$ is a differential field (with $\mathbb{C}$ as field of
constants). We are interested in the integrability of equation \eqref{equ1} in agreement with definition \ref{integrability}.\\

We introduce the following notations.
\begin{itemize}
\item Denote by $\Lambda\subseteq{\mathbb{C}}$ the set of
eigenvalues $\lambda$ such that equation (\ref{equ1}) is
integrable according with definition \ref{integrability}.
\item Denote by $\Lambda_+$ the set $\{\lambda\in\Lambda\cap \mathbb{R}: \lambda\geq 0\}$ and by
$\Lambda_-$ the set $\{\lambda\in\Lambda\cap \mathbb{R}:
\lambda\leq 0\}$.
\item Denote by $L_\lambda$ the Picard-Vessiot extension of $\mathcal
L_\lambda$. Thus, the differential Galois group of $\mathcal
L_\lambda$ is denoted by $\mathrm{DGal}(L_\lambda/K)$.
\end{itemize}

 The set $\Lambda$ will be called \textit{the algebraic spectrum} (or alternatively \textit{the Liouvillian spectral set}) of
$H$. We remark that $\Lambda$ can be $\emptyset$, i.e.,
$\mathrm{DGal}(L_\lambda/K)={\rm SL}(2,\mathbb{C})$ $\forall
\lambda \in \mathbb{C}.$ On the other hand, by theorem \ref{LK},
if $\lambda_0\in \Lambda$ then
$(\mathrm{DGal}(L_{\lambda_0}/K))^0\subseteq\mathbb{B}$.\\

\begin{definition}[Algebraically Solvable and Quasi-Solvable Potentials] We say that the
potential $V(x)\in K$ is:
\begin{itemize}
\item an \textit{algebraically solvable potential} when $\Lambda$ is an infinite set, or
\item an \textit{algebraically quasi-solvable potential} when $\Lambda$ is a non-empty finite
set, or
\item an \textit{algebraically non-solvable potential} when $\Lambda=\emptyset$.
\end{itemize}
When $\textrm{Card}(\Lambda)=1$, we say that $V(x)\in K$ is a
\textit{trivial} algebraically quasi-solvable potential.
\end{definition}
\medskip

\begin{examples} Assume $K=\mathbb{C}(x)$.
\begin{enumerate}
\item If $V(x)=x$, then
$\Lambda=\emptyset$, $V(x)$ is algebraically non-solvable, see
\cite{ka,ko}.
\item If $V(x)=0$, then $\Lambda=\mathbb{C}$, i.e., $V(x)$ is algebraically solvable. Furthermore,
$${\mathrm{DGal}}(L_0/K)=e,\quad {\mathrm{DGal}}(L_{\lambda}/K)=\mathbb{G}_m,\quad \lambda\neq 0.$$
\item If $V(x)={x^2\over 4}+\frac12$, then $\Lambda=\{n:
n\in\mathbb{Z}\}$, $V(x)$ is algebraically solvable. This example
corresponds to the Weber's equation, see subsection
\ref{riemansection}.
\item If $V(x)=x^4-2x$, then $\Lambda=\{0\}$, $V(x)$ is
algebraically quasi-solvable (trivial).
\end{enumerate}
\end{examples}
\medskip

\begin{remark}
We can obtain algebraically solvable and quasi-solvable potentials
in the following ways.
\begin{itemize}
\item Giving the potential $V$, we try to solve the differential equation $\partial_x^2\Psi_0=V\Psi_0$ expecting to obtain the superpotential
 $W=\partial_x\ln(\Psi_0)$. If the superpotential exists in the Liouvillian class (the differential equation is integrable), then
we search $\Lambda$ into the Schr\"odinger equation
$H\Psi=\lambda\Psi$. With this method the algebraic spectrum can
be the empty set. We will illustrate this in section 2.2 and 2.3.

\item Giving the superpotential $W$, we construct the potential
$V=\partial_xW+W^2$. After we can search the algebraic spectrum
$\Lambda$ in $H\Psi=\lambda\Psi$. With this method we have that
$0\in\Lambda$, so at least we obtain trivial algebraically
quasi-solvable potentials. We will illustrate this in section 2.2
and 2.3.

\item Using integrable parameterized differential equations which can be transformed into
Schr\"odinger equations. In this case the parameter of the
differential equation should coincide with the eigenvalues of $H$.
With this method we know previously the algebraic spectrum
$\Lambda$, thus, we can obtain algebraically solvable or
algebraically quasi-solvable potentials depending on $\Lambda$. We
will illustrate this in section 2.3.
\end{itemize}
\end{remark}
\medskip

\noindent We are interested in the spectrum (analytic spectrum) of
the algebraically solvable and quasi-solvable potentials, that is,
$\mathrm{spec}(H)\cap \Lambda\neq\emptyset$. For example, the
potential $V(x)=|x|$ has point spectrum (see \cite{bana}) although
$V(x)$ is algebraically non-solvable. Thus, when
$\mathrm{spec}(H)\cap \Lambda$ is an infinite set, in the usual
physical terminology these potentials are called \textit{solvable
(or exactly solvable) potentials}, see Natanzon \cite{nat}. In
analogous way, when $\mathrm{spec}(H)\cap \Lambda$ is a finite
set, the usual definition in physics of these potentials is
\textit{quasi-exactly solvable (or quasi-solvable) potentials}
(Turbiner \cite{tu}, Bender \& Dunne \cite{bedu}, Bender \&
Boettcher \cite{bebo}, Saad et al. \cite{sahaci}, Gibbons \&
Vesselov \cite{give}).\\

\begin{definition}\label{deftriso}
Let be $\mathcal{L}$, $\widetilde{\mathcal{L}},$ pairs of linear
differential equations defined over differential fields $K$ and
$\widetilde{K}$ respectively, with Picard-Vessiot extensions $L$
and $\widetilde{L}$. Let $\varphi$ be the transformation such that
$\mathcal{L}\mapsto\widetilde{\mathcal{L}}$,
$K\mapsto\widetilde{K}$ and $L\mapsto \widetilde{L}$, we say that:
\begin{enumerate}
\item $\varphi$ is an {\it iso-Galoisian transformation} if
$$\mathrm{DGal}(L/K)=\mathrm{DGal}(\widetilde{L}/\widetilde{K}).$$
If $\widetilde{L}=L$ and $\widetilde{K}=K$, we say that $\varphi$
is a {\it strong iso-Galoisian transformation}.
\item $\varphi$ is a {\it virtually iso-Galoisian transformation} if $$(\mathrm{DGal}(L/K))^0=(\mathrm{DGal}(\widetilde{L}/\widetilde{K}))^0.$$
\end{enumerate}
\end{definition}

\begin{remark}\label{prop1} The Eigenrings of two operators $\mathfrak
L$ and $\widetilde{\mathfrak L}$ are preserved under iso-Galoisian
transformations.
\end{remark}

\begin{proposition}\label{rlig} Consider the
differential equations
$$\mathcal{L}:=\partial_x^2y+a\partial_xy+by=0,\quad \widetilde{\mathcal{L}}:=\partial_x^2\zeta=r\zeta,\quad a,\,b,\,r\,\in K.$$
Let $\kappa\in\mathbb{Q}$, $f\in K$, $a=2\kappa\partial_x(\ln f)$
and $\varphi$ be the transformation such that
$\mathcal{L}\mapsto\widetilde{\mathcal{L}}$. The following
statements holds:

\begin{enumerate}
\item $\varphi$ is a strong isogaloisian transformation for
$\kappa\in\mathbb{Z}$.
\item $\varphi$ is a virtually strong isogaloisian transformation for
$\kappa\in\mathbb{Q}\setminus\mathbb{Z}$.
\end{enumerate}
\end{proposition}

\begin{proof} Assume that $\mathcal{B}=\{y_1,y_2\}$ is a basis
of solutions and $L$ is the Picard-Vessiot extension of
$\mathcal{L}$, $\mathcal{B}'=\{\zeta_1,\zeta_2\}$ is a basis of
solutions and $\widetilde L$ is the Picard-Vessiot extension of
$\widetilde{\mathcal{L}}$. With the change of dependent variable
$y=\zeta e^{-\frac12\int a}$ we obtain $r=a^2/4+\partial_x a/2-b$
and for instance $K=\widetilde K$. Thus, the relationship between
$L$ and $\widetilde L$ depends on $a$:
\begin{enumerate}
\item If $\kappa=n\in\mathbb{Z}$, then $\mathcal{B}'=\{f^ny_1,f^ny_2\}$
which means that $L=\widetilde L$ and $\varphi$ is strong
isogaloisian.
\item If $\kappa=\frac{n}m$, with $\gcd(n,m)=1$, $\frac{n}m\notin\mathbb{Z}$, then $\mathcal{B}'=\{f^{n\over m} y_1,f^{n\over m}y_2\}$
which means that $\widetilde L$ is either an algebraic extension
of degree at most $m$ of $L$, and $\varphi$ is virtually strong
isogaloisian, or $L=\widetilde L$ when $f^{\frac{n}m}\in K$ which
means that $\varphi$ is strong isogaloisian.
\end{enumerate}
\end{proof}
\begin{remark} The transformation $\varphi$ in proposition \ref{rlig} is
not injective, there are a lot of differential equations
$\mathcal{L}$ that are transformed in the same differential
equation $\widetilde{\mathcal{L}}$.
\end{remark}
\medskip

As immediate consequence of the previous proposition we have the
following corollary.\\

\begin{corollary}[Sturm-Liouville] Let $\mathcal{L}$ be the differential equation $$\partial_x\left(a \partial_xy\right)=(\lambda b-\mu)y,\quad
a,b\in K,\quad \lambda,\mu\in\mathbb{C}$$ in where $L$,
$\widetilde L$, $\widetilde{\mathcal{L}}$ and $\varphi$ are given
as in proposition \ref{rlig}. Then either $\widetilde L$ is a
quadratic extension of $L$ which means that $\varphi$ is virtually
strong isogaloisian or $\widetilde L=L$ when $a^{\frac12}\in K$
which means that $\varphi$ is strong isogaloisian.
\end{corollary}

\section{Supersymmetric Quantum Mechanics with Rational
Potentials}\label{susyrapo} Along this section we consider as
differential field $K=\mathbb{C}(x)$.

\subsection{Polynomial Potentials}
 We start considering the Schr\"odinger equation
\eqref{equ1} with polynomial potentials, i.e.,
$V\in\mathbb{C}[x]$, see \cite{bo,va1}. For simplicity and without
lost of generality, we consider monic polynomials due to the
reduced second order linear differential equation with polynomial
coefficient $c_nx^n+\ldots+c_1x+ c_0$ can be transformed into the
reduced second order linear differential equation with polynomial
coefficient $x^n+\ldots+q_1x+q_0$ through the change of variable
$x\mapsto \sqrt[n+2]{c_n} x$.\\

When $V$ is a polynomial of odd degree, is well known that the
differential Galois group of the Schr\"odinger equation
\eqref{equ1} is
$\mathrm{SL}(2,\mathbb{C})$, see \cite{ko}.\\

We present here the complete result for the Schr\"odinger equation
\eqref{equ1} with non-constant polynomial potential (Theorem
\ref{polynint}), see also \cite[\S 2]{acbl}. The following lemma
is useful for our purposes.\\

\begin{lemma}[Completing Squares, \cite{acbl}]\label{cosq}
Every even degree monic polynomial of  can be written in one only
way completing squares, that is,
\begin{equation}\label{square}
Q_{2n}(x)=x^{2n}+\sum_{k=0}^{2n-1}q_kx^k=\left(x^n+\sum_{k=0}^{n-1}a_kx^k\right)^2+\sum_{k=0}^{n-1}b_kx^k,
\end{equation}
where
$$a_{n-1}={q_{2n-1}\over 2},\quad a_{n-2}={q_{2n-2}-a^2_{n-1}\over 2},\quad a_{n-3}={q_{2n-3}-2a_{n-1}a_{n-2}\over 2},\cdots,$$

$$a_0={q_n-2a_1a_{n-1}-2a_2a_{n-2}-\cdots\over 2},\quad
b_0=q_0-a_0^2,\quad b_1=q_1-2a_0a_1,\quad \cdots,$$
$$b_{n-1}=q_{n-1}-2a_0a_{n-1}-2a_1a_{n-2}-\cdots.$$
\end{lemma}
\medskip
\begin{proof} See lemma 2.4 in \cite[p. 275]{acbl}.

\end{proof}
\medskip

We remark that $V(x)$ as in equation \eqref{square} can be written
in terms of the superpotential $W(x)$, i.e.,
$V(x)=W^2(x)-\partial_xW(x)$, when
$$nx^{n-1}+\sum_{k=1}^{n-1}ka_kx^{k-1}=-\sum_{k=0}^{n-1}b_kx^k$$
and $W(x)$ is given by $$x^n+\sum_{k=0}^{n-1}a_kx^k.$$

The following theorem also can be found in \cite[\S 2]{acbl}, see
also \cite{acbl2}. Here we present a quantum mechanics adapted
version.\\

\begin{theorem}[Polynomial potentials and Galois groups, \cite{acbl}]\label{polynint}
Let us consider the Schr\"odinger equation \eqref{equ1}, with
$V(x)\in\mathbb{C}[x]$ a polynomial of degree $k>0$. Then,  its
differential Galois group $\mathrm{DGal}(L_{\lambda}/K)$ falls in
one of the following cases:
\begin{enumerate}
\item $\mathrm{DGal}(L_{\lambda}/K)=\mathrm{SL}(2,\mathbb{C})$,
\item $\mathrm{DGal}(L_{\lambda}/K)=\mathbb{B}$,
\end{enumerate}
and the Eigenring of $H-\lambda$ is trivial, i.e., $\mathcal
E(H-\lambda)=\mathrm{Vect}(1)$. Furthermore,
$\mathrm{DGal}(L_{\lambda}/K)=\mathbb{B}$ if and only if the
following conditions hold:
\begin{enumerate}
\item $V(x)-\lambda$ is a polynomial of degree $k=2n$ writing in the form of equation
\eqref{square}.
\item $b_{n-1}-n$ or $-b_{n-1}-n$ is a positive even number $2m$, $m\in\mathbb{Z}_+$.
\item There exist a monic polynomial $P_m$ of degree $m$,
satisfying {\small \begin{displaymath}
\partial_x^2P_m + 2\left(x^n+\sum_{k=0}^{n-1}a_kx^k\right)\partial_xP_m +
\left(nx^{n-1}+\sum_{k=0}^{n-2}(k+1)a_{k+1}x^{k} -
\sum_{k=0}^{n-1}b_kx^k\right)P_m = 0,
\end{displaymath}}
or {\small \begin{displaymath}
\partial_x^2P_m - 2\left(x^n+\sum_{k=0}^{n-1}a_kx^k\right)\partial_xP_m -
\left(nx^{n-1}+\sum_{k=0}^{n-2}(k+1)a_{k+1}x^{k} +
\sum_{k=0}^{n-1}b_kx^k\right)P_m = 0.
\end{displaymath}}

\end{enumerate}
In such cases, the only possibilities for eigenfunctions with
rational superpotentials are given by
$$
\Psi_\lambda=P_me^{f(x)},\quad \textit{or}\quad
\Psi_\lambda=P_me^{-f(x)},\quad \textit {   where
}f(x)={x^{n+1}\over n+1}+\sum_{k=0}^{n-1}{a_kx^{k+1}\over k+1}.
$$
\end{theorem}
\medskip
\begin{proof} See theorem 2.5 in \cite[p. 276]{acbl}.

\end{proof}
\medskip
An easy consequence of the above theorem is the following.\\

\begin{corollary} Assume that $V(x)$ is an algebraically solvable polynomial
potential. Then $V(x)$ is of degree $2$.
\end{corollary}
\medskip
\begin{proof} Writing $V(x)-\lambda$ in the form of equation
\eqref{square} we see that $b_{n-1}-n=2m$ or $-b_{n-1}-n=2m$,
where $m\in\mathbb{Z}_+$. Thus, the integrability of the
Schr\"odinger equation  with $\mathrm{Card}(\Lambda)>1$ is
obtained when $b_{n-1}$ is constant, so $n=1$.
\end{proof}
\medskip

\begin{remark}\label{rembs} Given a polynomial potential $V(x)$
 such that $\mathrm{spec}_p(H)\cap\Lambda\neq \emptyset$, we can obtain bound
states and normalized wave functions if and only if the potential
$V(x)$ is a polynomial of degree $4n+2$. Furthermore, one
integrability condition of $H\Psi=\lambda\Psi$ for
$\lambda\in\Lambda$ is that $b_{2n}$ must be an odd integer. In
particular, if the potential
$$V(x)=x^{4n+2n}+\mu x^{2n},\quad n>0$$ is
a quasi-exactly solvable, then $\mu$ is an odd integer. For this
kind of potentials, we obtain bound states only when $\mu$ is a
negative odd integer.\\

On another hand, the non-constant polynomial potentials $V(x)$ of
degree $4n$ are associated to non-hermitian Hamiltonians and
$\mathcal{PT}$ invariance which are not considered here, see
\cite{bebo}. Furthermore, one integrability condition of
$H\Psi=\lambda\Psi$ for $\lambda\in\Lambda$ is that $b_{2n-1}$
must be an even integer. In particular, if the Schr\"odinger
equation
$$H\Psi=\lambda\Psi, \quad V(x)=x^{4n}+\mu x^{2n-1}, \quad \lambda\in\Lambda$$ is
integrable, then $\mu$ is an even integer.
\end{remark}
\medskip

 We present the following examples to illustrate the
previous
theorem and remark.\\

\noindent \textbf{Weber's Equation and Harmonic Oscillator.} The
Schr\"odinger equation with potential $V(x)=x^2+q_1x+q_0$
corresponds to the Rehm's form of the Weber's equation
\eqref{rehm}, which has been studied in section
\ref{riemansection}. By lemma \ref{cosq} we have
$$V(x)-\lambda=(x+a_0)^2+b_0,\quad a_0=q_1/2, \quad
b_0=q_0-q_1^2/4-\lambda.$$ So that we obtain $\pm b_0-1=2m$, where
$m\in\mathbb{Z}_+$. If $b_0$ is an odd integer, then
$$\mathrm{DGal}(L_{\lambda}/K)=\mathbb{B}, \, \mathcal E(H-\lambda)=\mathrm{Vect}(1),\,
\lambda\in\Lambda=\{\pm(2m+1)+q_0-q_1^2/4:m\in\mathbb{Z}_+\}$$ and
the set of eigenfunctions is either
$$\Psi_\lambda=P_me^{\frac12(x^2+q_1x)},\textit{ or, }
\Psi_\lambda=P_me^{-\frac12(x^2+q_1x)}.$$ In the second case we
have bound state wave function and
$\mathrm{spec}_p(H)\cap\Lambda=\mathrm{spec}_p(H)=\{E_m=2m+1+q_0-q_1^2/4:m\in\mathbb{Z}_+\}$,
which is infinite. The polynomials $P_m$ are related with the
Hermite polynomials $H_m$, \cite{ch,is,niuv}.\\

In particular we have the harmonic oscillator potential, which is
given in the list (\ref{eqshin}) and where $H\Psi=E\Psi$. Through
the change of independent variable $x\mapsto\sqrt{2\over \omega}x$
we obtain $V(x)=x^2-1$ and $\lambda={2\over\omega}E$, that is,
$q_1=0$ and $q_0=-1$. In this way $\Lambda=\{\pm(
2m+1)-1:m\in\mathbb{Z}_+\}$ and the set of eigenfunctions is
either
$$\Psi_\lambda=P_me^{\frac12x^2},\textit{ or, }
\Psi_\lambda=P_me^{-\frac12x^2},$$ where as below,
$\mathrm{DGal}(L_{\lambda}/K)=\mathbb{B}$ and
$\mathcal{E}(H-\lambda)=\{1\}$ for all $\lambda\in\Lambda$. In the
second case we have bound state wave function,
$\mathrm{spec}_p(H)\cap\Lambda=\mathrm{spec}_p(H)=\Lambda_+=\{2m:m\in\mathbb{Z}_+\}$
and $P_m=H_m$. The wave functions of $H\Psi=E\Psi$ for the
harmonic oscillator potential are given by

$$\Psi_m=H_m\left(\sqrt{2\over \omega}x\right)\Psi_0,\quad \Psi_0=e^{-{\omega\over4}x^2},\quad E=E_m=m\omega.$$
\medskip

\noindent \textbf{Quartic and Sextic Anharmonic Oscillator.} The
Schr\"odinger equation with potential
$V(x)=x^4+q_3x^3+q_2x^2+q_1x+q_0$ can be obtained through
transformations of confluent Heun's equation, which is not
considered here.  By lemma \ref{cosq} we have
$$V(x)-\lambda=(x^2+a_1x+a_0)^2+b_1x+b_0,$$ where $a_1=q_3/2$, $a_0=q_2/2-a_1^2/2$, $b_1=q_1-2a_0a_1$ and $b_0=q_0-a_0^2-\lambda.$
 So that we obtain $\pm b_1-2=2m$, where
$m\in\mathbb{Z}_+$. If $\Lambda\neq\emptyset$, then $b_1$ is an
even integer, $P_m$ satisfy the relation \eqref{recu1} and
$\mathrm{DGal}(L_{\lambda}/K)=\mathbb{B}$ for all
$\lambda\in\Lambda$. The set of eigenfunctions is either
$$\Psi_\lambda=P_me^{\frac{x^3}3+\frac{a_1x^2}2+a_0x},\textit{ or, }
\Psi_\lambda=P_me^{-\left(\frac{x^3}3+\frac{a_1x^2}2+a_0x\right)},$$
where $\lambda$ and $m$ are related, which means that $\Lambda$ is
finite, i.e., the potential is algebraically quasi-solvable. In
particular for $q_3=2\imath l$, $q_2=l^2-2k$, $q_1=2\imath(lk-J)$
and $q_0=0$, we have the quartic anharmonic oscillator potential,
which can be found in \cite{bebo}.\\

Now, considering the potentials $V(x,\mu)=x^4+4x^3+2x^2-\mu x$,
again by lemma \ref{cosq} we have that
$$V(x,\mu)-\lambda=(x^2+2x-1)^2+(4-\mu)x-1-\lambda,$$ so that $\pm(4-\mu)-2=2n$, where $n\in\mathbb{Z}_+$ and in consequence $\mu\in2\mathbb{Z}$. Such $\mu$ can be either
$\mu=2-2n$ or $\mu=2n+6$, where $n\in\mathbb{Z}_+$. By theorem
\ref{polynint}, there exists a monic polynomial $P_n$ satisfying
respectively
$$\partial_x^2P_n+(2x^2+4x-2)\partial_xP_n+((\mu-2)x+3+\lambda)P_n=0,\quad \mu=2-2n,\quad\textrm{or}$$
$$\partial_x^2P_n-(2x^2+4x-2)\partial_xP_n+((\mu-6)x-1+\lambda)P_n=0,\quad\mu=2n+6$$
 for $\Lambda\neq\emptyset$. This algebraic relation between the coefficients of $P_n$, $\mu$ and $\lambda$ give us the set
 $\Lambda$ in the following way:

 \begin{enumerate}
 \item Write
 $P_n=x^n+c_{n-1}x^{n-1}+\ldots+c_0$, where $c_i$ are unknown.
 \item Pick $\mu$ and replace $P_n$ in the algebraic relation \eqref{recu1} to obtain a polynomial of degree $n$ with $n+1$ undetermined coefficients
involving $c_0,\ldots c_{n-1}$ and $\lambda$. Each of such
coefficients must be zero.\item  The term $n+1$ is linear in
$\lambda$ and $c_{n-1}$, thus we write $c_{n-1}$ in terms of
$\lambda$. After of the elimination of the term $n+1$, we replace
$c_{n-1}$ in the term $n$ to obtain a quadratic polynomial in
$\lambda$ and so on until arrive to the constant term which is a
polynomial of degree $n+1$ in $\lambda$ ($Q_{n+1}(\lambda)$). In
this way, $\Lambda=\{\lambda:Q_{n+1}(\lambda)=0\}$ and
$c_0,\ldots,c_{n-1}$ are determined for each value of $\lambda$.
\end{enumerate}

For $\mu=2n+6$, we have:

$$\begin{array}{llll}
n=0,& V(x,6),&P_0=1,& \Lambda=\{1\}\\
n=1,& V(x,8),&P_1=x+1\mp\sqrt2,& \Lambda=\{3\pm2\sqrt2\}\\
\vdots& & &
 \end{array}$$ and the set of eigenfunctions is
$$\Psi_{\lambda,\mu}=P_ne^{-\frac13x^3-x^2+x}.$$

In the same way, we can obtain $\Lambda$, $P_n$ and
$\Psi_{\lambda,\mu}$ for $\mu=2-2n$. However, we have not bound
states, $\mathrm{spec}_p(H)\cap\Lambda=\emptyset$,
$\mathrm{DGal}(L_{\lambda}/K)=\mathbb{B}$ and
$\mathcal{E}(H-\lambda)=\mathrm{Vect}(1)$ for all $\lambda\in\Lambda$.\\

The well known \textit{sextic anharmonic oscillator}
$x^6+q_5x^5+\cdots+q_1x+q_0$ can be treated in a similar way,
obtaining bound states wave functions and the Bender-Dunne
orthogonal polynomials, which corresponds to $Q_{n+1}(\lambda)$,
i.e., we can have the same results of \cite{bedu,give,sahaci}. The
Schr\"odinger equation with this potential, under suitable
transformations, also falls in a confluent Heun's equation.\\

\subsection{Rational Potentials and Kovacic's Algorithm}
In this subsection we apply Kovacic's algorithm to solve the
Schr\"odinger equation with rational potentials listed in the
previous chapter (equation \eqref{eqshin}).\\

 \textbf{Three dimensional harmonic oscillator
potential:}
$$V(r)={1\over 4}\omega^2r^2+{\ell(\ell+1)\over r^2}-\left(\ell+{3\over 2}\right)\omega,\quad \ell\in\mathbb{Z},$$ we
can see that Schr\"odinger equation (equation (\ref{hams})) for
this case can be written as
$$\partial_r^2\Psi=\left(\left(\frac12\omega
r\right)^2+{\ell(\ell+1)\over
r^2}-\left(\ell+\frac32\right)\omega-E\right)\Psi.$$ By the change
$r\mapsto\left(\sqrt{2\over\omega}\right)r$ we obtain the
Schr\"odinger equation
$$\partial_r^2\Psi=\left(r^2+{\ell(\ell+1)\over r^2}-(2\ell+3)-\lambda\right)\Psi,\quad \lambda={2\over
\omega}E$$ and in order to apply Kovacic's algorithm, we denote:
$$R=r^2+{\ell(\ell+1)\over r^2}-(2\ell+3)-\lambda.$$

We can see that this equation could fall in case 1, in case 2 or
in case 4 (of Kovacic's algorithm). We start discarding the case 2
because by step 1 (of Kovacic's algorithm) we should have
conditions $c_2$ and $\infty_3$, in this way we should have
$E_c=\{2,4+4\ell,-4\ell\}$ and $E_{\infty}=\{-2\}$, and by step 2,
we should have that $n=-4\notin\mathbb{Z}_+$, so that
$D=\emptyset$, that is, this Schr\"odinger equation never falls in
case 2. Now, we only work with case 1; by step 1, conditions $c_2$
and $\infty_3$ are satisfied, so that
$$\left[ \sqrt {R}\right] _{c}=0,\quad\alpha_{c}^{\pm}={1\pm(2\ell+1)\over 2}, \quad\left[  \sqrt{R}\right]  _{\infty}=r,\quad \alpha_{\infty}^{\pm }={\mp(\lambda+2\ell+3)-1\over 2}.$$ By
step 2 we have the following possibilities for $n\in\mathbb{Z}_+$
and for $\lambda\in\Lambda$:
$$
\begin{array}{lll}
\Lambda_{++})\quad& n=\alpha^+_\infty - \alpha^+_0=-{1\over
2}\left(4\ell +6+\lambda\right),& \lambda=-2n-4\ell-6,\\ & &\\
\Lambda_{+-}) & n=\alpha^+_\infty - \alpha^-_0=-{1\over 2}\left(4
+\lambda\right),& \lambda=-2n-4, \\& &\\
\Lambda_{-+}) & n=\alpha^-_\infty - \alpha^+_0={\lambda\over 2},&
\lambda=2n,\\& & \\ \Lambda_{--}) & n=\alpha^-_\infty -
\alpha^-_0={1\over 2}\left(4\ell +2+\lambda\right),&
\lambda=2n-4\ell-2,
\end{array}
$$
where
$\Lambda_{++}\cup\Lambda_{+-}\cup\Lambda_{-+}\cup\Lambda_{--}=\Lambda$,
which means that $\lambda=2m,$ $m\in\mathbb{Z}$. Now, for
$\lambda\in\Lambda$, the rational function $\omega$ in Kovacic's
algorithm is given by:
$$
\begin{array}{lll}
\Lambda_{++})\quad& \omega=r+{\ell+1\over r},&
R_n=r^2+{\ell(\ell+1)\over r^2}+(2\ell+3)+2n,\\& & \\
\Lambda_{+-}) & \omega=r-{\ell\over r}, &
R_n=r^2+{\ell(\ell+1)\over r^2}-(2\ell-1)+2n,\\& & \\
\Lambda_{-+}) & \omega=-r+{\ell+1\over r},&
R_n= r^2+{\ell(\ell+1)\over r^2}-(2\ell+3)-2n,\\& &\\
\Lambda_{--}) & \omega=-r-{\ell\over r},&
R_n=r^2+{\ell(\ell+1)\over r^2}+(2\ell-1)-2n,
\end{array}
$$
where $R_n$ is the coefficient of the differential equation
$\widetilde{\mathcal L}_n:=\partial_r^2\Psi=R_n\Psi$, which is
integrable for every $n$ and for every $\lambda\in\Lambda$ we can
see that
$\mathrm{DGal}(\widetilde{L}_n/K)=\mathrm{DGal}(L_\lambda/K)$,
where $\mathcal L_\lambda:=H\Psi=\lambda\Psi$ and
$\lambda\in\Lambda$.

By step 3, there exists a polynomial of degree $n$ satisfying the
relation (\ref{recu1}):
$$
\begin{array}{lllll}
\Lambda_{++})\quad& \partial_r^2P_n+2\left(r+{\ell+1\over
 r}\right)\partial_rP_n-2nP_n&=&0,&\lambda\in\Lambda_-,\\& & \\
\Lambda_{+-}) & \partial_r^2P_n+2\left(r-{\ell\over r}\right)\partial_rP_n-2nP_n&=&0,&\lambda\in\Lambda_-,\\& &\\
\Lambda_{-+}) & \partial_r^2P_n+2\left(-r+{\ell+1\over r}\right)\partial_rP_n+2nP_n&=&0,&\lambda\in\Lambda_+,\\ & &\\
\Lambda_{--}) & \partial_r^2P_n+2\left(-r-{\ell\over
 r}\right)\partial_rP_n+2nP_n&=&0,&\lambda\in\Lambda.
\end{array}
$$

These polynomials exists for all $\lambda\in\Lambda$ when their
degrees are $n\in 2\mathbb{Z}$, while for $n\in 2\mathbb{Z}+1$,
they exists only for the cases $\Lambda_{-+})$ and $\Lambda_{--})$
with special conditions. In this way, we have obtained the
algebraic spectrum $\Lambda= 2\mathbb{Z}$, where
$\Lambda_{++}=4\mathbb{Z}_-$, $\Lambda_{+-}=2\mathbb{Z}_-$,
$\Lambda_{-+}=4\mathbb{Z}_+$, $\Lambda_{--}=2\mathbb{Z}$.

The possibilities for eigenfunctions, considering only $\lambda\in
4\mathbb{Z}$, are given by
$$
\begin{array}{lllll}
\Lambda_{++})\quad& \Psi_n(r)&=&r^{\ell+1}P_{2n}(r)e^{r^2\over 2},
&\lambda\in\Lambda_-,\\& &\\
\Lambda_{+-}) &\Psi_n(r)&=&r^{-\ell}P_{2n}(r)e^{r^2\over
2},&\lambda\in\Lambda_-,\\& & \\
\Lambda_{-+}) & \Psi_n(r)&=&r^{\ell+1}P_{2n}(r)e^{-r^2\over 2},
&\lambda\in\Lambda_+,\\& &\\ \Lambda_{--}) &
\Psi_n(r)&=&r^{-\ell}P_{2n}(r)e^{-r^2\over 2}, &\lambda\in\Lambda.
\end{array}
$$
To obtain the point spectrum, we look $\Psi_n$ satisfying the
bound state conditions (equation (\ref{condqm})) which is in only
true for $\lambda\in\Lambda_{-+}$. With the change
$r\mapsto\sqrt{\omega\over 2}r$, the point spectrum and ground
state of the Schr\"odinger equation with the 3D-harmonic
oscillator potential are respectively
$\mathrm{spec}_p(H)=\{E_n:n\in\mathbb{Z}_+\},$ where
$E_n=2n\omega$, being $\omega$ the angular velocity, and
$$\Psi_0=\left(\sqrt{\omega\over 2}r\right)^{\ell +
1}e^{-{\omega\over4}r^2}.$$

The bound state wave functions are obtained as
$\Psi_n=P_{2n}\Psi_0.$ Now, we can see that
$\mathrm{DGal}({L_0}/K)=\mathbb{B}$ and $\mathcal
E(H)=\mathrm{Vect}(1)$. Since $\Psi_n=P_{2n}\Psi_0,$ for all
$\lambda\in\Lambda$ we have that
$\mathrm{DGal}(L_\lambda/K)=\mathbb{B}$ and $\mathcal
E(H-\lambda)=\mathrm{Vect}(1)$. In particular,
$\mathrm{DGal}(L_\lambda/K)=\mathbb{B}$ and $\mathcal
E(H-\lambda)=\mathrm{Vect}(1)$ for all
$\lambda\in\mathrm{spec}_p(H)$, where $\lambda=\frac2{\omega}E$.
\\

We remark that the Schr\"odinger equation with the 3D-harmonic
oscillator potential, through the changes $r\mapsto \frac12\omega
r^2$ and $\Psi\mapsto \sqrt{r}\Psi$, fall in a Whittaker
differential equation (equation \eqref{whittaker}) in where the
parameters are given by $$\kappa={(2\ell+3)\omega +2E\over
4\omega},\quad \mu=\frac12\ell+\frac14.$$ Applying theorem
\ref{thmarram}, we can see that for integrability, $\pm\kappa\pm
\mu$ must be a half integer. These conditions coincides with our
four sets $\Lambda_{\pm\pm}$.
\\

 \textbf{Coulomb potential:}
$$V(r)=-{\mathrm{e}^2\over r}+{\ell(\ell+1)\over r^2}+{\mathrm{e}^4\over4(\ell+1)^2},\quad \ell\in\mathbb{Z}.$$ we
can see that the Schr\"odinger equation (equation (\ref{hams}))
for this case can be written as
$$\partial_r^2\Psi=\left({\ell(\ell+1)\over r^2}-{{\rm e}^2\over r} +\frac{{\rm
e}^4}{4(\ell+1)^2}-E\right)\Psi.$$ By the change
$r\mapsto{2(\ell+1)\over {\rm e}^2}r$ we obtain the Schr\"odinger
equation
$$\partial_r^2\Psi=\left({\ell(\ell+1)\over r^2}-{2(\ell+1)\over r}+1-\lambda\right)\Psi,\quad \lambda={4(\ell+1)^2\over {\rm
e}^4}E$$ and in order to apply Kovacic algorithm, we denote
$$R={\ell(\ell+1)\over r^2}-{2(\ell+1)\over r}+1-\lambda.$$

Firstly we analyze the case for $\lambda=1$: we can see that this
equation only could fall in case 2 or in case 4 of Kovacic's
algorithm. We start discarding the case 2 because by step 1 we
should have conditions $c_2$ and $\infty_3$. In this way we should
have $E_c=\{2,4+4\ell,-4\ell\}$ and $E_{\infty}=\{1\}$ and by step
2, we should have that $n\notin\mathbb{Z}$. Thus,
$n\notin\mathbb{Z}_+$ and $D=\emptyset$, that is, the differential
Galois group of this Schr\"odinger equation for $\lambda=1$ is
${\rm SL}(2,\mathbb{C})$.
\\

Now, we analyze the case for $\lambda\neq 1$: we can see that this
equation could fall in case 1, in case 2 or in case 4. We start
discarding the case 2 because by step 1 we should have conditions
$c_2$ and $\infty_3$, so that we should have
$E_c=\{2,4+4\ell,-4\ell\}$ and $E_{\infty}=\{0\}$. By step 2, we
should have that $n=2\ell\in\mathbb{Z}_+$, so that $D=\{2\ell\}$
and the rational function $\theta$ is $\theta={-2\ell\over x}$,
but we discard this case because only could exists one polynomial
of degree $2\ell$ for a fixed $\ell$, and for instance, only could
exist one eigenstate and one eigenfunction for the Schr\"odinger
equation.
\\

Now, we only work with case 1, by step 1, conditions $c_2$ and
$\infty_3$ are satisfied. Thus,
$$\left[ \sqrt {R}\right] _{c}=0,\quad\alpha_{c}^{\pm}={1\pm(2\ell+1)\over 2}, \quad\left[  \sqrt{R}\right]  _{\infty}=\sqrt{1-\lambda},\quad \alpha_{\infty}^{\pm }=\mp{\ell+1\over \sqrt{1-\lambda}}.$$ By
step 2 we have the following possibilities for $n\in\mathbb{Z}_+$
and for $\lambda\in\Lambda$:
$$
\begin{array}{lll}
\Lambda_{++})\quad& n=\alpha^+_\infty -
\alpha^+_0=-(\ell+1)\left(1+{1\over\sqrt{1-\lambda}}\right), &
\lambda=1-\left({\ell+1\over\ell+1+n}\right)^2,\\& &\\
\Lambda_{+-}) & n=\alpha^+_\infty - \alpha^-_0=-{\ell+1\over
\sqrt{1-\lambda}}+\ell,&
\lambda=1-\left({\ell+1\over\ell-n}\right)^2, \\& &\\
\Lambda_{-+}) & n=\alpha^-_\infty -
\alpha^+_0=(\ell+1)\left({1\over \sqrt{1-\lambda}}-1\right),&
\lambda=1-\left({\ell+1\over\ell+1+n}\right)^2,\\& &\\
\Lambda_{--}) & n=\alpha^-_\infty - \alpha^-_0={\ell+1\over
\sqrt{1-\lambda}}+\ell,&
\lambda=1-\left({\ell+1\over\ell-n}\right)^2.
\end{array}
$$
We can see that $\lambda\in\Lambda_-$ when $\lambda\leq 0$, while
$\lambda\in\Lambda_+$  when $0\leq \lambda< 1$. Furthermore:
$$
\begin{array}{lll}
\Lambda_{++})\quad& \ell\leq -1,
\quad&\lambda\in\Bigg\{{\begin{array}{ll}
 \Lambda_-,\quad& \ell\leq {-n-2\over 2}\\&\\
  \Lambda_+,&
{-n-2\over 2}\leq\ell\leq-1
\end{array}}\\& &\\
\Lambda_{+-}) & \ell>0, &\lambda\in\Bigg\{{\begin{array}{ll}
 \Lambda_-,\quad& \ell\geq {n-1\over 2}\\&\\
  \Lambda_+,&
0\leq\ell\leq{n-1\over 2}
\end{array}}\\& &\\
\Lambda_{-+}) &
\ell\in\mathbb{Z},&\lambda\in\Bigg\{{\begin{array}{ll}
 \Lambda_-,\quad& \ell\leq {-n-2\over 2}\\&\\
  \Lambda_+,&
\ell\geq-1
\end{array}}\\& &\\
\Lambda_{--}) & \ell>0, &\lambda\in\Bigg\{{\begin{array}{ll}
 \Lambda_-,\quad& \ell\geq {n-1\over 2}\\&\\
  \Lambda_+,&
0\leq\ell\leq {n-1\over 2}
\end{array}}
\end{array}
$$
 In this way, the possible algebraic spectrum can be
$\Lambda=\Lambda_{++}\cup\Lambda_{+-}\cup\Lambda_{-+}\cup\Lambda_{--}$,
that is
\begin{equation}\label{EQL}
\Lambda=\left\{1-\left({\ell+1\over\ell+1+n}\right)^2:
n\in\mathbb{Z}_+\right\}\cup
\left\{1-\left({\ell+1\over\ell-n}\right)^2:
n\in\mathbb{Z}_+\right\},
\end{equation}

Now, for $\lambda\in\Lambda$, the rational function $\omega$ is
given by: {\small
$$
\begin{array}{llll}
\Lambda_{++})\quad& \omega={\ell+1\over\ell+1+n}+{\ell+1\over r},&\lambda\in\Lambda_{++},& R_n={\ell(\ell+1)\over r^2}-{2(\ell+1)\over r}+\left({\ell+1\over\ell+1+n}\right)^2,\\& &\\
\Lambda_{+-}) & \omega={\ell+1\over\ell-n}-{\ell\over r},&\lambda\in\Lambda_{+-},& R_n={\ell(\ell+1)\over r^2}-{2(\ell+1)\over r}+\left({\ell+1\over\ell-n}\right)^2,\\& &\\
\Lambda_{-+}) & \omega=-{\ell+1\over\ell+1+n}+{\ell+1\over
r},&\lambda\in\Lambda_{-+},& R_n={\ell(\ell+1)\over
r^2}-{2(\ell+1)\over r}+\left({\ell+1\over\ell+1+n}\right)^2,\\&
&\\ \Lambda_{--}) & \omega=-{\ell+1\over\ell-n}-{\ell\over
r},&\lambda\in\Lambda_{--},& R_n={\ell(\ell+1)\over
r^2}-{2(\ell+1)\over r}+\left({\ell+1\over\ell-n}\right)^2,
\end{array}
$$}
\noindent where $R_n$ is the coefficient of the differential
equation $\partial_r^2\Psi=R_n\Psi$, which is integrable for every
$n$.
\\

\noindent By step 3, there exists a polynomial of degree $n$
satisfying the relation (\ref{recu1}),
$$
\begin{array}{llll}
\Lambda_{++})\quad&
\partial_r^2P_n+2\left({\ell+1\over\ell+1+n}+{\ell+1\over r}\right)\partial_rP_n+{2(\ell+1)\over r}\left(1+{\ell+1\over\ell+1+n}\right)P_n&=&0,\\& & &\\
\Lambda_{+-}) & \partial_r^2P_n+2\left({\ell+1\over\ell-n}-{\ell\over r}\right)\partial_rP_n+{2(\ell+1)\over r}\left(1-{\ell+1\over\ell-n}\right)P_n&=&0,\\& & &\\
\Lambda_{-+}) & \partial_r^2P_n+2\left(-{\ell+1\over\ell+1+n}+{\ell+1\over r}\right)\partial_rP_n+{2(\ell+1)\over r}\left(1-{\ell+1\over\ell+1+n}\right)P_n&=&0,\\& & &\\
\Lambda_{--}) &
\partial_r^2P_n+2\left(-{\ell+1\over\ell-n}-{\ell\over
r}\right)\partial_rP_n+{2(\ell+1)\over
r}\left(1+{\ell+1\over\ell-n}\right)P_n&=&0.
\end{array}
$$

These polynomials exists for every $\lambda\in\Lambda$ when $n\in
\mathbb{Z}$, but $P_0=1$ is satisfied only for
$\lambda\in\Lambda_{-+}$. In this way, we have confirmed that the
algebraic spectrum $\Lambda$ is given by equation (\ref{EQL}).\\

The possibilities for eigenfunctions are given by

$$
{\tiny
\begin{array}{llllll}
\Lambda_{++})\quad&
\Psi_n(r)&=&r^{\ell+1}P_n(r)f_n(r)e^{r},&f_n(r)=e^{{-nr\over
\ell+1+n}}, &\lambda\in\Bigg\{{\begin{array}{ll}
 \Lambda_-,& \ell\leq {-n-2\over 2}\\&\\
  \Lambda_+,&
{-n-2\over 2}\leq\ell\leq-1
\end{array}}\\&&&&&\\
\Lambda_{+-}) &
\Psi_n(r)&=&r^{-\ell}P_n(r)f_n(r)e^{r},&f_n(r)=e^{{n+1\over
\ell-n}r}, &\lambda\in\Bigg\{{\begin{array}{ll}
 \Lambda_-,& \ell\geq {n-1\over 2}\\&\\
  \Lambda_+,&
0\leq\ell\leq {n-1\over 2}
\end{array}}\\&&&&&\\
\Lambda_{-+}) &
\Psi_n(r)&=&r^{\ell+1}P_n(r)f_n(r)e^{-r},&f_n(r)=e^{{nr\over
\ell+1+n}}, &\lambda\in\Bigg\{{\begin{array}{ll}
 \Lambda_-,& \ell\leq {-n-2\over 2}\\&\\
  \Lambda_+,&
\ell\geq-1
\end{array}}\\&&&&&\\
\Lambda_{--}) &
\Psi_n(r)&=&r^{-\ell}P_n(r)f_n(r)e^{-r},&f_n(r)=e^{{n+1\over
n-\ell}r}, &\lambda\in\Bigg\{{\begin{array}{ll}
 \Lambda_-,& \ell\geq {n-1\over 2}\\&\\
  \Lambda_+,&
0\leq\ell\leq {n-1\over 2}
\end{array}}
\end{array}}
$$
but $\Psi_n$ should satisfy the bound state conditions (equation
(\ref{condqm})) which is only true for $\lambda\in\Lambda_{-+}\cap
\Lambda_+$, so that we choose $\Lambda_{-+}\cap \Lambda_+={\rm
spec}_p(H)$, that is
$$
{\rm spec}_p(H)=\left\{1-\left(
{\ell+1\over\ell+n+1}\right)^2:\quad n\in\mathbb{Z}_+,\quad
\ell\geq -1 \right\}.
$$
By the change $r\mapsto{\rm e
 ^2\over 2(\ell+1)}r$, the point spectrum and ground state of the Schr\"odinger equation with Coulomb potential are respectively
$$\mathrm{spec}_p(H)=\{E_n:n\in\mathbb{Z}_+\},\quad E_n={{\rm e^4}\over 4}\left({1\over (\ell+1)^2}-{1\over
(\ell+1+n)^2}\right)$$ and
$$\Psi_0=\left({\rm e
 ^2\over 2(\ell+1)}r\right)^{\ell + 1}e^{-{\rm e
 ^2\over 2(\ell+1)}r}.$$
The eigenstates are given by $\Psi_n=P_{n}f_n\Psi_0,$ where $$
f_n(r)=e^{{n{\rm e}^2r\over 2(\ell+1+n)(\ell+1)}}.$$ Now, we can
see that $\mathrm{DGal}({L_0}/K)=\mathbb{B}$ and $\mathcal
E(H)=\mathrm{Vect}(1)$. Since $\Psi_n=P_{2n}f_n\Psi_0,$ for all
$\lambda\in\Lambda$ we have that
$\mathrm{DGal}({L_\lambda}/K)=\mathbb{B}$ and $\mathcal
E(H-\lambda)=\mathrm{Vect}(1)$. In particular,
$\mathrm{DGal}(L_\lambda/K)=\mathbb{B}$ and $\mathcal
E(H-\lambda)=\mathrm{Vect}(1)$ for all $E\in\mathrm{spec}_p(H)$,
where $E=\frac{\mathrm{e}^4}{4(\ell +1)^2}\lambda$.
\\

We remark that, as in the three dimensional harmonic oscillator,
the Schr\"odinger equation with the Coulomb potential, through the
change $$r\mapsto {\frac {\sqrt {-4\, \left(\ell+1
\right)^{2}E+{\rm e}^{4 }}}{\ell+1}}r,$$ falls in a Whittaker
differential equation (equation \eqref{whittaker}) in where the
parameters are given by
$$\kappa={\frac {{\rm e}^{2
} \left( \ell+1 \right) }{\sqrt {-4\, \left( \ell+1 \right)
^{2}E+{\rm e}^ {4}}}} ,\quad \mu=\ell+\frac12.$$ Applying theorem
\ref{thmarram}, we can impose $\pm\kappa\pm \mu$ half integer, to
coincides with our four sets $\Lambda_{\pm\pm}$.
\\

\begin{remark}
By direct application of Kovacic's algorithm we have:
\begin{itemize}
\item The Schr\"odinger equation \eqref{equ1} with potential
$$V(x)=ax^2+{b\over x^2}$$ is
integrable for $\lambda\in\Lambda$ when

\begin{itemize}
\item $a=0,\quad b=\mu(\mu+1), \quad \mu\in \mathbb{C}, \quad \Lambda=\mathbb{C},$
\item $a=1,\quad b=0, \quad \lambda\in\Lambda=2\mathbb{Z}+1,$
\item $a=1,\quad b=\ell(\ell+1), \quad \ell\in \mathbb{Z}^*, \Lambda=2\mathbb{Z}+1.$
\end{itemize}
\item The only rational potentials (up to transformations) in which the elements of the algebraic spectrum are placed at the same
distance, belongs to the family of potentials given by
$$V(x)=\sum_{k=-\infty}^2 a_kx^k,\quad a_2\neq 0.$$ In particular, the set $\Lambda$ for the harmonic oscillator ($a=1$, $b=0$)
 and $3D$ harmonic oscillator ($a=1$, $b=\ell(\ell +1)$) satisfies this.

\end{itemize}
\end{remark}

\begin{proposition}\label{luckphyst} Let $\mathcal L_\lambda$ be the Schr\"odinger equation \eqref{equ1} with $K=\mathbb{C}(x)$ and Picard-Vessiot extension $L_\lambda$.
 If $\mathrm{DGal}(L_0/K)$ is finite primitive, then $\mathrm{DGal}(L_\lambda/K)$ is not finite primitive for
  all $\lambda\in\Lambda\setminus \{0\}$.
\end{proposition}

\begin{proof} Pick $V\in\mathbb{C}(x)$ such that $\mathcal L_0$ falls in case 3 of Kovacic
algorithm, then $\circ u_\infty\geq 2$. Assume $t,
s\in\mathbb{C}[x]$ such that $V=\frac{s}t$, then ${\rm deg}(t)\geq
{\rm deg}
 (s)+2$ and
$V-\lambda=\frac{s-\lambda t}{t}$. Now, for $\lambda\neq 0 $ we
have that $\mathrm{deg}(s-\lambda t)=\mathrm{deg}(t)$ and
therefore $\circ (V-\lambda)_\infty=0$. So that for $\lambda\neq
0$, the equation $\mathcal L_\lambda$ does not falls in case 3 of
Kovacic algorithm and therefore $\mathrm{DGal}(L_\lambda/K)$ is
not finite primitive.
\end{proof}

\begin{corollary}\label{luckyphysc}
 Let $\mathcal L_\lambda$ be the Schr\"odinger equation \eqref{equ1} with $K=\mathbb{C}(x)$ and Picard-Vessiot extension $L_\lambda$. If
 $\mathrm{Card}(\Lambda)>1$, then there is either zero or one
 value of $\lambda$ such that $\mathrm{DGal}(L_\lambda/K)$ is a finite primitive group.
\end{corollary}
\begin{proof} Assume that $\mathrm{Card}(\Lambda)>1$. Thus, by proposition \ref{luckphyst}, the
Schr\"odinger equation does not falls in case 3 of Kovacic's
algorithm.
\end{proof}
\medskip

It seems that the study of the differential Galois groups of the
Schr\"odinger equation with the Coulomb potential has been
analyzed by Jean-Pierre Ramis using his summability theory in the
eighties of the past century, see \cite{comramis}.

\subsection{Darboux Transformations}

Here we present a Galoisian approach to Darboux transformation,
Crum iteration and shape invariant potentials. We denote by
$W(y_1,\ldots,y_n)$ the Wronskian
$$W(y_1,\ldots,y_n)=\left|\begin{matrix}y_1&\cdots& y_n\\
 \vdots& & \vdots \\ \partial^{n-1}_xy_1&\cdots&
 \partial^{n-1}_xy_n\end{matrix}\right|,$$ by $\mathrm{DT}$ the
 Darboux transformation, by $\mathrm{DT}_n$ the $n$ iteration of $\mathrm{DT}$ and by $\mathrm{CI}_n$ the Crum iteration. Also we use
 the notation of subsection \ref{susysection}. We recall that
 $K=\mathbb{C}(x)$ and in case of  other differential fields, as usually is considered along this thesis,
  we mean the smallest differential containing the coefficients of the linear differential equations.
\begin{theorem}[Galoisian version of DT]\label{darth} Assume $H_\pm=\partial_x^2+ V_{\pm}(x)$ and $\Lambda\neq\emptyset$. Let $\mathcal{L}_\lambda$
 given by the Schr\"odinger equation $H_-\Psi^{(-)}=\lambda\Psi^{(-)}$
with $V_-(x)\in K$ and $\widetilde{\mathcal{L}}_\lambda$ given by
the Schr\"odinger equation $H_+\Psi^{(+)}=\lambda\Psi^{(+)}$ with
$V_+(x)\in\widetilde K$. Let $\mathrm{DT}$ be the transformation
such that $\mathcal{L}\mapsto\widetilde{\mathcal{L}}$, $V_-\mapsto
V_{+}$,
 $\Psi^{(-)}\mapsto\Psi^{(+)}$. Then the following statements
 holds:
\begin{description}
\item[i)] $\mathrm{DT}(V_-)=V_+=\Psi^{(-)}_{\lambda_1}\partial_x^2\left({1\over
\Psi^{(-)}_{\lambda_1}}\right)+\lambda_1=V_--2\partial_x^2(\ln\Psi^{(-)}_{\lambda_1}),$\\
$\mathrm{DT}(\Psi^{(-)}_{\lambda_1})={\Psi}^{(+)}_{\lambda_1}={1\over\Psi^{(-)}_{\lambda_1}}$,
 where $\Psi^{(-)}_{\lambda_1}$ is a particular solution of $\mathcal{L}_{\lambda_1}$, $\lambda_1\in\Lambda$.

\item[ii)] $\mathrm{DT}(\Psi^{(-)}_\lambda)=\Psi^{(+)}_{\lambda}=\partial_x\Psi^{(-)}_{\lambda}-\partial_x(\ln\Psi^{(-)}_{\lambda_1})\Psi^{(-)}_{\lambda}
={W(\Psi^{(-)}_{\lambda_1},\Psi^{(-)}_{\lambda})\over
W(\Psi^{(-)}_{\lambda_1})}$, $\lambda\neq\lambda_1$, where
$\Psi^{(-)}_{\lambda}$ is the general solution of
$\mathcal{L}_\lambda$ for
$\lambda\in\Lambda\setminus\{\lambda_1\}$ and
$\Psi^{(+)}_{\lambda}$ is the general solution of
$\widetilde{\mathcal{L}}_\lambda$ also for
$\lambda\in\Lambda\setminus\{\lambda_1\}$.
\end{description}
\end{theorem}
In agreement with the previous theorem we obtain the following
results.

\begin{proposition}\label{darbiso}
$\mathrm{DT}$ is isogaloisian and virtually strong isogaloisian.
Furthermore, if $\partial_x(\ln\Psi^{(-)}_{\lambda_1})\in K$, then
$\mathrm{DT}$ is strong isogaloisian.
\end{proposition}

\begin{proof}
 Let $K$, ${L_\lambda}$ be the differential field and
the Picard-Vessiot extension of the equation
$\mathcal{L}_\lambda$. Let $\widetilde{{K}}$,
$\widetilde{{L}}_\lambda$ the differential field and the
Picard-Vessiot extension of the equation
$\widetilde{\mathcal{L}}_\lambda$. Due to
$\mathrm{DT}(V_-)=V_+=2W^2-V_--2\lambda_1$, where
$W=-\partial_x(\ln\Psi^{(-)}_{\lambda_1})$, we have
$\widetilde{{K}}=K\left\langle
\partial_x(\ln\Psi^{(-)}_{\lambda_1})\right\rangle.$ By theorem
\ref{singer} we have that the Riccati equation
$\partial_xW=V_--W^2$ has one algebraic solution, in this case
$W=-\partial_x(\ln \Psi^{(-)}_{\lambda_1})$. Let $\langle
\Psi^{(-)}_{(1,\lambda)},\Psi^{(-)}_{(2,\lambda)}\rangle$ be a
basis of solutions for equation $\mathcal{L}_\lambda$ and $\langle
\Psi^{(+)}_{(1,\lambda)},\Psi^{(+)}_{(2,\lambda)}\rangle$ a basis
of solutions for equation $\widetilde{\mathcal{L}}_\lambda$. Since
the differential field for equation
$\widetilde{\mathcal{L}}_\lambda$ is $\widetilde
K=K\langle\partial_x(\ln \Psi^{(-)}_{\lambda_1})\rangle$, we have
that $L=K\langle
\Psi^{(-)}_{(1,\lambda)},\Psi^{(-)}_{(2,\lambda)}\rangle$ and
$$\widetilde L=\widetilde K\langle \Psi^{(+)}_{(1,\lambda)},\Psi^{(+)}_{(2,\lambda)}\rangle=K\langle
\Psi^{(+)}_{(1,\lambda)},\Psi^{(+)}_{(2,\lambda)},\partial_x(\ln
\Psi^{(-)}_{\lambda_1})\rangle$$ $$=K\langle
\Psi^{(-)}_{(1,\lambda)},\Psi^{(-)}_{(2,\lambda)},\partial_x(\ln
\Psi^{(-)}_{\lambda_1})\rangle=\widetilde K\langle
\Psi^{(-)}_{(1,\lambda)},\Psi^{(-)}_{(2,\lambda)}\rangle,$$ for
$\lambda=\lambda_1$ and for $\lambda\neq\lambda_1$. Since
$\partial_x(\ln \Psi^{(-)}_{\lambda_1})$ is algebraic over $K$,
then
$$(\mathrm{DGal}(L_\lambda/K))^0=(\mathrm{DGal}(\widetilde L_\lambda/K))^0,\quad \mathrm{DGal}(L_\lambda/K)
=\mathrm{DGal}(\widetilde L_\lambda/\widetilde K),$$
which means that $\mathrm{DT}$ is a virtually strong and isogalosian transformation.\\
In the case $\partial_x(\ln \Psi^{(-)}_{\lambda_1})\in K$, then
$\widetilde K=K$ and $\widetilde L=L$, which means that
$\mathrm{DT}$ is a strong isogalosian transformation.
\end{proof}

\begin{proposition}\label{dareige} Consider $\mathfrak{L}_\lambda:=H_--\lambda$ and $\widetilde{
\mathfrak{L}}_\lambda:=H_+-\lambda$ such that
$\mathrm{DT}(H_--\lambda)=H_+-\lambda$. The eigenrings of
$\mathfrak{L}_\lambda$ and $\widetilde{\mathfrak{L}}_\lambda$ are
isomorphic.

\end{proposition}

\begin{proof}
Assume $\mathcal E(\mathfrak L_\lambda)$ and $\mathcal
E(\widetilde{\mathfrak L}_\lambda)$ the eigenrings of
$\mathfrak{L}_\lambda$ and $\widetilde{\mathfrak{L}}_\lambda$
respectively. By proposition \ref{darbiso} the connected identity
component of the Galois group is preserved by Darboux
transformation and for instance the eigenrings is preserved by
Darboux transformation. Now, suppose that $T\in\mathcal E
(\mathfrak L_\lambda)$, $\mathrm{Sol}(\mathfrak L_\lambda)$ and
$\mathrm{Sol}(\widetilde{\mathfrak{L}}_\lambda)$ the solutions
space for $\mathfrak L_\lambda\Psi^{(-)}=0$ and
$\widetilde{\mathfrak L}_\lambda\Psi^{(-)}=0$ respectively. To
transform $\mathcal E (\mathfrak L_\lambda)$ into $\mathcal E
(\widetilde{\mathfrak{L}}_\lambda)$ we follows the diagram:

 $$\xymatrix{\mathrm{Sol}(\mathfrak L_\lambda) \ar[r]^-{T} & \mathrm{Sol}(\mathfrak L_\lambda) \\
   \mathrm{Sol}(\widetilde{\mathfrak L}_\lambda)  \ar[r]_{\widetilde T} \ar[u]^{A^\dagger} & \ar@{<-}[u]_{A} \mathrm{Sol}(\widetilde{\mathfrak L}_\lambda) }
   \qquad \begin{array}{c}\\ \\ \\
   \Rightarrow \widetilde{T}=ATA^\dagger{\rm mod}(\widetilde{\mathfrak L}_\lambda)\in\mathcal E(\widetilde{\mathfrak L}_\lambda),\end{array}$$
where $A^\dagger$ and $A$ are the raising and lowering operators.
\end{proof}

\begin{example}
Assume the Schr\"odinger equation $\mathcal L_\lambda$ with
potential $V=V_-=0$, which means that $\Lambda=\mathbb{C}$. If we
choose $\lambda_1=0$ and as particular solution
$\Psi_{0}^{(-)}=x$, then for $\lambda\neq 0$ the general solution
is given by
$$\Psi^{(-)}_{\lambda}=c_1e^{\sqrt{-\lambda}x}+c_2e^{-\sqrt{-\lambda}x}.$$
Applying the Darboux transformation $\mathrm{DT}$, we have
$\mathrm{DT}(\mathcal L_\lambda)=\widetilde{\mathcal L}_\lambda$,
where
$$\mathrm{DT}(V_-)=V_+={2\over x^2}$$ and for $\lambda\neq
0$
$$\mathrm{DT}(\Psi^{(-)}_{\lambda})=\Psi^{(+)}_{\lambda}={c_1(\sqrt{-\lambda}x-1)e^{\sqrt{-\lambda}x}\over x}-{c_2(\sqrt{-\lambda}x+1)e^{-\sqrt{-\lambda}x}\over x}.$$
We can see that $\widetilde{K}=K=\mathbb{C}(x)$ for all
$\lambda\in \Lambda$ and the Picard-Vessiot extensions are given
by $L_0=\widetilde{L}_0=\mathbb{C}(x)$,
$L_\lambda=\widetilde{L}_\lambda=\mathbb{C}(x,e^{\sqrt{\lambda}x})$
for $\lambda\in\mathbb{C}^*$. In this way, we have that
$\mathrm{DGal}(L_0/K)=\mathrm{DGal}(\widetilde{L}_0/K)=e$; for
$\lambda\neq0$, we have
$\mathrm{DGal}(L_\lambda/K)=\mathrm{DGal}(\widetilde{L}_\lambda/K)=\mathbb{G}_m$.
The eigenrings of the operators $\mathfrak L_0$ and $\mathfrak
L_\lambda$ are given by
$$\mathcal E(\mathfrak
L_0)=\mathrm{Vect}\left(1,x\partial_x,x\partial_x-1,x^2\partial_x-x\right),$$
$$\mathcal E(\widetilde{\mathfrak
L}_0)=\mathrm{Vect}\left(1,x\partial_x-1,x^4\partial_x-2x^3,{\partial_x\over
x^2}+{1\over x^3}\right)$$ and for $ \lambda\neq 0$
$$\mathcal E(\mathfrak L_{\lambda})=\mathrm{Vect}\left(1,\partial_x\right), \quad \mathcal
E(\widetilde{\mathfrak
L}_{\lambda})=\mathrm{Vect}\left(1,-\left(\lambda +{1\over x^2}
\right)\partial_x-{1\over x^3}\right),$$ where $\mathcal
L_{\lambda}:=\mathfrak L_{\lambda}\Psi^{(-)}=0$ and
$\widetilde{\mathcal L}_{\lambda}:=\widetilde{\mathfrak
L}_{\lambda}\Psi^{(+)}=0$.

\end{example}
\medskip

Applying iteratively the Darboux transformation, theorem
\ref{darth}, and by propositions \ref{darbiso}, \ref{dareige}
we have the following results.\\

\begin{proposition}[Galoisian version of DT$_n$]\label{darbit1}
 Let be $\Lambda\neq\emptyset$, $\mathcal{L}^{(n)}_\lambda$ given by $H^{(n)}\Psi^{(n)}=\lambda\Psi^{(n)},$
$V_n\in{K}_n,$ $K_0=K$, $V_0=V_-$, $H^{(0)}=H_-$,
$\Psi^{(0)}=\Psi^{(-)}$ and $L_\lambda^{(n)}$ the Picard-Vessiot
extension of $\mathcal{L}^{(n)}_\lambda$. Let
$\mathcal{L}_\lambda^{(n+1)}$ given by
$H^{(n+1)}\Psi^{(n+1)}=\lambda\Psi^{(n+1)},$ $V_{n+1}\in K_{n+1}$.
Let $\mathrm{DT}_n$  such that
$\mathcal{L}_\lambda^{(n)}\mapsto\mathcal{L}_\lambda^{(n+1)}$,
$V_n\mapsto V_{n+1}$,
 $\Psi^{(n)}_{\lambda}\mapsto\Psi^{(n+1)}_\lambda$ and $L_\lambda^{(n+1)}$ the Picard-Vessiot
extension of $\mathcal{L}^{(n+1)}_\lambda$. Then the following
statements
 holds:
 \begin{description}
\item[i)] $\mathrm{DT}_n(V_-)=\mathrm{DT}(V_n)=V_{n+1}=V_n-2\partial_x^2\left(\ln
\Psi_{\lambda}^{(n)}\right)=V_--2\sum_{k=0}^{n}\partial_x^2\left(\ln
\Psi_{\lambda_k}^{(k)}\right)$, where $\Psi_{\lambda_k}^{(k)}$ is
a particular solution for $\lambda=\lambda_k$, $k=0,\ldots, n$. In
particular, if $\lambda_n=\lambda_0$ and $\Lambda=\mathbb{C}$,
then there exists $\Psi^{(n)}_{\lambda_n}$ such that $V_{n}\neq
V_{n-2}$, with $n\geq 2$.
\item[ii)] $\mathrm{DT}(\Psi_{\lambda}^{(n)})=\mathrm{DT}_n(\Psi^{(-)}_{\lambda})=\Psi^{(n+1)}_{\lambda}=
\partial_x\Psi^{(n)}_{\lambda}-\Psi^{(n)}_{\lambda}{\partial_x\Psi_{\lambda_n}^{(n)}\over
\Psi_{\lambda_n}^{(n)}}={W(\Psi^{(n)}_{\lambda_n},\Psi^{(n)}_{\lambda})\over
W(\Psi^{(n)}_{\lambda_n})} $ where $\Psi^{(n)}_{\lambda}$ is a
general solution for $\lambda\in\Lambda\setminus\{\lambda_n\}$ of
$\mathcal{L}^{(n)}_\lambda$.

\item[iii)] $K_{n+1}=K_n\left\langle\partial_x(\ln
\Psi^{(n)}_{\lambda_n}) \right\rangle$.

\item[iv)] $\mathrm{DT}_n$ is isogaloisian and virtually strongly isogaloisian. Furthermore, if $\partial_x(\ln\Psi^{(n)}_{\lambda_n})\in K_n$ then $\mathrm{DT}_n$
is strongly isogaloisian.
\item[v)] The eigenrings of $H^{(n)}-\lambda$ and $H^{(n+1)}-\lambda$
are isomorphic.
\end{description}
\end{proposition}
\begin{proof} By induction on theorem \ref{darth} we obtain i) and
ii). By induction on proposition \ref{darbiso} we obtain iii) and
iv). By induction on proposition \ref{dareige} we obtain v).

\end{proof}
\medskip

\begin{example}
Starting with $V=0$, the following potentials can be obtained
using Darboux iteration $\mathrm{DT}_n$ (see \cite{beev,blbose}).
$$
I)\,\,V_n={n(n-1)b^2\over(bx+c)^2},\quad
II)\,\,V_n={m^2n(n-1)(b^2-a^2)\over (a\cosh(mx)+b\sinh(mx))^2},
$$
$$
III)\,\,V_n={-4abm^2n(n-1)\over (ae^{mx}+be^{-mx})^2},\quad
IV)\,\,V_n={m^2n(n-1)(b^2+a^2)\over (a\cos(mx)+b\sin(mx))^2}.
$$ In particular for the rational potential given in \textit{I}), we have $K=K_n=\mathbb{C}(x)$
and for $\lambda_n=\lambda=0$, we have $$\Psi_0^{(n)}={c_1\over
(bx+c)^n}+c_2(bx+c)^{n+1}, \textrm{   so that   }
\mathrm{DGal}(L_0/K)=\mathrm{DGal}(L_0^{(n)}/K)=e,$$
$$\mathcal E(H^{(n)})=\mathrm{Vect}\left(1,x\partial_x-1,x^{2n+2}\partial_x-(n+1)x^{2n+1},{\partial_x\over
x^{2n} }+{n\over x^{2n+1}}\right),$$ whilst for $\lambda\neq 0$
and $\lambda_n=0$, the general solution $\Psi_\lambda^{(n)}$ is
given by
$$\Psi_\lambda^{(n)}(x)=c_1f_n(x,\lambda)h_n(\sin(\sqrt{\lambda}x)+c_2g_n(x,\lambda)j_n(\cos(\sqrt{\lambda}x),$$
where $f_n,g_n,h_n,j_n\in\mathbb{C}(x)$, so that
$$ \mathrm{DGal}(L_\lambda/K)=\mathrm{DGal}(L^{(n)}_\lambda/K)=\mathbb{G}_m,$$ and
$$\dim_{\mathbb{C}}\mathcal E(H-\lambda)=
\dim_{\mathbb{C}}\mathcal E(H^{(n)}-\lambda)=2.$$

\end{example}
\begin{proposition}[Galoisian version of CI$_n$]\label{crumit1}
Consider  $\mathcal L_\lambda$ given by $H\Psi=\lambda\Psi$,
$H=-\partial_x^2+V$, $V\in K$, such that
$\mathrm{Card}(\Lambda)>n$ for a fixed $n\in\mathbb{Z}_+$. Let
$\mathcal L_\lambda^{(n)}$ be given by
$H^{(n)}\Psi^{(n)}=\lambda\Psi^{(n)}$, where
$H^{(n)}=\partial_x^2+V_n$, $V_n\in K_n$. Let $\mathrm{CI}_n$ be
the transformation such that
$\mathcal{L}_\lambda\mapsto\mathcal{L}_\lambda^{(n)}$, $V\mapsto
V_n$, $(\Psi_{\lambda_1},\ldots,
\Psi_{\lambda_n},\Psi_\lambda)\mapsto\Psi_\lambda^{(n)}$, where
for $k=1,\ldots, n$ and the equation $\mathcal L_\lambda$,
 the function $\Psi_\lambda$ is   the general solution  for
$\lambda\neq\lambda_k$ and $\Psi_{\lambda_k}$ is a particular
solution for $\lambda=\lambda_k$. Then the following statements
 holds:
 \begin{description}
\item[i)] $\mathrm{CI}_n(\mathcal{L}_\lambda)=\mathcal L_\lambda^{(n)}$
where $\mathrm{CI}_n(V)=V_n=V-2\partial_x^2\left(\ln
W(\Psi_{\lambda_1},\ldots,\Psi_{\lambda_n})\right)$ and
$$\mathrm{CI}_n(\Psi_\lambda)=\Psi_\lambda^{(n)}={W(\Psi_{\lambda_1},\ldots,\Psi_{\lambda_n},\Psi_\lambda)\over
W(\Psi_{\lambda_1},\ldots,\Psi_{\lambda_n})},$$ where
$\Psi_\lambda^{(n)}$ is the general solution of $\mathcal
L_\lambda^{(n)}$.

\item[ii)] $K_{n}=K\langle \partial_x\left(\ln
W(\Psi_{\lambda_1},\ldots,\Psi_{\lambda_n})\right)\rangle$.
\item[iii)] $\mathrm{CI}_n$ is isogaloisian and virtually strongly isogaloisian. Furthermore, if $$\partial_x\left(\ln
W(\Psi_{\lambda_1},\ldots,\Psi_{\lambda_n})\right)\in K_n,$$ then
$\mathrm{CI}_n$ is strongly isogaloisian.
\item[iv)] The eigenrings of $H-\lambda$ and
$H^{(n)}-\lambda$ are isomorphic.
\end{description}
\end{proposition}
\begin{proof} By induction on theorem \ref{darth} we obtain i). By induction on proposition \ref{darbiso} we obtain ii) and
iii). By induction on proposition \ref{dareige} we obtain iv).

\end{proof}
\medskip

\begin{example}
To illustrate the Crum iteration with rational potentials, we
consider $V=\frac2{x^2}$. The general solution of $\mathcal
L_\lambda:= H\Psi=\lambda\Psi$ is $${c_1e^{kx}(kx-1)\over
x}+{c_2e^{-kx}(kx+1)\over x},\quad \lambda=-k^2,$$ the
eigenfunctions for $\lambda_1=-1$, and $\lambda_2=-4$, are
respectively given by $$\Psi_{-1}={e^{-x}(x+1)\over x},\quad
\Psi_{-4}={e^{-2x}(2x+1)\over 2x}.$$ Thus, we obtain
$$\mathrm{CI}_2(V)=V_2=\frac8{(2x+3)^2}$$ and the
general solution of $\mathcal L
_\lambda^{(2)}:=H^{(2)}\Psi^{(2)}=\lambda\Psi^{(2)}$ is
$$\mathrm{CI}_2(\Psi_\lambda)=\Psi_\lambda^{(2)}={\frac{c_1\left( k \left( 2\,x+3
 \right) -2 \right) {e^{kx}}}{2\,x+3}}+{\frac {c_2 \left( 2+k
 \left( 2\,x+3 \right)  \right) {e^{-kx}}}{4\,x+6}},\quad
 \lambda=-k^2.$$ The differential Galois groups and eigenrings are
 given by:
$$\mathrm{DGal}(L_0/K)=\mathrm{DGal}(L^{(2)}_0/K)=e,\quad \dim_{\mathbb{C}}\mathcal E(H)=\dim_{\mathbb{C}}\mathcal E(H^{(2)})=4,$$ and for
$\lambda\neq 0$
$$\mathrm{DGal}(L_\lambda/K)=\mathrm{DGal}(L^{(2)}_\lambda/K)=\mathbb{G}_m,\quad \dim_{\mathbb{C}}\mathcal
E(H-\lambda)=\dim_{\mathbb{C}}\mathcal E(H^{(2)}-\lambda)=2.$$
\end{example}
\begin{proposition}\label{sspp}
The supersymmetric partner potentials $V_\pm$ are rational
functions if and only if the superpotential $W$ is a rational
function.
\end{proposition}

\begin{proof} The supersymmetric partner potentials $V_\pm$ are
written as $V_\pm=W^2\pm\partial_xW$. We start considering the
superpotential $W\in\mathbb{C}(x)$, so trivially we have that
$V_\pm\in\mathbb{C}(x)$. Now assuming that $V_\pm\in\mathbb{C}(x)$
we have that $\partial_xW\in\mathbb{C}(x)$ and
$W^2\in\mathbb{C}(x)$, which implies that
$W\partial_xW\in\mathbb{C}(x)$ and therefore $W\in\mathbb{C}(x)$.

\end{proof}
\medskip

\begin{corollary} The superpotential $W\in\mathbb{C}(x)$ if and only if $\mathrm{DT}$ is strong isogaloisian.
\end{corollary}

\begin{proof} Assume that the superpotential $W\in\mathbb{C}(x)$.
Thus, by proposition \ref{darbiso}, $\mathrm{DT}$ is strong
isogaloisian. Now, assume that $\mathrm{DT}$ is strong
isogaloisian. Thus, $V_\pm\in\mathbb{C}(x)$ and by proposition
\ref{sspp} we have that $W\in\mathbb{C}(x)$.

\end{proof}
\medskip

The following definition is a partial Galoisian adaptation of the
original definition given in \cite{ge} ($K=\mathbb{C}(x)$). The
complete Galoisian adaptation is given when $K$ is any
differential field.\\

\begin{definition}[Rational Shape Invariant Potentials] Assume
$V_\pm(x;\mu)\in\mathbb{C}(x;\mu)$, where $\mu$ is a family of
parameters. The potential $V=V_-\in\mathbb{C}(x)$ is said to be
rational shape invariant potential with respect to $\mu$ and
$E=E_n$ being $n\in \mathbb{Z}_+$, if there exists $f$ such that
$$V_+(x;a_0)=V_-(x;a_1)+R(a_1),\quad a_1=f(a_0),\quad E_n=\sum_{k=2}^{n+1}R(a_k),\quad E_0=0.$$
\end{definition}
\medskip

\begin{remark}\label{remshape} We propose the following steps to check whether $V\in\mathbb{C}(x)$ is shape
invariant.

\begin{description}
\item[Step 1.] Introduce parameters in $W(x)$ to obtain $W(x;\mu)$,
write $V_\pm(x;\mu)=W^2(x;\mu)\pm\partial_xW(x;\mu)$, and replace
$\mu$ by $a_0$ and $a_1$.
\item [Step 2.] Obtain polynomials $\mathcal P\in\mathbb{C}[x;a_0,a_1]$ and $\mathcal
Q\in\mathbb{C}[x;a_0,a_1]$ such that
$$\partial_x(V_+(x;a_0)-V_-(x;a_1))={\mathcal P(x;a_0,a_1)\over \mathcal Q(x;a_0,a_1)}.$$
\item [Step 3.] Set $\mathcal P(x;a_0,a_1)\equiv 0$, as polynomial in $x$, to obtain $a_1$ in function of $a_0$, i.e.,
$a_1=f(a_0)$. Also obtain $R(a_1)=V_+(x;a_0)-V_-(x;a_1)$ and
verify that exists $k\in\mathbb{Z}^+$ such that
$R(a_1)+\cdots+R(a_k)\neq 0$.
\end{description}

\end{remark}
\medskip

\begin{example} Consider the superpotential of the three
dimensional harmonic oscillator $W(r;\ell)=r-\frac{\ell+1}{r}$. By
step 1, the supersymmetric partner potentials are
$$V_-(r;\ell)=r^2+\frac{\ell(\ell+1)}{r^2}-2\ell-3,\quad
V_+(r;\ell)=r^2+\frac{(\ell+1)(\ell+2)}{r^2}-2\ell-1.$$ By step 2,
we have $\partial_r(V_+(r;a_0)-V_-(r;a_1))=- 2{a_0^2 + 3a_0 -
a_1^2 - a_1 + 2\over r^3}$. By step 3, $(a_0+1)(a_0+2)
=a_1(a_1+1)$, so that $a_1=f(a_0)=a_0+1$, $a_n=f(a_{n-1})=a_0+n$,
$R(a_1)=2$. Thus, we obtain the energy levels $E_n=2n$ and the
wave functions $\Psi^{(-)}_n(r;\ell)=A^\dagger(r;\ell)\cdots
A^\dagger(r;\ell+n-1)\Psi^{(-)}_0(r;\ell+n)$, compare with \cite{duka}.\\

\end{example}
By theorem \ref{darth} and propositions \ref{darbiso},
\ref{dareige} and \ref{sspp} we have the following result.

\begin{theorem}\label{thsip} Consider $\mathcal
L_n:=H\Psi^{(-)}=E_n\Psi^{(-)}$ with Picard-Vessiot extension
$L_n$, where $n\in\mathbb{Z}_+$. If $V=V_-\in\mathbb{C}(x)$ is a
shape invariant potential with respect to $E=E_n$, then
$$\mathrm{DGal}(L_{n+1}/K)=\mathrm{DGal}(L_{n}/K),\quad \mathcal E(H-E_{n+1})\simeq \mathcal E(H-E_n),\quad n>0.$$
\end{theorem}
\medskip

\begin{remark} The differential automorphisms $\sigma$ commutes
with the raising and lowering operators $A$ and $A^\dagger$ due to
$W\in\mathbb{C}(x)$. Furthermore the wave functions $\Psi^{(-)}_n$
can be written as $\Psi^{(-)}_n=P_nf_n\Psi^{(-)}_0$, where $P_n$
is a polynomial of degree $n$ in $x$ and $f_n$ is a sequence of
functions being $f_0(x)=1$ such as was shown in the case of
Harmonic oscillators and Coulomb potentials.
\end{remark}

\section[Algebrization in Supersymmetric Quantum Mechanics]{The Role of the Algebrization in Supersymmetric Quantum Mechanics}
In supersymmetric quantum mechanics, there exists potentials which
are not rational functions and, for this reason, it is difficult
to apply our Galoisian approach such as in section \ref{susyrapo}.
In this section we give a solution to this problem presenting some
results concerning differential equations with non-rational
coefficients. For these differential equations it is useful, when
is possible, to replace it by a new differential equation over the
Riemann sphere $\mathbb{P}^1$ (that is, with rational
coefficients). To do this, we can use a change of variables. The
equation over $\mathbb{P}^1$ is called the \emph{algebraic form}
or \emph{algebrization} of the original equation.

\medskip

This algebraic form dates back to the 19th century (Liouville,
Darboux), but the problem of obtaining the algebraic form (if it
exists) of a given differential equation is in general not an easy
task. Here we develop a new method using the concept of
\emph{Hamiltonian change of variables}. This change of variables
allow us to compute the algebraic form of a large number of
differential equations of different types. In particular, for
second order linear differential equations, we can apply
\textit{Kovacic's algorithm} over the algebraic form to solve the
original equation.

\medskip
The following definition can be found in \cite{ber,howe,howe2}.

\begin{definition}[Pullbacks of differential equations] Let $\mathfrak L_1\in
K_1[\partial_z]$ and $\mathfrak L_2\in K_2[\partial_x]$ be
differential operators, the expression $\mathfrak L_2\otimes
(\partial_x+v)$ refers to the operator whose solutions are the
solutions of $\mathfrak L_2$ multiplied by the solution $e^{-\int
vdx}$ of $\partial_x+v$.

\begin{itemize}
\item $\mathfrak L_2$ is a \emph{proper pullback} of $\mathfrak L_1$ by means of $f\in K_2$ if the change of variable $z=f(x)$ changes
$\mathfrak L_1$ into $\mathfrak L_2$.
\item $\mathfrak L_2$ is a \emph{pullback} (also known as weak pullback) of $\mathfrak L_1$ by means of $f\in K_2$ if there exists $v\in K_2$
such that $\mathfrak L_2\otimes (\partial_x+v)$ is a proper
pullback of $\mathfrak L_1$ by means of $f$.
\end{itemize}
\end{definition}

In case of compact Riemann surfaces, the geometric mechanism
behind the algebrization is a ramified covering of compact Riemann
surfaces, see \cite{mo2,mo}.

\subsection{Second Order Linear Differential Equations}\label{sec21}
Some results presented in this subsection also can be found in
\cite[\S 2 ]{acbl}.
\begin{proposition}[Change of the independent variable, \cite{acbl}]\label{pr1}
  Let us consider the following equation, with coefficients in
$\mathbb{C}(z)$:
\begin{equation}\label{eq1} \mathcal L_z:=
\partial_z^2 y+a(z)\partial_zy+b(z)y=0,
\end{equation}
and $\mathbb{C}(z)\hookrightarrow  L$ the corresponding
Picard-Vessiot extension. Let $(K,\delta)$ be a differential field
with $\mathbb C$ as field of constants. Let $\theta\in K$ be a
non-constant element. Then, by the change of variable $z =
\theta(x)$, equation \eqref{eq1} is transformed in
\begin{equation}\label{eq2} \mathcal L_x:=
\partial_x^2{r}+\left(a(\theta)\partial_x{\theta}-{\partial_x^2{\theta}\over
\partial_x{\theta}}\right)\partial_x{r}+b(\theta)(\partial_x{\theta})^2r=0,
\quad
 \partial_x=\delta, \quad r=y\circ\theta.
\end{equation}
  Let $K_0\subset K$ be the smallest differential field containing
$\theta$ and $\mathbb C$. Then equation \eqref{eq2} is a
differential equation with coefficients in $K_0$. Let
$K_0\hookrightarrow L_0$ be the corresponding Picard-Vessiot
extension. Assume that
  $$\mathbb C(z) \to K_0,\quad z\mapsto \theta$$
is an algebraic extension, then
$$(\mathrm{DGal}(L_0 / K_0))^0 = (\mathrm{DGal}(L / \mathbb C(z)))^0.$$
\end{proposition}

\begin{proposition}\label{propjaw}Assume $\mathcal L_x$ and $\mathcal L_z$ as in proposition \ref{pr1}. Let $\varphi$ be the transformation given by
$$\varphi:\begin{array}{l}z\mapsto \theta(x)\\ \\
\partial_z\mapsto\frac1{\partial_x\theta}\delta.\end{array}$$ Then
$\mathrm{DGal}(L_0/K_0)\simeq\mathrm{DGal}(L/K_0\cap L)\subset
\mathrm{DGal}(L/\mathbb{C}(z))$. Furthermore, if $K_0\cap L$ is
algebraic over $\mathbb{C}(z)$, then
$(\mathrm{DGal}(L_0/K_0))^0\simeq
(\mathrm{DGal}(L/\mathbb{C}(x)))^0.$
\end{proposition}

\begin{proof} By Proposition \ref{pr1}, the transformation $\varphi$
leads us to
$$\mathbb{C}(z)\simeq\varphi(\mathbb{C}(z))\hookrightarrow
K_0,$$ that is, we identify $\mathbb{C}(z)$ with
$\varphi(\mathbb{C}(z))$, and so that we can view $\mathbb{C}(z)$
as a subfield of $K_0$ and then by the Kaplansky's diagram (see
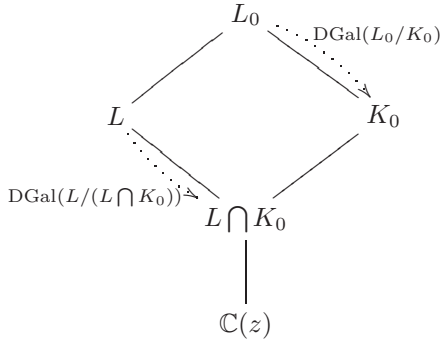
\cite{ka,we2}),

 \xymatrix{
      & L_0 \ar@{-}[dl] \ar@{-}[dr] \ar@/^/@{.>}[dr]^{\mathrm{DGal}(L_0/K_0)} & \\
L \ar@{-}[dr] \ar@/_/@{.>}[dr]_{\mathrm{DGal}(L/(L\bigcap K_0)) }
& & K_{0}
\ar@{-}[dl] \\
         & L\bigcap K_0  \ar@{-}[d]  &  \\
         & \mathbb{C}(z)             &
}

\noindent so that we have
$$\mathrm{DGal}(L_0/K_0)\simeq\mathrm{DGal}(L/K_0\cap
L)\subset \mathrm{DGal}(L/\mathbb{C}(z))$$ and if $K_0\cap L$ is
algebraic over $\mathbb{C}(z)$, then
$$(\mathrm{DGal}(L_0/K_0))^0\simeq
(\mathrm{DGal}(L/\mathbb{C}(z)))^0.$$
\end{proof}

Along the rest of this section we write $z=z(x)$ instead of
$\theta$.\\

\begin{remark}[Hard Algebrization]\label{hardal}  The proper pullback from
equation \eqref{eq2} to equation \eqref{eq1} is an algebrization
process. Therefore, we can try to algebrize any second order
linear differential equations with non-rational coefficients
(proper pullback) if we can put it in the form of equation
\eqref{eq2}. To do this, which will be called \textit{hard
algebrization}, we use the following steps.
\begin{description}
\item[Step 1.] Find $(\partial_x z)^2$ in the coefficient of
$y$ to obtain $\partial_xz$ and $z$.

\item[Step 2.] Divide by $(\partial_xz)^2$ to the coefficient of
$y$ to obtain $b(z)$ and check whether $b\in\mathbb{C}(z)$.

\item[Step 3.] Add $(\partial_x^2 z)/\partial_xz$ and divide
by $\partial_xz$ to the coefficient of $\partial_xy$ to obtain
$a(z)$ and check whether $a\in\mathbb{C}(z)$.

\end{description}
\end{remark}
\medskip

\noindent To illustrate this method, hard algebrization, we
present the following example.\\

\begin{example} In \cite[p. 256]{si3}, Singer presents the second order linear differential equation
$$\partial_x^2 r-{1\over x(\ln x+1)}\partial_x r-(\ln x+1)^2r=0.$$ To algebrize
this differential equation we choose $(\partial_xz)^2=(\ln
x+1)^2$, so that $\partial_xz=\ln x+1$ and for instance
$$z=\int (\ln x+1 )dx=x\ln x,\quad b(z)=-1.$$ Now we find $a(z)$
in the expression
$$a(z)(\ln x+1)-{1\over x(\ln x+1)}=-{1\over x(\ln x+1)},$$
obtaining $a(z)=0$. So that the new differential equation is given
by $\partial_z^2y-y=0$, in which $y(z(x))=r(x)$ and one basis of
solutions of this differential equation is given by $\langle
e^z,e^{-z}\rangle$. Thus, the respective basis of solutions of the
first differential equation is given by $\langle e^{x\ln x}
,e^{-x\ln x}\rangle$.
\end{example}
\medskip

In general, this method is not clear because the quest of $z=z(x)$
in $b(z)(\partial_xz)^2$ can be purely a lottery, or simply there
is not exists $z$ such that $a(z),b(z)\in\mathbb{C}(z)$. For
example, the equations presented by Singer in \cite[p. 257, 261,
270]{si3} and given by
$$\partial_x^2 r+{\mp 2x\ln^2x\mp 2x\ln x-1\over x\ln
x+x}\partial_x r+{-2x\ln^2 x-3x\ln x-x\mp 1\over x\ln x+ x }r=0,$$
$$\partial_x^2 r+{4x\ln x+ 2x\over 4x^2\ln x}\partial_x
r-{1\over 4x^2\ln x}r=0,\quad (x^2\ln^2x)\partial_x^2
r+(x\ln^2x-3x\ln x)\partial_x r+3r=0,$$ cannot be algebrized
systematically with this method, although it corresponds to
pullbacks (not proper pullback) of differential equations with
constant
coefficients.\\

In \cite{brfr}, Bronstein and Fredet developed and implemented an
algorithm to solve differential equations over
$\mathbb{C}(x,e^{\int_{}^{}f})$ without algebrizing it, see also
\cite{fr}. As an application of proposition \ref{pr1} we have the
following result\footnote{This result is given in \cite[\S
2]{acbl}, but we include here the proof for completeness.}.\\

\begin{proposition}[Linear differential equation over
$\mathbb{C}(x,e^{\int_{}^{}f})$, \cite{acbl}]\label{brofe}

Let $f\in\mathbb{C}(x)$ be a rational function. Then, the
differential equation

\begin{equation}\label{exp}
\partial_x^2{r}-\left(f+{\partial_x{f}\over f}
-fe^{\int_{}^{}f}a\left(e^{\int_{}^{}f}\right)\right)\partial_x{r}+\left(f\left(e^{\int_{}^{}f}\right)\right)^2b\left(e^{\int_{}^{}f}\right)r=0,
\end{equation}
is algebrizable by the change $z=e^{\int_{}^{}f}$ and its
algebraic form is given by
$$\partial_z^2y+a(z)\partial_zy+b(z)y=0, \quad
r(x)=y(z(x)).$$
\end{proposition}
\begin{proof}
Assume that $r(x)=y(z(x)),$  and $z=z(x)=e^{\int f dx}$. We can
see that
$$\partial_xz=fz,\quad \partial_zy={\partial_x r\over fz},\quad \partial_z^2y={1\over fz}\partial_x\left({\partial_x r\over fz}\right)=
{1\over \left(fe^{\int_{}^{}f}\right)^2}\left(\partial_x^2
r-f+\left({\partial_x{f}\over f}\right)\right)\partial_x r,$$
replacing in $\partial_z^2y+a(z)\partial_zy+b(z)y=0$ we obtain
equation \eqref{exp}.
\end{proof}

\begin{example}
The differential equation $$ \partial_x^2{r}-\left(x+{1\over x}
-2xe^{x^2}\right)\partial_x{r}+\lambda x^2e^{x^2}r=0,
$$
is algebrizable by the change $z=e^{x^2\over 2}$ and its algebraic
form is given by
$$\partial_z^2y+2z\partial_zy+\lambda y=0.$$

\end{example}
\begin{remark}
According to proposition \ref{brofe}, we have the following cases.

\begin{enumerate}
\item $f=n{\partial_xh\over h}$, for a rational function $h$, $n\in\mathbb{Z}_+$, we have the trivial case, both equations
are over the Riemann sphere and they have the same differential
field, so that does not need to be algebrized.

\item $f={1\over n}{\partial_xh\over h}$, for a rational function $h$,
$n\in\mathbb{Z}^+$, \eqref{exp} is defined over an algebraic
extension of $\mathbb{C}(x)$ and so that this equation is not
necessarily over the Riemann sphere.

\item $f\neq q{\partial_xh\over h}$, for any rational function $h$,  $q\in\mathbb{Q}$, \eqref{exp} is defined over a
transcendental extension of $\mathbb{C}(x)$ and so that this
equation is not over the Riemann sphere.
\end{enumerate}
\end{remark}

To algebrize second order linear differential equations is easier
when the term in $\partial_x r$ is absent, that is, in the form of
equation \eqref{LDE} and the change of variable is
{\it{Hamiltonian}}.

\begin{definition}[Hamiltonian change of variable, \cite{acbl}]\label{def2} A change of
variable $z=z(x)$ is called \textit{Hamiltonian} if
$(z(x),\partial_xz(x))$ is a solution curve of the autonomous
classical one degree of freedom Hamiltonian system
$$\begin{array}{l}\partial_xz=\partial_wH\\ \partial_xw=-\partial_zH\end{array}\quad \textrm{with}\quad H=H(z,w)={w^2\over 2}+V(z),$$
for some $V\in K$.
\end{definition}
\begin{remark}
  Assume that we algebrize equation \eqref{eq2} through a Hamiltonian
change of variables $z = z(x)$, i.e., $V\in\mathbb{C}(z)$. Then,
$K_0 = \mathbb C(z,
\partial_xz, \ldots)$, but, we have the algebraic relation,
  $$(\partial_xz) ^2 = 2h - 2V(z), \quad h = H(z,\partial_xz) \in \mathbb C,$$
so that $K_0 = \mathbb C(z,\partial_xz)$ is an algebraic extension
of $\mathbb C(z)$. By proposition \ref{pr1} the identity connected
component of the differential Galois group is conserved. On the
other hand, we can identify a Hamiltonian change of variable
$z=z(x)$ when there exists $\alpha\in K$ such that $(\partial_x
z)^2= \alpha(z)$. Thus, we introduce the {\it Hamiltonian
algebrization}, which corresponds to the algebrization process
done through a Hamiltonian change of variable.

\end{remark}

The following result, which also can be found in \cite[\S
2]{acbl}, is an example of Hamiltonian algebrization and
correspond to the case of reduced second order linear differential
equations.

\begin{proposition}[Hamiltonian Algebrization, \cite{acbl}]\label{pr2}
The differential equation
$$\partial_x^2{r}=q(x)r$$
is algebrizable through a Hamiltonian change of variable $z=z(x)$
if and only if there exist $f,\alpha$ such that
$${\partial_z\alpha\over\alpha},\quad {f\over \alpha}\in \mathbb{C}(z),\text{ where } f(z(x))=q(x),\quad \alpha(z)=2(H-V(z))=(\partial_xz)^2.$$
Furthermore, the algebraic form of the equation
$\partial_x^2{r}=q(x)r$ is
\begin{equation}\label{eq4}
\partial_z^2y+{1\over2}{\partial_z\alpha\over \alpha}\partial_zy-{f\over\alpha}y=0,\quad
r(x)=y(z(x)).
\end{equation}
\end{proposition}

\begin{remark}[Using the Algebrization Method]
The goal is to algebrize the differential equation $\partial_x^2
r=q(x)r$, so that we propose the following steps.

\begin{description}
\item[Step 1] Find a \emph{Hamiltonian change of variable} $z=z(x)$ and two functions  $f$ and $\alpha$ such that $q(x)=f(z(x))$ and
$(\partial_xz (x))^2=\alpha(z(x))$.
\item[Step 2] Verify whether or not $f(z)/\alpha (z)\in\mathbb{C}(z)$ and $\partial_z\alpha(z)/\alpha
(z)\in\mathbb{C}(z)$ to see if the equation $\partial_x^2 r=q(x)r$
is algebrizable.
\item[Step 3] If the equation $\partial_x^2
r=q(x)r$ is algebrizable, its algebrization is
$$\partial_z^2y+{1\over2}{\partial_z\alpha\over \alpha}\partial_zy-{f\over\alpha}y=0,\quad y(z(x))=r(x).$$
\end{description}
When we have algebrized the differential equation $\partial_x^2
r=q(x)r$, we study its integrability, Eigenring and its
differential Galois group.
\end{remark}
\begin{examples} Consider the following examples.
\begin{itemize}
\item Given the differential equation $\partial_x^2 r=f(\tan x)r$ with $f\in\mathbb{C}(\tan x)$, we can choose
 $z=z(x)=\tan x$ to obtain $\alpha(z)=(1+z^2)^2$, so that $z=z(x)$ is a Hamiltonian change of variable. We can
see that ${\partial_z\alpha\over\alpha}, {f\over
\alpha}\in\mathbb{C}(z)$ and the algebraic form of the
differential equation $\partial_x^2 r=f(\tan x)r$ with this
Hamiltonian change of variable is
$$\partial_z^2y+{2z\over 1+z^2}\partial_zy-{f(z)\over (1+z^2)^2}y=0,\quad y(\tan x)=r(x).$$

\item Given the differential equation $$\partial_x^2 r={\sqrt{1+x^2}+x^2\over 1+x^2}r,$$ we can choose
 $z=z(x)=\sqrt{1+x^2}$ to obtain $$f(z)={z^2+z-1\over z^2},\quad \alpha(z)={z^2-1\over z^2},$$ so that $z=z(x)$ is a Hamiltonian change of variable. We can
see that ${\partial_z\alpha\over\alpha}, {f\over
\alpha}\in\mathbb{C}(z)$ and the algebraic form for this case is
$$\partial_z^2y+{1\over z(z^2-1)}\partial_zy-{z^2+z-1\over z^2-1}y=0,\quad y(\sqrt{1+x^2})=r(x).$$
\end{itemize}
\end{examples}

We remark that in general the method of Hamiltonian algebrization
is not an algorithm, because the problem is to obtain a suitable
Hamiltonian $H$ satisfying definition \ref{def2}. We present now a
particular case of Hamiltonian algebrization considered as an
algorithm\footnote{Proposition \ref{cor1} is a slight improvement
of a similar result given in \cite[\S 2]{acbl}. Furthermore, we
include the proof here for completeness.}.

\begin{proposition}[Hamiltonian Algebrization Algorithm, \cite{acbl}]\label{cor1}
Consider $q(x)=g(z_1,\cdots, z_n)$, where $z_i=e^{\lambda_i x}$,
$\lambda_i\in\mathbb{C}^*$. The equation $\partial_x^2{r}=q(x)r$
is algebrizable if and only if.
$${\lambda_i\over \lambda_j}\in \mathbb{Q^*},\quad 1\leq i\leq n,\, 1\leq j\leq n,\quad g\in \mathbb{C}(z).$$
Furthermore, $\lambda_i=c_i\lambda$, where
$\lambda\in\mathbb{C}^*$ and $c_i\in \mathbb{Q}^*$ and for the
Hamiltonian change of variable
\begin{displaymath}
z=e^{\lambda x\over q},\text{ where } c_i={p_i\over q_i},\,
p_i,q_i\in\mathbb{Z}^*,\text{ }\gcd(p_i,q_i)=1 \text{ and }
q=\mathrm{lcm}(q_1,\cdots,q_n),
\end{displaymath} the algebrization of the differential equation
$\partial_x^2{r}=q(x)r$ is $$\partial_z^2y+{1\over
z}\partial_zy-q^2{g(z^{m_1},\ldots, z^{m_n})\over
\lambda^2z^2}y=0,\quad m_i={qp_i\over q_i},\quad y(z(x))=r(x).$$
\end{proposition}
\begin{proof} Assuming $\lambda_i/\lambda_j=c_{ij}\in\mathbb{Q}^*$ we can see that there exists $\lambda\in\mathbb{C}^*$
 and $c_i\in\mathbb{Q}^*$ such that $\lambda_i=\lambda c_i$, so that $$e^{\lambda_ix}=e^{c_i\lambda x}=e^{{p_i\over q_i}\lambda x}=
 \left(e^{{\lambda\over q}x}\right)^{qp_i\over q_i},\, p_i,q_i\in\mathbb{Z}^*,\, \gcd(p_i,q_i)=1,\, \mathrm{lcm}(q_1,\ldots,q_n)=q.$$ Now,
 setting $z=z(x)=e^{{\lambda\over q}x}$ we can see that
$$f(z)=g(z^{m_1},\ldots, z^{m_n}),\quad m_i={qp_i\over q_i},\quad \alpha={\lambda^2z^2\over q^2}.$$ Due
to $q|q_i$, we have that $m_i\in\mathbb{Z}$, so that
$${\partial_z\alpha\over\alpha},\quad {f\over\alpha}\in\mathbb{C}(z)$$ and
the algebraic form is given by $$\partial_z^2y+{1\over
z}\partial_zy-q^2{g(z^{m_1},\ldots, z^{m_n})\over
\lambda^2z^2}y=0,\quad y(z(x))=r(x).$$
\end{proof}
\begin{remark} Propositions \ref{pr2} and \ref{cor1} allow the algebrization
of a large number of second order differential equations, see for
example \cite{poza}. In particular, under the assumptions of
proposition \ref{cor1}, we can algebrize automatically
differential equations with trigonometrical or hyperbolic
coefficients.
\end{remark}
\begin{examples} Consider the following examples.
\begin{itemize}
\item Given the differential equation $$\partial_x^2 r={e^{\frac12x}+3e^{-\frac23x}-2e^{\frac54x}\over e^x+e^{-\frac32x}}r,\,
 \lambda_1=\frac12,\, \lambda_2=-\frac23,\, \lambda_3=\frac54,\, \lambda_4=1,\,\lambda_5=-\frac32,$$
we see that $\lambda_i/\lambda_j\in\mathbb{Q}$, $\lambda=1$,
$q=\mathrm{lcm}(1,2,3,4)=12$ and the Hamiltonian change of
variable for this case is $z=z(x)=e^{\frac1{12}x}$. We can see
that
$$\alpha(z)={1\over 144}z^2,\quad f(z)={z^6+3z^{-8}-2z^{15}\over
z^{12}+z^{-18}},\quad {\partial_z\alpha\over\alpha}, {f\over
\alpha}\in\mathbb{C}(z)$$ and the algebraic form is given by
$$\partial_z^2y+{1\over z}\partial_zy-144{z^6+3z^{-8}-2z^{15}\over z^{14}+z^{-16}}y=0,\quad y(e^{{1\over 12}x})=r(x).$$

\item Given the differential equation $$\partial_x^2 r=(e^{2\sqrt2x}+e^{-\sqrt2x}-e^{3x})r,\, \lambda_1=2\sqrt2,\, \lambda_2=-\sqrt2,\,\lambda_3=3,$$
we see that $\lambda_1/\lambda_2\in\mathbb{Q}$, but
$\lambda_1/\lambda_3\notin\mathbb{Q}$, so that this differential
equation cannot be algebrized.
\end{itemize}
\end{examples}

We remark that it is possible to use the algebrization method to
transform differential equations, although either the starting
equation has rational coefficients or the transformed equation has
not rational coefficients.

\begin{examples} As illustration we present the following examples.
\begin{itemize}
\item Consider the following differential equation
$$\partial_x^2r={x^4+3x^2-5\over x^2+1}r=0,$$ we can choose
$z=z(x)=x^2$ so that $\alpha=4z$ and the new differential equation
is $$\partial_z^2y+{1\over2 z}\partial_zy-{z^2+2z-5\over
4z(z+1)}y=0,\quad y(x^2)=r(x).$$
\item Consider the Mathieu's differential
equation $\partial_x^2 r=(a+b\cos(x))r$, we can choose
$z(x)=\ln(\cos(x))$ so that $\alpha=e^{-2z}-1$ and the new
differential equation is
$$\partial_z^2y-{1\over 1-e^{2z}}\partial_zy-{ae^{2z}+be^{3z}\over 1-e^{2z}}y=0,\quad y(\ln(\cos (x)))=r(x).$$
\end{itemize}
\end{examples}

Recently, the Hamiltonian algebrization (propositions \ref{pr2}
and \ref{cor1}) has been applied in \cite{ac,acalde,acbl} to
obtain non-integrability in the framework of \textit{Morales-Ramis
theory} \cite{mo2,mo}.

\subsection{The Operator $\widehat{\partial_z}$ and the Hamiltonian Algebrization}
The generalization of proposition \ref{pr1} to higher order linear
differential equations is difficult. But, it is possible to obtain
generalizations of proposition \ref{pr2} by means of Hamiltonian
change of variable. We recall that $z=z(x)$ is a Hamiltonian
change of variable if there exists $\alpha$ such that
$(\partial_xz)^2=\alpha(z)$. More specifically, if $z=z(x)$ is a
Hamiltonian change of variable, we can write
$\partial_xz=\sqrt{\alpha}$, which leads us to the following
notation: $\widehat{\partial}_z=\sqrt{\alpha}\partial_z$.

We can see that $\widehat{\partial}_z$ is a \textit{derivation}
because satisfy
$\widehat{\partial}_z(f+g)=\widehat{\partial}_zf+\widehat{\partial}_zg$
and the Leibnitz rules $$\widehat{\partial}_z(f\cdot
g)=\widehat{\partial}_zf\cdot g +f\cdot\widehat{\partial}_zg,\quad
\widehat{\partial}_z\left(\frac{f}{g}\right)=\frac{\widehat{\partial}_zf\cdot
g -f\cdot\widehat{\partial}_zg}{g^2}.$$ We can notice that the
chain rule is given by $\widehat{\partial}_z(f\circ
g)=\partial_gf\circ g\widehat{\partial}_z(g)\neq
\widehat{\partial}_gf\circ g\widehat{\partial}_z(g)$. The
iteration of $\widehat{\partial}_z$ is given by
$$\widehat{\partial}_z^0=1,\quad
\widehat{\partial}_z=\sqrt{\alpha}\partial_z,\quad
\widehat{\partial}_z^n=\sqrt{\alpha}\partial_z\widehat{\partial}^{n-1}_z=
\underbrace{\sqrt{\alpha}\partial_z\left(\ldots
\left(\sqrt{\alpha}\partial_z\right)\right)}_{n\text{ times }
\sqrt{\alpha}\partial_z}.$$ We say that a Hamiltonian change of
variable is rational when the potential $V\in\mathbb{C}(x)$ and
for instance $\alpha\in\mathbb{C}(x)$. Along the rest of this
thesis, we understand
$\widehat{\partial}_z=\sqrt{\alpha}\partial_z$ where $z=z(x)$ is a
Hamiltonian change of variable and $\partial_xz=\sqrt{\alpha}$. In
particular, $\widehat{\partial}_z=\partial_z=\partial_x$ if and
only if $\sqrt{{\alpha}}=1$, i.e., $z=x$.\\

\begin{theorem}\label{propjawpa}  Consider the systems of linear differential equations
$[A]$ and $[\widehat{A}]$ given respectively by
$$\partial_x\mathbf{Y}=-A\mathbf{Y},\quad
\widehat{\partial}_z\widehat{\mathbf{Y}}=-\widehat{A}\widehat{\mathbf{Y}},\quad
A=[a_{ij}],\quad \widehat{A}=[\widehat{a}_{ij}],\quad
\mathbf{Y}=[y_{i1}], \quad
\widehat{\mathbf{Y}}=[\widehat{y}_{i1}],$$ where $a_{ij}\in
K=\mathbb{C}(z(x),\partial_x(z(x)))$,
$\widehat{a}_{ij}\in\mathbb{C}(z)\subseteq
\widehat{K}=\mathbb{C}(z,\sqrt{\alpha})$, $1\leq i\leq n$, $1\leq
j\leq n$, $a_{ij}(x)=\widehat{a}_{ij}(z(x))$ and
$y_{i1}(x)=y_{i1}(z(x))$. Suppose that $L$ and $\widehat{L}$ are
the Picard-Vessiot extensions of $[A]$ and $[\widehat{A}]$
respectively. If the transformation $\varphi$ is given by
$$\varphi:\begin{array}{l}x\mapsto z\\ a_{ij}\mapsto
\widehat{a}_{ij}\\y_{i1}(x)\mapsto\widehat{y}_{i1}(z(x))\\
 \partial_x\mapsto\widehat{\partial}_z\end{array},$$
then the following statements hold.
\begin{itemize}
\item $K\simeq \widehat{K}$,\quad $(K,\partial_x)\simeq (\widehat{K},\widehat{\partial}_z)$.
\item $\mathrm{DGal}(L/K)\simeq\mathrm{DGal}(\widehat{L}/\widehat{K})\subset
\mathrm{DGal}(\widehat{L}/{\mathbb{C}(z)})$.
\item $(\mathrm{DGal}(L/K))^0\simeq
(\mathrm{DGal}(\widehat{L}/{\mathbb{C}(z)})^0.$
\item $\mathcal E(A)\simeq \mathcal E(\widehat{A}).$
\end{itemize}
\end{theorem}

\begin{proof} We proceed as in the proof of proposition \ref{propjaw}. As $z=z(x)$ is a rational Hamiltonian change of variable, the transformation $\varphi$ leads us to
$$\mathbb{C}(z)\simeq\varphi(\mathbb{C}(z))\hookrightarrow
K,\quad K\simeq \widehat{K},\quad \mathbb{C}(z)\hookrightarrow
\widehat{K},\quad (K,\partial_x)\simeq
(\widehat{K},\widehat{\partial}_z)$$ that is, we identify
$\mathbb{C}(z)$ with $\varphi(\mathbb{C}(z))$, and so that we can
view $\mathbb{C}(z)$ as a subfield of $K$ and then, by the
Kaplansky's diagram (see \cite{ka,we2}),

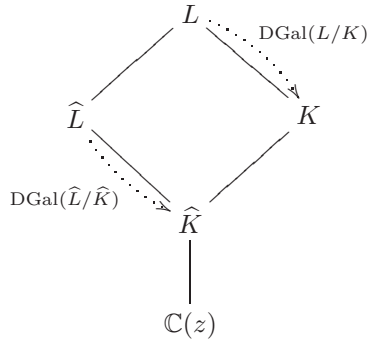
 \xymatrix{
      & L \ar@{-}[dl] \ar@{-}[dr] \ar@/^/@{.>}[dr]^{\mathrm{DGal}(L/K)} & \\
\widehat{L} \ar@{-}[dr]
\ar@/_/@{.>}[dr]_{\mathrm{DGal}(\widehat{L}/{\widehat{K}})} & & K
\ar@{-}[dl] \\
         & \widehat{K}  \ar@{-}[d]  &  \\
         & \mathbb{C}(z)             &
}

\noindent so that we have
$\mathrm{DGal}(L/K)\simeq\mathrm{DGal}(\widehat{L}/{\widehat{K}})\subset
\mathrm{DGal}(\widehat{L}/{\mathbb{C}(z)})$,
$(\mathrm{DGal}(L/K))^0\simeq
(\mathrm{DGal}(\widehat{L}/{\mathbb{C}(z)}))^0,$ and $\mathcal
E(A)\simeq \mathcal E(\widehat{A}).$
\end{proof}
\medskip

We remark that the transformation $\varphi$, given in theorem
\ref{propjawpa}, is virtually strong isogaloisian when
$\sqrt{\alpha}\notin\mathbb{C}(z)$ and for
$\sqrt{\alpha}\in\mathbb{C}(z)$, $\varphi$ is strong isogaloisian.
Furthermore, by cyclic vector method (see \cite{vasi}), we can
write the systems $[A]$ and $[\widehat{A}]$ in terms of the
differential equations ${\mathcal L}$ and $\widehat{\mathcal L}$.
Thus, $\widehat{\mathcal L}$ is the proper pullback of $\mathcal
L$ and $\mathcal E(\mathfrak L)\simeq \mathcal
E(\widehat{\mathfrak L}).$\medskip

\begin{example} Consider the system $$[A]:=\begin{array}{l}\partial_x\gamma_1 =-{\frac
{2\sqrt{2}}{{e^{x}}+{e^{-x}}}}\gamma_3,\\
\\\partial_x\gamma_2=\frac{e ^{x}-e^{-x}}{e^{x}+e^{-x}}\gamma_3,\\ \\
\partial_x\gamma_3 =\frac {2 \sqrt {2}}{e^{x}+e^{-x}}\gamma_1-\frac {e^{x}- e^{-x}}
{e^{x}+e^{-x}}\gamma_2,
\end{array}
$$ which through the Hamiltonian change of variable $z=e^x$, and for instance $\sqrt{\alpha}=z$, it is transformed in the system

$$
\widehat{[A]}:=\begin{array}{l}\partial_z\widehat{\gamma}_1 =-\frac{2\sqrt{2}}{z^2+1}\widehat{\gamma}_3,\\
\\\partial_z\widehat{\gamma}_2=\frac{z^2-1}{z(z^2+1)}\widehat{\gamma}_3,\\ \\
\partial_z\widehat{\gamma}_3 =\frac {2 \sqrt {2}}{z^2+1}\widehat{\gamma}_1-\frac{z^2-1}
{x(x^2+1)}\widehat{\gamma}_2.
\end{array}
$$
One solution of the system $\widehat{[A]}$ is given by
$${1\over z^2+1}\begin{pmatrix}{\sqrt{2}\over 2}(1 - z^2)\\ z\\
-z\end{pmatrix},$$ and for instance, $${1\over e^{2x}+1}\begin{pmatrix}{\sqrt{2}\over 2}(1 - e^{2x})\\ e^x\\
-e^x\end{pmatrix}$$ is the corresponding solution for the system
$[A]$.
\end{example}
\medskip

\begin{remark}
The algebrization given in proposition \ref{pr2} is an example of
how the introduction of the new derivative $\widehat{\partial}_z$
simplifies the proofs and computations. Such proposition is
naturally extended to $\partial_x^2y+a\partial_xy+by=0$, using
$\varphi$ to obtain
$\widehat{\partial}_z^2\widehat{y}+\widehat{a}\widehat{\partial}_z\widehat{y}+\widehat{b}\widehat{y}=0$,
which is equivalent to
\begin{equation}\label{eqalgen}
\alpha\partial_z^2\widehat{y}+\left({\partial_x\alpha\over
2}+{\sqrt{\alpha}\widehat{a}
}\right)\partial_z\widehat{y}+\widehat{b}\widehat{y}=0,
\end{equation} where $y(x)=\widehat{y}(z(x))$, $\widehat{a}(z(x))=a(x)$ and
$\widehat{b}(z(x))=b(x)$.\\

In general, for $y(x)=\widehat{y}(z(x))$, the equation
$F(\partial_x^ny,\ldots,y,x)=0$ with coefficients given by
$a_{i_k}(x)$ is transformed in the equation
$\widehat{F}(\widehat{\partial}_z^n\widehat{y},\ldots,\widehat{y},z)=0$
with coefficients given by $\widehat{a}_{i_k}(z)$, where
$a_{i_k}(x)=\widehat{a}_{i_k}(z(x))$. In particular, for
$\sqrt{\alpha}, \widehat{a}_{i_k}\in\mathbb{C}(z)$, the equation
$\widehat{F}(\widehat{\partial}_z^n\widehat{y},\ldots,\widehat{y},z)=0$
is the Hamiltonian algebrization of
$F(\partial_x^ny,\ldots,y,x)=0$. Now, if each derivation
$\partial_x$ has order even, then $\alpha$ and $\widehat{a}_{i_k}$
can be rational functions to algebrize the equation
$F(\partial_x^ny,\ldots,y,x)=0$, where
$a_{i_k}\in\mathbb{C}(z(x),\partial_xz(x))$. for example, that
happens for linear differential equations given by
$$\partial_x^{2n}y+a_{n-1}(x)\partial_x^{2n-2}y+\ldots+a_2(x)\partial_x^4y+a_1(x)\partial_x^2y+a_0(x)y=0.$$
Finally, the algebrization algorithm given in proposition
\ref{cor1} can be naturally extended to any differential equation
$$F(\partial_x^ny,\partial_x^{n-1}y,\ldots,\partial_xy,y,
e^{\mu t})=0,$$ that by means of the change of variable $z=e^{\mu
x}$ is transformed into
$$\widehat{F}(\widehat{\partial}_z^n\widehat{y},\widehat{\partial}_z^{n-1}\widehat{y},\ldots,\widehat{\partial}_z\widehat{y},y,
z)=0.$$ Particularly, we consider the algebrization of Riccati
equations, higher order linear differential equations and systems.
\end{remark}
\medskip

\begin{examples} The following corresponds
to some examples of algebrizations for differential equations
given in \cite[p. 258, 266]{si3}.

\begin{enumerate}
\item The equation $\mathcal L:=\partial_x^2 y+(-2e^x-1)\partial_x y+e^{2x}y=0$ with the Hamiltonian
change of variable $z=e^x$, $\sqrt{\alpha}=z$, $\widehat{a}=-2z-1$
and $\widehat{b}=z^2$ is transformed in the equation
$\widehat{\mathcal
L}:=\partial_z^2\widehat{y}-2\partial_z\widehat{y}+\widehat{y}=0$
which can be easily solved. A basis of solutions for $\mathcal L$
and $\widehat{\mathcal L}$ are given by $\langle e^z,ze^z\rangle$
and $\langle e^{e^x},e^xe^{e^x}\rangle$ respectively. Furthermore
$K=\mathbb{C}(e^x)$, $\widehat K=\mathbb{C}(z)$, $L$ and
$\widehat{L}$ are the Picard-Vessiot extensions of $\mathcal L$
and $\widehat{\mathcal L}$ respectively. Thus,
$\mathrm{DGal}(L/K)=\mathrm{DGal}(\widehat{L}/{\widehat{K}})$.

\item The differential equation $$\mathcal L:= \partial_x^2 y+{-24e^x-25\over 4e^x+5}\partial_x y+{20e^x\over 4e^x+5}y=0$$ with the Hamiltonian
change of variable $z=e^x$, $\sqrt{\alpha}=z$,
$$\widehat{a}={-24z-25\over 4z+5} \textrm{  and   } \widehat{b}={20z\over
4z+5}$$ is transformed in the equation
$$\widehat{\mathcal L}:=\partial_z^2\widehat{y}+{-20(z+1)\over x(4z+5)}\partial_z\widehat{y}+{20\over z(4z+5)}\widehat{y}=0,$$
which can be solved with Kovacic algorithm. A basis of solutions
for $\widehat{\mathcal L}$ is $\langle z+1,z^5\rangle$, so that a
basis for $\mathcal L$ is $\langle e^x+1,e^{5x}\rangle$.
Furthermore $K=\mathbb{C}(e^x)$, $\widehat{K}=\mathbb{C}(z)$, $L$
and $\widehat{L}$ are the Picard-Vessiot extensions of $\mathcal
L$ and $\widehat{\mathcal L}$ respectively. Thus,
$\mathrm{DGal}(L/K)=\mathrm{DGal}(\widehat L/ {\widehat{K}})=e$.
\end{enumerate}

\end{examples}
\medskip

\begin{remark}[Algebrization of the Riccati equation]\label{alric} The Riccati equation
\begin{equation}\label{eqalric0}\partial_xv=a(x)+b(x)v+c(x)v^2\end{equation} through
the Hamiltonian change of variable $z=z(x)$, becomes in the
Riccati equation
\begin{equation}\label{eqalric}
\partial_z\widehat{v}={1\over
\sqrt{\alpha}}(\widehat{a}(z)+\widehat{b}(z)\widehat{v}+\widehat{c}(z)\widehat{v}^2),
\end{equation}
where $v(x)=\widehat{v}(z(x))$, $\widehat{a}(z(x))=a(x)$,
$\widehat{b}(z(x))=b(x)$, $\widehat{c}(z(x))=c(x)$ and
$\sqrt{\alpha(z(x))}=\partial_xz(x)$. Furthermore, if
$\sqrt{\alpha}$, $\widehat{a}$, $\widehat{b}$, $\widehat{c}\in
\mathbb{C}(x)$, equation \eqref{eqalric} is the algebrization of
equation \eqref{eqalric0}.
\end{remark}
\medskip

\begin{example} Consider the Riccati differential equation
$$\mathcal L:=\partial_x v=\left(\tanh x-{1\over \tanh x}\right)v+\left(3\tanh x-3\tanh^3x\right)v^2,$$
which through the Hamiltonian change of variable $z=\tanh x$, for
instance $\sqrt{\alpha}=1-z^2$, is transformed into the Riccati
differential equation $$\widehat{\mathcal L}:=\partial_zv=-{1\over
z}v+3zv^2.$$ One solution for the equation $\widehat{\mathcal L}$
is
$$-{1\over z(3z-c)}, \text{  being }c \text{ a constant  },$$ so
that the corresponding solution for equation $\mathcal L$ is
$$-{1\over \tanh x(3\tanh x-c)}.$$
\end{example}

The following result is the algebrized version of the relationship
between the Eigenrings of systems and operators.

\begin{proposition}\label{aleigen} Consider the differential fields $K$, $\widehat{K}$ and consider the systems $[A]$ and $[\widehat{A}]$ given by
$$\partial_x{\mathbf{X}}=-A\mathbf{X}, \,
\widehat{\partial}_z{\widehat{\mathbf{X}}}=-\widehat{A}\widehat{\mathbf{X}},
\, \widehat{\partial}_z=\sqrt{\alpha}\partial_z,\, A=[a_{ij}],\,
\widehat{A}=[\widehat{a}_{ij}], \,a_{ij}\in K, \,
\widehat{a}_{ij}\in \widehat{K},$$ where $z=z(x)$,
$\mathbf{X}(x)=\widehat{\mathbf{X}}(z(x))$,
$\widehat{a}_{ij}(z(x))=a_{ij}(x)$, then $\mathcal{E}(A)\simeq
\mathcal{E}(\widehat{A})$. In particular, if we consider the
linear differential equations $$\mathcal L:=\partial_x^ny+
\sum_{k=0}^{n-1}a_k\partial_x^k y=0\quad \textrm{and} \quad
\widehat{\mathcal L}:=\widehat{\partial}_z^n\widehat{y}+
\sum_{k=0}^{n-1}\widehat{a}_k\widehat{\partial}_z^k
\widehat{y}=0,$$ where $z=z(x)$, $y(x)=\widehat{z}((x))$,
$\widehat{a}_{k}(z(x))=a_{k}(x)$, $a_k\in K$, $\widehat{a}_{k}\in
\widehat{K}$, then $\mathcal{E}(\mathfrak L)\simeq
\mathcal{E}(\widehat{\mathfrak L})$, where $\mathcal L:=\mathfrak
L y=0 $ and $\widehat{\mathcal L}:=\widehat{\mathfrak L}
\widehat{y}=0 $. Furthermore, assuming
$$P=\begin{pmatrix}p_{11}&\dots&p_{1n}\\ \vdots
\\p_{n1}&\ldots&p_{nn}\end{pmatrix},\quad
A=\begin{pmatrix}0&1&\dots&0\\ \vdots
\\-a_0&-a_1&\ldots&-a_{n-1}\end{pmatrix},$$ then
$$\mathcal E(\mathfrak L)=\left\{\sum_{k=1}^np_{1k}\partial_x^{k-1}:\,
\partial_xP=PA-AP,\, p_{ik}\in K\right\},$$
if and only if $$\mathcal E(\widehat{\mathfrak
L})=\left\{\sum_{k=1}^n\widehat{p}_{1k}\widehat{\partial}_z^{k-1}:\,
\widehat{\partial}_z\widehat{P}=\widehat{P}\widehat{A}-\widehat{A}\widehat{P},
\, \widehat{p}_{ik}\in \widehat{K} \right\}.$$
\end{proposition}

\begin{proof} By theorem \ref{propjawpa} we have that $K\simeq \widehat{K}$, $\mathcal{E}(A)\simeq
\mathcal{E}(\widehat{A})$ and $\mathcal{E}(\mathfrak L)\simeq
\mathcal{E}(\widehat{\mathfrak L})$. Using the derivation
$\widehat{\partial}_z$ and by induction on lemma \ref{propjaw} we
complete the proof.

\end{proof}

\begin{examples} We consider two different examples to illustrate the previous proposition.

\begin{itemize}\item Consider the differential equation
$\mathcal L_1:=\partial_x^2y-(1+\cos x-\cos^2x)y=0$. By means of
the Hamiltonian change of variable $z=z(x)=\cos x$, with
$\sqrt{\alpha}=-\sqrt{1-z^2}$, $\mathcal L_1$ is transformed into
the differential equation
$$\widehat{\mathcal L}_1:=\partial_z^2\widehat{y}-{z\over 1-z^2}\partial_z\widehat{y}-{1+z-z^2\over
1-z^2}\widehat{y}=0.$$ Now, computing the eigenring of
$\widehat{\mathfrak L}_1$ we have that $\mathcal
E(\widehat{\mathfrak L}_1)=\mathrm{Vect}(1)$, therefore the
eigenring of $\mathfrak L_1$ is given $\mathcal E(\mathfrak
L_1)=\mathrm{Vect}(1)$.

\item Now we consider the differential equation
$\mathcal L_2:=\partial_x^2y=\left(e^{2x}+\frac94\right)y$. By
means of the Hamiltonian change of variable $z=e^x$, with
$\sqrt{\alpha}=x$, $\mathcal L_2$ is transformed into the
differential equation
$$\widehat{\mathcal L}_2:=\partial_z^2\widehat{y}+{1\over z}\partial_z\widehat{y}-\left(1+{9\over
4x^2}\right)\widehat{y}=0.$$ Now, computing the eigenring of
$\widehat{\mathfrak L}_2$ we have that {\small{$$\mathcal
E(\widehat{\mathfrak L}_2)=\mathrm{Vect}\left(1,-2\left(\frac
{z^2-1}{z^2}\right)\partial_z-\frac{z^2-3}{z^3}
\right)=\mathrm{Vect}\left(1,-2\left(\frac
{z^2-1}{z^3}\right)\widehat{\partial}_z-\frac{z^2-3}{z^3}
\right),$$}} therefore the eigenring of $\mathfrak{L}_2$ is given
by
$$\mathcal E(\mathfrak{L}_2)=\mathrm{Vect}\left(1,-2\left(\frac
{e^{2x}-1}{e^{3x}}\right)\partial_x-\frac{e^{2x}-3}{e^{3x}}
\right).$$ The same result is obtained via matrix formalism, where
$$A=\begin{pmatrix}0&1\\e^{2x}+\frac94&0\end{pmatrix},\,
\widehat{A}=\begin{pmatrix}0&1\\z^{2}+\frac94&0\end{pmatrix},\,
\partial_xP=PA-AP,\,
\widehat{\partial}_z\widehat{P}=\widehat{P}\widehat{A}-\widehat{A}\widehat{P},$$
with $P\in Mat(2,\mathbb{C}(e^x))$ and $\widehat{P}\in
Mat(2,\mathbb{C}(z))$.

\end{itemize}
\end{examples}

\subsection{Applications in Supersymmetric Quantum Mechanics}
In this subsection we apply the derivation $\widehat{\partial}_z$
to the Schr\"odinger equation $H\Psi=\lambda\Psi$, where
$H=-\partial_x^2+V(x)$, $V\in K$. Assume that $z=z(x)$ is a
rational Hamiltonian change of variable for $H\Psi=\lambda\Psi$,
then $K=\mathbb{C}(z(x),\partial_xz(x))$. Thus, the
\textit{algebrized Schr\"odinger equation} is written as
\begin{equation}\label{eqalgs1}\widehat{H}\widehat{\Psi}=\lambda\widehat{\Psi},\quad
\widehat{H}=-\widehat{\partial}_z^2+\widehat{V}(z), \quad
\widehat{\partial}_z^2=\alpha\partial_z^2+\frac12\partial_z\alpha\partial_z,\quad
\widehat{K}=\mathbb{C}(z,\sqrt{\alpha}).\end{equation} The
\textit{reduced algebrized Schr\"odinger equation}, obtained
through equation \eqref{redsec}, is given by

\begin{equation}\label{eqalgs2}\widehat{\mathbf{H}}\Phi=\lambda\Phi,\quad \widehat{\mathbf{H}}=\alpha(z)\left(-{\partial}_z^2+\widehat{\mathbf
V}(z)\right),\quad \begin{array}{l}\widehat{\mathbf
V}(z)=\mathcal{V}+{\widehat{V}(z)\over \alpha},\\ \\ \mathcal{V}=
\partial_z\mathcal{W}+\mathcal{W}^2,\\ \\
\mathcal{W}={1\over 4}{\partial_z\alpha(z)\over
\alpha(z)}.\end{array}\end{equation} The eigenfunctions $\Psi$,
$\widehat{\Psi}$ and $\Phi$ corresponding to the operators $H$,
$\widehat{H}$ and $\widehat{\mathbf{H}}$  are related respectively
as
$$\Phi(z(x))=\sqrt[4]{\alpha}\widehat{\Psi}(z(x))=\sqrt[4]{\alpha}\Psi(x).$$

In order to apply the Kovacic's algorithm we only consider the
algebrized operator $\widehat{\mathbf{H}}$, whilst the eigenrings
will be computed on $\widehat{H}$. Also it is possible to apply
the version of Kovacic's algorithm given in reference \cite{ulwe}
to the algebraized operator $\widehat{H}$. The following results
are obtained by applying Kovacic's algorithm to the reduced
algebrized Schr\"odinger equation (equation \eqref{eqalgs2})
$\widehat{\mathbf{H}}\Phi=\lambda\Phi$.

\begin{proposition}
Let $\widehat{\mathbf{L}}_\lambda$ be the Picard-Vessiot extension
of the reduced algebrized Schr\"odinger equation
$\widehat{\mathbf{H}}\Phi=\lambda\Phi$ with
$\alpha,\widehat{V}\in\mathbb{C}[z]$. If $\deg\alpha < 2+\deg
\widehat{V}$, then
$\mathrm{DGal}(\widehat{\mathbf{L}}/{\widehat{K}})$ is a not
finite primitive for every $\lambda\in\Lambda$.
\end{proposition}
\begin{proof} Suppose that $\deg \alpha=n$ and $\deg \widehat{V}=m$. The reduced algebrized Schr\"odinger equation
$\widehat{\mathbf{H}}\Phi=\lambda\Phi$ can be written in the form
$$\partial_z^2\Phi=r\Phi,\quad r={4\alpha\partial_z^2\alpha-3(\partial_z\alpha)^2+16\alpha(\widehat{V}-\lambda)\over
16\alpha^2}.$$ Due to $m>n-2$ we have that $\circ
(r_\infty)=n-m<2$, which does not satisfy the condition ($\infty$)
of the case 3 of Kovacic's algorithm, therefore
$\mathrm{DGal}(\widehat{\mathbf{L}}/{\widehat{K}})$ is a not
finite primitive for every $\lambda\in\Lambda$.

\end{proof}
\begin{proposition}
Let $\widehat{\mathbf{L}}_\lambda$ be the Picard-Vessiot extension
of the reduced algebrized Schr\"odinger equation
$\widehat{\mathbf{H}}\Phi=\lambda\Phi$ with
$\alpha\in\mathbb{C}[z]$, $\widehat{V}\in\mathbb{C}(z)$. If
$\circ(\widehat{V})_\infty <2-\deg\alpha$, then
$\mathrm{DGal}(\widehat{\mathbf{L}}_\lambda/{\widehat{K}})$ is a
not finite primitive for every $\lambda\in\Lambda$.
\end{proposition}
\begin{proof}  Suppose that $\widehat{V}=s/t$, being $s$ and $t$ co-primes polynomials in $\mathbb{C}(z)$. Assume that $\deg \alpha=n$,
$\deg s=m$ and $\deg t=p$. The reduced algebrized Schr\"odinger
equation $\widehat{\mathbf{H}}\Phi=\lambda\Phi$ can be written in
the form
$$\partial_z^2\Phi=r\Phi,\quad r={4t\alpha\partial_z^2\alpha-3t(\partial_z\alpha)^2+16\alpha(s-\lambda t)\over
16t\alpha^2}.$$ Due to $m>n+p-2$ we have that $\circ
(r_\infty)=p+n-m<2$, which does not satisfy the condition
($\infty$) of the case 3 of Kovacic's algorithm, therefore
$\mathrm{DGal}(\widehat{\mathbf{L}}/{\widehat{K}})$ is a not
finite primitive for every $\lambda\in\Lambda$.

\end{proof}
\medskip

\begin{remark}
In a natural way, we obtain the algebrized versions of Darboux
transformation, i.e., the \textit{algebrized Darboux
transformation}, denoted by $\widehat{\mathrm{DT}}$. By
$\widehat{\mathrm{DT}}_n$ we denote the $n$ iteration of
$\widehat{\mathrm{DT}}$, and by $\widehat{\mathrm{CI}}_n$ we
denote the \textit{algebrized Crum iteration}, where the \textit{algebrized wronskian} is given by $$\, \widehat{W}(\widehat{y}_1,\ldots,\widehat{y}_n)=\left|\begin{matrix}\widehat{y}_1&\cdots& \widehat{y}_n\\
 \vdots& & \vdots \\ \widehat{\partial}^{n-1}_z\widehat{y}_1&\cdots& \widehat{\partial}^{n-1}_z\widehat{y}_n\end{matrix}\right|.$$
In the same way, we define \textit{algebrized shape invariant
potentials}, \textit{algebrized superpotential} $\widehat{W}$,
\textit{algebrized supersymmetric Hamiltonians} $\widehat{H}_\pm$,
algebrized supersymmetric partner potentials $\widehat{V}_\pm$,
\textit{algebrized ground state}
$\widehat{\Psi}_0^{(-)}=e^{-\int{\widehat{W}\over
\sqrt{\alpha}}dz}$,\textit{ algebrized wave functions}
$\widehat{\Psi}_\lambda^{(-)}$, \textit{algebrized raising and
lowering operators} $\widehat{A}$ and $\widehat{A}^\dagger$. Thus,
we can rewrite entirely the section \ref{susyrapo} using the
derivation $\widehat{\partial}_z$.
\end{remark}
\medskip

The following theorem show us the relationship between the
algebrization and Darboux transformation.\\

\begin{theorem}  Given the Schr\"odinger equation $\mathcal L_\lambda:=H_-\Psi^{(-)}=\lambda\Psi^{(-)}$, the relationship
 between the algebrization $\varphi$ and Darboux
transformations $\mathrm{DT}$, $\widehat{\mathrm{DT}}$ with
respect to $\mathcal L_\lambda$ is given by
$\widehat{\mathrm{DT}}\varphi=\varphi\mathrm{DT}$, that is
$\widehat{\mathrm{DT}}\varphi(\mathcal L
)=\varphi\mathrm{DT}(\mathcal L)$. In other words, the Darboux
transformations $\mathrm{DT}$ and $\widehat{\mathrm{DT}}$ are
\emph{intertwined} by the algebrization $\varphi$.
\end{theorem}
\begin{proof} Assume the equations $\mathcal L_{\lambda}:=H_-\Psi^{(-)}=\lambda\Psi^{(-)}$ , $\widehat{\mathcal L}_\lambda:=
\widehat{H}_-\widehat{\Psi}^{(-)}=\lambda\widehat{\Psi}^{(-)}$,
$\widetilde{\mathcal L}_\lambda:=H_+\Psi^{(+)}=\lambda\Psi^{(+)}$
and $\widetilde{\widehat{\mathcal L}}:=
\widehat{H}_+\widehat{\Psi}^{(+)}=\lambda\widehat{\Psi}^{(+)}$,
where the Darboux transformations $\mathrm{DT}$ and
$\widehat{\mathrm{DT}}$ are given by $\mathrm{DT}(\mathcal
L)=\widetilde{\mathcal L}$,
$\widehat{\mathrm{DT}}(\widehat{\mathcal
L})=\widetilde{\widehat{\mathcal L}}$,
$$ \mathrm{DT}:
 \begin{array}{l} V_-\mapsto V_+\\ \\
  \Psi_\lambda^{(-)}\mapsto \Psi_\lambda^{(+)}\end{array},\quad \widehat{\mathrm{DT}}:\begin{array}{l}
   \widehat{V}_-\mapsto \widehat{V}_+\\ \\ \widehat{\Psi}_\lambda^{(-)}\mapsto\widehat{\Psi}_\lambda^{(+)},\end{array}$$
   and $\varphi(\mathcal
L_\lambda)=\widehat{\mathcal L_\lambda}$, where the algebrization
$\varphi$ is given as in theorem \ref{propjawpa}. Then the
following diagram commutes
 $$\xymatrix{\mathcal L_\lambda \ar[r]^-{\mathrm{DT}} &\widetilde{\mathcal L}_\lambda \\
   \widehat{\mathcal L}_\lambda  \ar[r]_{\widehat{\mathrm{DT}}} \ar@{<-}[u]^{\varphi} & \ar@{<-}[u]_{\varphi} \widetilde{\widehat{\mathcal L}}_\lambda }
   \qquad \begin{array}{c}\\ \\ \\
   \Rightarrow \widehat{\mathrm{DT}}\varphi(\mathcal L
)=\varphi\mathrm{DT}(\mathcal
L)\Leftrightarrow\widetilde{\widehat{\mathcal
L}}=\widehat{\widetilde{\mathcal L}}.\end{array}$$
\end{proof}
To illustrate $\widehat{\mathrm{DT}}$ we present the following
examples.\\

\begin{examples} Consider the algebrized Schr\"odinger equation $\widehat{H}\widehat{\Psi}^{(-)}=\lambda\widehat{\Psi}^{(-)}$ with:
\begin{itemize}
\item $\sqrt{\alpha(z)}=\sqrt{z^2-1}$ and $\widehat{V}_-(z)=\frac{z}{z-1}$. Taking
$\lambda_1=1$ and $\widehat{\Psi}_1^{(-)}=\sqrt{z+1\over z-1}$, we
have that
$\widehat{\mathrm{DT}}(\widehat{V}_-)=\widehat{V}_+(z)=\frac{z}{z+1}$
and
$$\widehat{\mathrm{DT}}(\widehat{\Psi}^{(-)}_{\lambda})=\widehat{\Psi}^{(+)}_{\lambda}=\sqrt{z^2-1}\partial_z\widehat{\Psi}^{(-)}_\lambda+{1\over
\sqrt{z^2-1}}\widehat{\Psi}_\lambda^{(-)},$$ where
$\widehat{\Psi}_{\lambda}^{(-)}$ is the general solution of
$\widehat{H}_-\widehat{\Psi}^{(-)}=\lambda\widehat{\Psi}^{(-)}$
for $\lambda\neq 1$.

The original potential corresponding to this example is given by
$V_-(x)=\frac{\cosh x}{\cosh x -1}$ and for $\lambda_1=1$ the
particular solution $\Psi_1^{(-)}$ is given by ${\sinh x\over
\cosh x-1}$. Applying $\mathrm{DT}$ we have that
$\mathrm{DT}(V_-)=V_+(x)=\frac{\cosh x}{\cosh x +1}$ and
$\mathrm{DT}(\Psi_\lambda^{(-)})=\Psi_\lambda^{(+)}=\partial_x\Psi^{(-)}_\lambda+\frac1{\sinh
x}\Psi^{(-)}_\lambda$.

\item $\sqrt{\alpha}=-z$, $\widehat{V}_-(z)=z^2-z$. Taking $\lambda_1=0$ and $\widehat{\Psi}^{(-)}_0=e^{-z}$
 we have that $\widehat{\mathrm{DT}}(\widehat{V}_-)=V_+=z^2+z$ and
$\widehat{\mathrm{DT}}(\widehat{\Psi}^{(-)}_{\lambda})=\widehat{\Psi}^{(+)}_{\lambda}=-z\partial_z\widehat{\Psi}^{(-)}_\lambda-z\widehat{\Psi}^{(-)}_\lambda$,
where $\Psi_{\lambda}^{(-)}$ is the general solution of
$\widehat{H}_-\widehat{\Psi}^{(-)}=\lambda\widehat{\Psi}^{(-)}$
for $\lambda\neq 0$. This example corresponds to the Morse
potential $V_-(x)=e^{-2x}-e^{-x}$, introduced in the list
\eqref{eqshin}.
\end{itemize}
\end{examples}
\medskip

To illustrate $\widehat{\mathrm{CI}}_n$ we present the following
example, which is related with the Chebyshev polynomials.\\

\begin{example}

Now, considering $\sqrt{\alpha}=-\sqrt{1-z^2}$, $V=0$ with
eigenvalues and eigenfunctions $\lambda_1=1$, $\lambda_2=4$,
$\widehat{\Psi}_{1}=z$, $\widehat{\Psi}_{4}=2z^2-1$,
$\widehat{\Psi}_{n^2}=T_n(z)$, where $T_n(z)$ is the Chebyshev
polynomial of first kind of degree $n$. The algebrized Wronskian
for $n=2$ is
$$\widehat{W}(z,2z^2-1)=-\sqrt{1-z^2}(2z^2+1),$$
 and by algebrized Crum iteration we obtain the potential $$\widehat{\mathrm{CI}}_2(\widehat{V})=\widehat{V}_2=((2z^2-1)\partial_z^2+z\partial_z)\ln \widehat{W}(z,2z^2-1)$$
 and the algebrized wave functions
$$
\widehat{\mathrm{CI}}_2(\widehat{\Psi}_\lambda)=\widehat{\Psi}_\lambda^{(2)}={\widehat{W}(z,2z^2-1,T_n)\over
\widehat{W}(z,2z^2-1)}.$$

\end{example}
\medskip

In a natural way we introduce the concept of algebrized shape
invariant potentials
$\widehat{V}_{n+1}(z,a_n)=\widehat{V}_n(z,a_{n+1})+R(a_n)$, where
the energy levels for $n>0$ are given by $E_n=R(a_1)+\cdots
R(a_n)$ and the algebrized eigenfunctions are given by
$\widehat{\Psi}_n(a_1)=\widehat{A}^\dagger(z,a_1)\cdots
\widehat{A}^\dagger(z,a_n)\Psi_0(z,a_n)$. To illustrate the
algebrized shape invariant potentials and the operators $\widehat
A$ and $\widehat A^\dagger$, we present the following example.\\

\begin{example} Assume $\sqrt{\alpha}=1-z^2$ and the algebrized super
potential $\widehat{W}(z)=z$. Following the method proposed in
remark \ref{remshape}, step 1, we introduce $\mu\in\mathbb{C}$ to
obtain $\widehat{W}(z;\mu)=\mu z$, and
$$\widehat{V}_\pm(z;\mu)=\widehat{W}^2(z;\mu)\pm\widehat{\partial}_z\widehat{W}(z;\mu)=\mu(\mu\mp1)z^2\pm\mu,$$ thus,
$\widehat{V}_+(z;a_0)=a_0(a_0-1)z^2+a_0$ and
$\widehat{V}_-(z;a_1)=a_1(a_1+1)z^2-a_1.$
 By step 2,  $$\widehat{\partial}_z(V_+(z;a_0)-V_-(z;a_1))=2z(1-z^2)(a_0(a_0-1)-a_1(a_1+1)).$$ By step 3, we obtain
$$a_1(a_1+1)=a_0(a_0-1),\quad a_1^2-a_0^2=-(a_1+a_0),$$ and assuming $a_1\neq\pm a_0$ we have
$a_1=f(a_0)=a_0-1$ and
$R(a_1)=2a_0+1=(a_0+1)^2-a_0^2=a_1^2-a_0^2$. This means that the
potentials $\widehat{V}_\pm$ are algebrized shape invariant
potentials where $E=E_n$ is easily obtained,
$$E_n=\sum_{k=1}^nR(a_k)=\sum_{k=1}^n\left(a_{k}^2-a_{k-1}^2\right)=a_n^2-a_0^2,\quad a_n=f^{n}(a_0)=a_0+n.$$ Now, the algebrized ground state wave
function of $\widehat{V}_-(z,a_0)$ is
$$\widehat{\Psi}_0=e^{\int{a_0z\over
1-z^2}dz}=\frac1{\left(\sqrt{1-z^2}\right)^{a_0}}.$$ Finally, we
can obtain the rest of eigenfunctions using the algebrized raising
operator:
$$\widehat{\Psi}_n(z,a_0)=\widehat{A}^\dagger(z,a_0)\widehat{A}^\dagger(z,a_1)\cdots\widehat{A}^\dagger(z,a_{n-1})\widehat{\Psi}_0(z,a_n).$$
This example corresponds to P\"oschl-Teller potential introduced
in the list \eqref{eqshin}.
\end{example}
\medskip

Now to illustrate the power of Kovacic's algorithm with the
derivation $\widehat{\partial}_z$, we study some Schr\"odinger
equations for non-rational shape invariant potentials given in
list \eqref{eqshin}. We work with specific values of these
potentials, although we can apply our machinery (algebrization
method and Kovacic's algorithm) using all the parameters of such
potentials.
\\

 \textbf{Morse potential:} $V(x)=e^{-2x}-e^{-x}.$\\

The Schr\"odinger equation $H\Psi=\lambda\Psi$ is
$$\partial_x^2\Psi=\left(e^{-2x}-e^{-x}-\lambda\right)\Psi.$$

By the Hamiltonian change of variable $z=z(x)=e^{-x}$, we obtain
$$\alpha(z)=z^2,\quad \widehat V(z)=z^2-z,\quad \widehat{\mathbf{V}}(z)={z^2-z-\frac14\over
z^2}.$$ Thus, $\widehat{K}=\mathbb{C}(z)$ and $K=\mathbb{C}(e^x)$.
In this way, the algebrized Schr\"odinger equation
$\widehat{H}\widehat{\Psi}=\lambda\widehat{\Psi}$ is
$$z^2\partial_z^2\widehat{\Psi}+z\partial_z\widehat{\Psi}-(z^2-z-\lambda)\widehat{\Psi}=0$$ and the reduced
algebrized Schr\"odinger equation
$\widehat{\mathbf{H}}\widehat{\Phi}=\lambda\widehat{\Phi}$ is
\begin{displaymath}\partial_z^2\Phi=r\Phi,\quad r={ z^2- z-\frac14-\lambda\over
z^2}.\end{displaymath}
 This equation only could fall in
case 1, in case 2 or in case 4 (of Kovacic's algorithm). We start
analyzing the case 1: by conditions $c_2$ and $\infty_3$ we have
that

$$\left[ \sqrt {r}\right] _{0}=0,\quad\alpha_{0}^{\pm}={1\pm 2\sqrt{-\lambda}\over 2}, \quad\left[  \sqrt{r}\right]  _{\infty}=1,\quad
{\rm and}\quad \alpha_{\infty}^{\pm }=\mp{1\over 2}.$$ By step 2
we have the following possibilities for $n\in\mathbb{Z}_+$ and for
$\lambda\in\Lambda$:
$$
\begin{array}{lll}
\Lambda_{++})\quad& n=\alpha^+_\infty -
\alpha^+_0=-1-\sqrt{-\lambda}, &
\lambda=-\left(n+1\right)^2,\\&&\\
\Lambda_{+-}) & n=\alpha^+_\infty -
\alpha^-_0=-1+\sqrt{-\lambda},& \lambda=-\left(n+1\right)^2,
\\&&\\
\Lambda_{-+}) & n=\alpha^-_\infty - \alpha^+_0=-\sqrt{-\lambda},&
\lambda=-n^2,\\&&\\ \Lambda_{--}) & n=\alpha^-_\infty -
\alpha^-_0=\sqrt{-\lambda},& \lambda=-n^2.
\end{array}
$$
We can see that $\lambda\in\Lambda_-=\{-n^2:n\in\mathbb{Z}_+\}$.
Now, for $\lambda\in\Lambda$, the rational function $\omega$ is
given by:
$$
\begin{array}{llll}
\Lambda_{++})\quad& \omega=1+{3+2n\over 2
z},&\lambda\in\Lambda_{++},& r_n={4n^2+8n+3\over 4 z^2} +
{2n+3\over z}+1,\\&&&\\
\Lambda_{+-}) & \omega=1-{1+2n\over 2 z},&\lambda\in\Lambda_{+-},&
r_n={4n^2+8n+3\over 4 z^2} -
{2n+1\over z}+1,\\&&&\\
\Lambda_{-+}) & \omega=-1+{1+2n\over 2
z},&\lambda\in\Lambda_{-+},& r_n={4n^2-1\over 4 z^2} -{2n+1\over
z}+1,\\&&&\\ \Lambda_{--}) & \omega=-1+{1-2n\over 2
z},&\lambda\in\Lambda_{--},& r_n={4n^2-1\over 4 z^2} + {2n-1\over
z}+1,
\end{array}
$$
where $r_n$ is the coefficient of the differential equation
$\partial_z^2\Phi=r_n\Phi$.
\\

\noindent By step 3, there exists a polynomial of degree $n$
satisfying the relation (\ref{recu1}),
$$
\begin{array}{llll}
\Lambda_{++})\quad& \partial_z^2 \widehat{P}_n+2\left(1+{3+2n\over
2 z}\right)\partial_z \widehat{P}_n+{2(n+2)\over  z}\widehat{P}_n&=&0,\\&&&\\
\Lambda_{+-}) & \partial_z^2 \widehat{P}_n+2\left(1-{1+2n\over
2 z}\right)\partial_z \widehat{P}_n+{2(-n)\over  z}\widehat{P}_n&=&0,\\&&&\\
\Lambda_{-+}) & \partial_z^2 \widehat{P}_n+2\left(-1+{1+2n\over 2
z}\right)\partial_z \widehat{P}_n+{2(-n)\over  z}\widehat{P}_n&=&0,\\&&&\\
\Lambda_{--}) &
\partial_z^2 \widehat{P}_n+2\left(-1+{1-2n\over 2 z}\right)\partial_z \widehat{P}_n+{2n\over  z}\widehat{P}_n&=&0.
\end{array}
$$

These polynomials only exists for $n=\lambda=0$, with
$\lambda\in\Lambda_{-+}\cup\Lambda_{--}$. So that the solutions of
$H\Psi=0$, $\widehat{H}\widehat{\Psi}=0$ and
$\widehat{\mathbf{H}}\Phi=0$ are given by
$$
\Phi_0=\sqrt{ z}e^{- z},\quad \widehat{\Psi}_0=e^{- z},\quad
\Psi=e^{- e^{-x}}.
$$
The wave function $\Psi_0$ satisfy the conditions (\ref{condqm}),
which means that is ground state (see \cite{duka}) and
$0\in\mathrm{spec}_p(H)$. Furthermore, we have

$$\mathrm{DGal}(L_0/{K})=\mathrm{DGal}(\widehat{L}_0/\widehat{K})=\mathrm{DGal}(\widehat{\mathbf{L}}_0/{\mathbb{C}(z)})=\mathbb{B},$$
$$\mathcal{E}(H)=\mathcal{E}(\widehat{H})=\mathcal{E}(\widehat{\mathbf{H}})=\mathrm{Vect}(1).$$

We follow with the case two. The conditions $c_2$ and $\infty_3$
are satisfied, in this way we have
$$E_c=\left\{2,2+4\sqrt{-\lambda},2-4\sqrt{-\lambda}\right\}\quad
\textrm{and}\quad E_{\infty}=\{0\},$$ and by step two, we have
that $2\pm\sqrt{-\lambda}=m\in\mathbb{Z}_+$, so that
$\lambda=-\left(\frac{m+1}{2}\right)^2$ and the rational function
$\theta$ has the following possibilities
$$\theta_+={2+m\over  z},\quad \theta_-=-{m\over  z}.$$

\noindent By step three, there exist a monic polynomial of degree
$m$ satisfying the recurrence relation (\ref{recu2}):
$$
\begin{array}{llll}
\theta_{+})& \partial_z^3 \widehat{P}_m+{3m+6\over  z}\partial_z^2
\widehat{P}_m-{4· z^2 - 4 z - 2·m^2 - 7·m -6\over  z^2}\partial_z
\widehat{P}_m-{ 4·m· z +8· z - 4·m - 6\over
 z^2}\widehat{P}_m&=&0,\\&&&\\ \theta_{-}) & \partial_z^3 \widehat{P}_m-{3m\over  z}\partial_z^2 \widehat{P}_m-{4· z^2
- 4· z - 2m^2-m\over  z^2}\partial_z \widehat{P}_m+{ 4·m· z - 4·m
- 2\over
 z^2}\widehat{P}_m&=&0.
\end{array}
$$
We can see that for $m=1$ the polynomial exists only for the case
$\theta_-$, being $\widehat{P}_1=z-1/2$. In general, these
polynomials could exist only for the case $\theta_-$ with
$m=2n-1$, $n\geq 1$, that is $\lambda\in\{-n^2:n\geq 1\}$.
\\

For instance, by case one and case two, we obtain
$\Lambda=\{-n^2:n\geq 0\}=\mathrm{spec}_p(H)$. Now, the rational
function $\phi$ and the quadratic expression for $\omega$ are
$$
\phi=-{2n-1\over z}+{\partial_z \widehat{P}_{2n-1}\over
\widehat{P}_{2n-1}},\quad
\omega^2+M\omega+N=0,\quad\omega=\frac{-M\pm\sqrt{M^2-4N}}{2},
$$
where the coefficients $M$ and $N$ are given by
$$
M={2n-1\over z}-{\partial_z \widehat{P}_{2n-1}\over
\widehat{P}_{2n-1}},\quad N={n^2-n+\frac14\over
z^2}-{(2n-1){\partial_z \widehat{P}_{2n-1}\over
\widehat{P}_{2n-1}}-2\over z}+{\partial_z^2
\widehat{P}_{2n-1}\over \widehat{P}_{2n-1}}-2.
$$
Now, $\triangle=M^2-4N\neq 0$, which means that
$\mathbf{\widehat{H}}\Phi=-n^2\Phi$ with $n\in\mathbb{Z}^+$ has
two solutions given by Kovacic's algorithm:
$$\Phi_{1,n}={\sqrt{ z}\widehat{P}_ne^{- z}\over  z^n},\quad \Phi_{2,n}={\sqrt{ z}\widehat{P}_{n-1}e^{ z}\over  z^n}.$$
The solutions of $\widehat{H}\widehat{\Psi}=-n^2\widehat{\Psi}$
are given by
$$\widehat{\Psi}_{1,n}={\widehat{P}_ne^{- z}\over  z^n},\quad \widehat{\Psi}_{2,n}={\widehat{P}_{n-1}e^{ z}\over  z^n},$$
and therefore, the solutions of the Schr\"odinger equation
$H\Psi=-n^2\Psi$ are

$$\Psi_{1,n}=P_ne^{-e^{-x}}e^{nx},\quad
 \Psi_{2,n}=P_{n-1}e^{e^{-x}}e^{nx},\quad P_n=\widehat{P}_n\circ z.$$

The wave functions $\Psi_{1,n}=\Psi_n$ satisfies the conditions of
bound state, and for $n=0$, this solution coincides with the
ground state presented above. Therefore we have
$$\Phi_n=\Phi_0\widehat{f}_n\widehat{P}_n,\quad \widehat{\Psi}_n=\widehat{\Psi}_0\widehat{f}_n\widehat{P}_n,\quad \widehat{f}_n( z)=\frac{1}{
z^{n}}.$$ Thus, the bound states wave functions are obtained as
$$\Psi_n=\Psi_0f_nP_n,\quad
f_n(x)=\widehat{f}_n(e^{-x})=e^{nx}.$$ The Eigenrings and
differential Galois groups for $n>0$ satisfies
$$\mathrm{DGal}(L_n/K)=\mathrm{DGal}(\widehat{L}_n/{\widehat{K}})=\mathrm{DGal}(\widehat{\mathbf{L}}_n/{\mathbb{C}(z)})=\mathbb{G}_m,$$
$$\dim_{\mathbb{C}}\mathcal E(\widehat{\mathbf{H}}+n^2)= \dim_{\mathbb{C}}\mathcal E(\widehat{{H}}+n^2)=\dim_{\mathbb{C}}\mathcal E(H+n^2)=2.$$

We remark that the Schr\"odinger equation with Morse potential,
under suitable changes of variables \cite{lali}, falls in a
Bessel's differential equation. Thus we can obtain its
integrability by means of corollary \ref{corbessel}.
\\

It is known that  Eckart, Rosen-Morse, Scarf and P\"oschl-Teller
potentials, under suitable transformations, fall in an
Hypergeometric equation which allows apply theorem \ref{kimurath}.
These potentials are inter-related by point canonical coordinate
transformations (see \cite[p. 314]{cokasu} ), so that
$\Lambda=\mathbb{C}$ due to P\"oschl-Teller potential is obtained
by means of Darboux transformations of $V=0$ (\cite{masa,ro}).
 We consider some particular cases of Eckart, Scarf and Poschl-Teller potentials applying only the case 1 of Kovacic's algorithm. The case 1 allow us
 to obtain the enumerable set $\Lambda_n\subset \Lambda$, which include the classical results obtained by means of supersymmetric quantum mechanics.
 Cases 2 and 3 of Kovacic algorithm also can be applied, but are not considered here.\\

 \textbf{Eckart potential:} $V(x)=4\coth(x)+5$, $x>0$.\\

The Schr\"odinger equation $H\Psi=\lambda\Psi$ is
$$\partial_x^2\Psi=\left(4\coth(x)+5-\lambda\right)\Psi.$$

By the Hamiltonian change of variable $z=z(x)=\coth(x)$, we obtain
$$\alpha(z)=(1-z^2)^2,\quad \widehat V(z)=4z+5,\quad \widehat{\mathbf{V}}(z)=\frac4{(z + 1)·(z - 1)^2}.$$
 Thus, $\widehat{K}=\mathbb{C}(z)$ and $K=\mathbb{C}(\coth(x))$.
In this way, the algebrized Schr\"odinger equation
$\widehat{H}\widehat{\Psi}=\lambda\widehat{\Psi}$ is
$$(1-z^2)^2\partial_z^2\widehat{\Psi}-2z(1-z^2)\partial_z\widehat{\Psi}-(4z+5-\lambda)\widehat{\Psi}=0$$ and the reduced
algebrized Schr\"odinger equation
$\widehat{\mathbf{H}}\widehat{\Phi}=\lambda\widehat{\Phi}$ is
\begin{displaymath}\partial_z^2\Phi=r\Phi,\quad r={\frac {4z+4-\lambda}{ \left(z-1 \right) ^{2} \left(z+1 \right) ^{2}}}=
 {\frac {2-\frac{\lambda}4}{\left(  z-1 \right) ^{2}}}+{\frac {\frac{\lambda}4-1}{(
 z-1)}}+
{\frac {-\frac{\lambda}4}{\left(  z+1 \right) ^{2}}}+{\frac
{1-\frac{\lambda}4}{( z+1)}}\end{displaymath}

We can see that this equation could fall in any case of Kovacic's
algorithm. Considering $\lambda=0$, the conditions
$\{c_1,c_2,\infty_1\}$  of case 1 are satisfied, obtaining $$
[\sqrt{r}]_{-1}=[\sqrt{r}]_{1}=[\sqrt{r}]_{\infty}=\alpha^+_{\infty}=0,
\quad\alpha^{\pm}_{-1}=\alpha^-_{\infty}=1,\quad
\alpha^+_{1}=2,\quad\alpha^-_{1}=-1. $$ By step 2, the elements of
$D$ are $0$ and $1$. The rational function $\omega$ for $n=0$ and
for $n=1$ must be
$$\omega={1\over  z+1}+{-1\over  z-1}.$$
By step 3 we search the monic polynomial of degree $n$ satisfying
the relation (\ref{recu1}). Starting with $n=0$ the only one
possibility is $\widehat{P}_0( z)=1$, which effectively satisfy
the relation (\ref{recu1}), while $\widehat{P}_1( z)= z+a_0$ does
not exists. In this way we have obtained one solution using
Kovacic algorithm:
$$\Phi_0={ z+1\over  z-1},\quad \widehat{\Psi}_0=\sqrt{ z+1\over ( z-1)^3},$$
this means that $0\in\Lambda_n$. We can obtain the second solution
using the first solution:
$$\Phi_{0,2}={\frac {{ z}^{2}+ z-4-4\,\ln  \left(  z+1 \right)  z-4\,\ln  \left(  z+1
 \right) }{ z-1}},\quad \widehat{\Psi}_{0,2}={\Phi_{0,2}\over \sqrt{
 z^2-1}}.$$ Furthermore the differential Galois groups and Eigenrings for $\lambda=0$ are
 $$\mathrm{DGal}(\widehat{\mathbf{L}}_0/{\mathbb{C}(z)})=\mathbb{G}_a,\quad
 \mathrm{DGal}(L_0/{K})=\mathrm{DGal}(\widehat{L}_0/{\widehat{K}})=\mathbb{G}^{\{2\}},$$
$$\mathcal E(\widehat{\mathbf{H}})=\mathrm{Vect} \left(1,{(z+1)^2\over (1-z)^2}\partial_z+{2(z+1)\over (1-z)^3}\right),$$
 $$ \mathcal E(\widehat{{H}})=\mathrm{Vect}\left(1,{(z+1)^2\over (1-z)^2}\partial_z-{z^2+3z+2\over (1-z)^3}\right),$$
 $$\mathcal E(H)=\mathrm{Vect}\left(1,{(\coth(x)+1)^2\over (1-\coth(x))^2(1-\coth^2(x))}\partial_x-{\coth^2(x)+3\coth(x)+2\over (1-\coth(x))^3}\right).$$

 Now, for $\lambda\neq 0$, the conditions $\{c_2,\infty_1\}$ of case 1 are
 satisfied:

$$\begin{array}{l}
[\sqrt{r}]_{-1}=[\sqrt{r}]_{1}=[\sqrt{r}]_{\infty}=\alpha^+_{\infty}=0,
\quad\alpha^-_{\infty}=1,\\ \\
\alpha^{\pm}_{-1}={1\pm\sqrt{1-\lambda}\over 2
},\quad\alpha^{\pm}_{1}={1\pm\sqrt{9-\lambda}\over 2 }.
\end{array}$$ By step 2 we have the following possibilities for
$n\in\mathbb{Z}_+$ and for $\lambda\in\Lambda$:
$$
\begin{array}{lll}
\Lambda_{++-}) & n=\alpha^+_\infty -
\alpha^+_{-1}-\alpha^-_{1}=-1-{\sqrt{1-\lambda}-\sqrt{9-\lambda}\over
2},& \lambda=4-\frac4{(n + 1)^2} -n^2 - 2·n, \\&&\\
\Lambda_{+--}) & n=\alpha^+_\infty -
\alpha^-_{-1}-\alpha^-_{1}=-1+{\sqrt{1-\lambda}+\sqrt{9-\lambda}\over
2},& \lambda=4-\frac4{(n + 1)^2} - n^2 - 2·n, \\&& \\
\Lambda_{-+-}) & n=\alpha^-_\infty -
\alpha^+_{-1}-\alpha^-_{1}={\sqrt{1-\lambda}+\sqrt{9-\lambda}\over
2},& \lambda=5-\frac4{n^2} -n^2, \\&&\\ \Lambda_{---}) &
n=\alpha^-_\infty -
\alpha^-_{-1}-\alpha^-_{1}={\sqrt{1-\lambda}+\sqrt{9-\lambda}\over
2},& \lambda=5-\frac4{n^2} - n^2.
\end{array}
$$
Therefore, we have that
$$\Lambda_n\subseteq\left\{4-\frac4{(n + 1)^2} - n^2- 2·n :n\in\mathbb{Z}_+\right\}\cup
\left\{5-\frac4{n^2} - n^2:n\in\mathbb{Z}_+\right\}.$$

Now, for $\lambda\in\Lambda$, the rational function $\omega$ is
given by: {\small$$
\begin{array}{lll}
\Lambda_{++-})\quad& \omega={ z·(n - 1) - n^2 - 2·n - 1\over (n +
1)·( z + 1)·( z - 1)},& r_n={-2· z^2·(n - 1) +4· z·(n + 1)^2 +(n +
1)·(n^3 + 3·n^2 + 2·n + 2)\over(n + 1)^2·( z +
1)^2·( z - 1)^2},\\&&\\
\Lambda_{+--}) & \omega={n· z·(n + 1) + 2\over (n + 1)·( z + 1)·(1
-  z)},& r_n={n· z^2·(n + 1)^3 + 4 z·(n + 1)^2 + n^3 +
2·n^2 + n + 4\over(n + 1)^2·( z + 1)^2·( z - 1)^2},\\&&\\
\Lambda_{-+-}) & \omega={ z·(n - 2) - n^2\over n·( z + 1)·( z -
1)},& r_n={-2· z^2·(n - 2) + 4·n^2· z +n·(n^3 - n + 2)\over n^2·(
z + 1)^2·( z - 1)^2},\\&&\\ \Lambda_{---}) & \omega={n· z·(n - 1)
+ 2\over n·( z + 1)·(1 -  z)},& r_n={n^3· z^2·(n - 1) + 4·n^2· z +
n^3 - n^2 + 4\over n^2·( z + 1)^2·( z - 1)^2},
\end{array}
$$}
where $r_n$ is the coefficient of the differential equation
$\partial_z^2\Phi=r_n\Phi$.
\\

\noindent By step 3, there exists a monic polynomial of degree $n$
satisfying the relation (\ref{recu1}),
$$
\begin{array}{llll}
\Lambda_{++-})\quad& \partial_z^2\widehat P_n+2\left({ z·(n - 1) -
n^2 - 2·n - 1\over (n + 1)·( z + 1)·( z -
1)}\right)\partial_z\widehat P_n+{2·(1 -
n)\over((n + 1)^2·( z + 1)·( z - 1)}\widehat{P}_n&=&0,\\&&&\\
\Lambda_{+--}) & \partial_z^2\widehat P_n+2\left({n· z·(n + 1) +
2\over (n + 1)·( z + 1)·(1 -  z)}\right)\partial_z\widehat
P_n+{n·(n + 1)\over ( z +
1)·( z - 1)}\widehat{P}_n&=&0,\\&&&\\
\Lambda_{-+-}) & \partial_z^2\widehat P_n+2\left({ z·(n - 2) -
n^2\over n·( z + 1)·( z - 1)}\right)\partial_z\widehat P_n+{2·(2 -
n)\over n^2·( z + 1)·( z - 1)}\widehat{P}_n&=&0,\\&&&\\
\Lambda_{---}) &
\partial_z^2\widehat P_n+2\left({n· z·(n - 1) + 2\over n·( z +
1)·(1 - z)}\right)\partial_z\widehat P_n+{n·(n - 1)\over ( z +
1)·( z - 1)}\widehat{P}_n&=&0.
\end{array}
$$
The only one case in which there exist the polynomial
$\widehat{P}_n$ of degree $n$ is for $\Lambda_{+--})$. The
solutions of the equation $\widehat{\mathbf{H}}\Phi=\lambda\Phi$,
with $\lambda\neq 0$, are:
$$
\begin{array}{llll}
\Lambda_{++-})\quad&
\Phi_n=\widehat{P}_n\widehat{f}_n\Phi_0,&\Phi_0={1\over z
-1}&\widehat{f}_n=( z - 1)^{n·(1 - n)´\over 2·(n + 1)}·( z +
1)^{n·(n +
3)\over 2·(n + 1)},\\&&&\\
\Lambda_{+--}) & \Phi_n=\widehat{P}_n\widehat{f}_n\Phi_0,&\Phi_0={
z+1\over z -1}&f_n=( z - 1)^{n·(1 - n)\over 2·(n + 1)}·( z + 1)^{-
n·(n +
3)\over 2·(n + 1)},\\&&&\\
\Lambda_{-+-}) &
\Phi_n=\widehat{P}_n\widehat{f}_n\Phi_1,&\Phi_1={1\over  z
-1}&\widehat{f}_n=( z + 1)^{n^2 + n - 2\over 2·n}·( z - 1)^{-n^2 +
3·n - 2 \over 2·n},\\&&&\\ \Lambda_{---}) &
\Phi_n=\widehat{P}_n\widehat{f}_n\Phi_1,&\Phi_1={ z+1\over  z
-1}&\widehat{f}_n=( z + 1)^{-n^2 - n + 2 \over 2·n}·( z - 1)^{-n^2
+ 3·n- 2 \over 2·n}.
\end{array}
$$
In any case $\widehat{\Psi}_n={\Phi_n\over 1- z^2}$, but the case
$\Lambda_{+--})$ includes the classical results obtained by means
of supersymmetric quantum mechanics. Thus, replacing $z$ by $\coth
(x)$ we obtain the eigenstates $\Psi_n$. The Eigenrings and
differential Galois groups for $n>0$ and $\lambda\in\Lambda_n$
satisfies
$$\mathrm{DGal}(L_\lambda/{K})\subseteq \mathbb{G}^{\{2m\}},\quad \mathrm{DGal}(\widehat{L}_\lambda/{\widehat{K}})\subseteq
\mathbb{G}^{\{2m\}},$$
$$\mathrm{DGal}(\widehat{\mathbf{L}}_\lambda/{\mathbb{C}(z)})=\mathbb{G}_m,$$
$$\dim_{\mathbb{C}(z)}\mathcal E(\widehat{\mathbf{H}}+\lambda)= 2,\quad \mathcal E(\widehat{{H}}+\lambda)=\mathcal E(H+\lambda)=\mathrm{Vect}(1).$$
\medskip

 \textbf{Scarf potential:} $V(x)={\sinh^2x-3\sinh x\over \cosh^2 x}.$\\

The Schr\"odinger equation $H\Psi=E\Psi$ is
$$\partial_x^2\Psi=\left({\sinh^2x-3\sinh x\over \cosh^2 x}-E\right)\Psi.$$

By the Hamiltonian change of variable $z=z(x)=\sinh(x)$, we obtain
$$\alpha(z)=1+z^2,\quad \widehat V(z)={z^2-3z\over 1+z^2}.$$
 Thus, $\widehat{K}=\mathbb{C}(z,\sqrt{1+z^2})$ and $K=\mathbb{C}(\sinh(x),\cosh(x))$.
In this way, the reduced algebrized Schr\"odinger equation
$\widehat{\mathbf{H}}{\Phi}=\lambda{\Phi}$ is
$$\partial_z^2=\left({\lambda z^2 -12 z
 +\lambda-1\over 4·( z^2 + 1)^2}\right)\Phi, \quad \lambda=3-4E.$$

Applying Kovacic's algorithm for this equation with $\lambda=0$,
we see that does not falls in case 1. We consider only
$\lambda\neq 0 $.  By conditions $\{c_2,\infty_2\}$ of case 1 we
have that

$$\begin{array}{l}
[\sqrt{r}]_{-\mathrm{i}}=[\sqrt{r}]_{\mathrm{i}}=[\sqrt{r}]_{\infty}=0,
\quad\alpha^\pm_{\infty}={1\pm\sqrt{1+\lambda}\over 2},\\ \\
\alpha^{+}_{-\mathrm{i}}=\frac54-\frac{\mathrm{i}}2,\quad\alpha^-_{-\mathrm{i}}=-\frac{1}4+\frac{\mathrm{i}}2\quad
\alpha^{+}_{\mathrm{i}}=\frac54+\frac{\mathrm{i}}2,\quad\alpha^-_{\mathrm{i}}=-\frac{1}4-\frac{\mathrm{i}}2
.
\end{array}$$ By step 2 we have the following possibilities for
$n\in\mathbb{Z}_+$ and for $\lambda\in\Lambda$:
$$
\begin{array}{lll}
\Lambda_{+++}) & n=\alpha^+_\infty -
\alpha^+_{-\mathrm{i}}-\alpha^+_{\mathrm{i}}={\sqrt{\lambda+1}-4\over
2},& \lambda=4n^2 + 16n + 15, \\&&\\ \Lambda_{+--}) &
n=\alpha^+_\infty -
\alpha^-_{-\mathrm{i}}-\alpha^-_{\mathrm{i}}={\sqrt{\lambda+1}+2\over
2},& \lambda=4n^2 - 8n + 3,
\end{array}
$$
obtaining in this way
$$\Lambda_n\subseteq\left\{4n^2 + 16n + 15 :n\in\mathbb{Z}_+\right\}\cup
\left\{4n^2 -8n+3:n\in\mathbb{Z}_+\right\}.$$

Now, the rational function $\omega$ is given by:
$$
\Lambda_{+++})\quad\quad \omega={5· z - 2\over 2·( z^2 +
1)},\qquad \Lambda_{+--})\quad {2 -  z \over 2·( z^2 + 1)}.
$$
By step 3, there exists $\widehat{P}_0=1$ and a polynomial of
degree $n\geq 1$ should satisfy either of the relation
(\ref{recu1}),
$$
\begin{array}{llll}
\Lambda_{+++})\quad& \partial_z^2\widehat P_n+{5· z - 2\over  z^2
+ 1}\partial_z\widehat P_n-{n· z^2·(n
+ 4) + 3· z + n^2 + 4·n - 3\over ( z^2 + 1)^2}\widehat{P}_n&=&0,\\&&&\\
\Lambda_{+--}) & \partial_z^2\widehat P_n+{2 -  z \over  z^2 +
1}\partial_z\widehat P_n-{n· z^2·(n - 2) + 3· z + n^2 - 2·n -
3\over ( z^2 + 1)^2}\widehat{P}_n&=&0.
\end{array}
$$
In both cases there exists the polynomial $\widehat{P}_n$ of
degree $n\geq 1$. Basis of solutions $\{\Phi_{1,n},\Phi_{2,n}\}$
of the reduced algebrized Schr\"odinger equation are:
$$
\begin{array}{llll}
\Lambda_{+++})\quad&
\Phi_{1,n}=\widehat{P}_n\widehat{f}_n\Phi_{1,0},&\Phi_{1,0}=(1+
z^2)^\frac54e^{-\arctan z},&\widehat{f}_n=1,\\&&&\\
\quad&
\Phi_{2,n}=\widehat{Q}_n\widehat{g}_n\Phi_{2,0},&\Phi_{2,0}={22+21x+12x^2+6x^3\over
\sqrt[4]{1+ z^2}}e^{-\arctan z},&\widehat{g}_n=1.\\&&&\\
\Lambda_{+--}) &
\Phi_{1,n}=\widehat{P}_n\widehat{f}_n\Phi_{1,0},&\Phi_{1,0}={1\over
\sqrt[4]{1+ z^2}}e^{\arctan z},&\widehat{f}_n=1,\\&&&\\
&
\Phi_{2,n}=\widehat{Q}_n\widehat{g}_n\Phi_{2,0},&\Phi_{2,0}={1\over
\sqrt[4]{1+ z^2}}e^{\arctan z}\int{1\over \sqrt{1+
z^2}}e^{-2\arctan z}d z,&\widehat{g}_n=1.
\end{array}
$$
\medskip

In both cases $\widehat{\Psi}={\Phi\over \sqrt[4]{1+ z^2}}$, but
the classical case (see references \cite{cokasu,duka}) is
$\Lambda_{+--})$, so that replacing $ z$ by $\sinh x$ and
$\lambda$ by $3-4E$ we obtain the eigenstates $\Psi_n$.
\\

The Eigenrings and differential Galois groups are
$$\mathcal E(H-\lambda)=\mathcal E(\widehat{H}-\lambda)=\mathcal E(\widehat{\mathbf{H}}-\lambda)=\mathrm{Vect}(1),$$
$$\mathrm{DGal}(L_\lambda/K)=\mathrm{DGal}(\widehat{L}_\lambda/{\widehat{K}})=\mathrm{DGal}(\widehat{\mathbf{L}}_\lambda/{\mathbb{C}(x)})=\mathbb{B}.$$
\\

\textbf{P\"oschl-Teller potential:}
$V(\mathrm{r})={\cosh^4(x)-\cosh^2(x) +2\over \sinh^2(x)\cosh^2
(x)},$ $x>0$. The reduced algebrized Schr\"odinger equation
$\widehat{\mathbf{H}}\Phi=E\Phi$ is
$$\partial_z^2\Phi=\left({\lambda z^4· -
(\lambda+3) z^2 + 8\over 4· z^2( z^2 - 1)^2}\right)\Phi, \quad
\lambda=3-4E.$$

Considering $\lambda=0$ and starting with the conditions
$\{c_2,\infty_1\}$ of case 1, we obtain
$$\begin{array}{l}
[\sqrt{r}]_{0}=[\sqrt{r}]_{-1}=[\sqrt{r}]_{1}=[\sqrt{r}]_{\infty}=\alpha^+_{\infty}=0,\quad
\alpha^-_{\infty}=1,
\\
\\
 \alpha^+_{-1}=\alpha^+_{1}=\frac54,\quad
 \alpha^-_{-1}=\alpha^-_{1}=-\frac14,\quad \alpha^+_{0}=2,\quad\alpha^-_{0}=-1.\end{array}$$ By step 2, the
elements of $D$ are $0$ and $1$. The rational function $\omega$
has the following possibilities for $n=0$ and for $n=1$:
$$\begin{array}{lll}
\Lambda_{++--}) & n=0,&\omega={5/4\over  z+1}+{-1/4\over
 z-1}+{-1\over  z},\\& &\\
\Lambda_{+-+-}) & n=0,&\omega={-1/4\over  z+1}+{5/4\over
 z-1}+{-1\over  z},\\ & & \\
\Lambda_{-+--}) & n=1,&\omega={5/4\over  z+1}+{-1/4\over
 z-1}+{-1\over  z},\\& &\\ \Lambda_{--+-}) &
n=1,&\omega={-1/4\over  z+1}+{5/4\over  z-1}+{-1\over  z}.
\end{array}
$$ By step 3 we search the monic polynomial of degree $n$
satisfying the relation (\ref{recu1}). Starting with $n=0$ the
only one possibility for $\Lambda_{++--})$ and $\Lambda_{+-+-})$
is $\widehat{P}_0( z)=1$, which does not satisfy the relation
(\ref{recu1}) in both cases, while $\widehat{P}_1( z)= z+a_0$
effectively does exists, in where $a_0=-{2\over 3}$ for
$\Lambda_{-+--})$ and $a_0={2\over 3}$ for $\Lambda_{--+-})$. In
this way we have obtained two solutions ($\Phi_{1,0}$,
$\Phi_{2,0}$) using Kovacic's algorithm:
$$\begin{array}{ll}\Phi_{1,0}=\left(1-\frac2{3 z}\right)\sqrt[4]{( z+1)^5\over  z-1},&\widehat{\Psi}_{1,0}=\left(1-\frac2{3 z}\right){ z+1\over \sqrt{ z-1}},\\
& \\
{\Phi}_{2,0}=\left(1+\frac2{3 z}\right)\sqrt[4]{( z-1)^5\over
 z+1},&\widehat{\Psi}_{2,0}=\left(1+\frac2{3 z}\right){ z-1\over
\sqrt{ z+1}}, \end{array}$$ this means that $0\in\Lambda_n$.
Furthermore,
$$\mathrm{DGal}(\widehat{\mathbf{L}}_0/{\mathbb{C}(x)})=\mathbb{G}^{[4]},\quad \mathrm{DGal}(\widehat{{L}}_0/{\widehat{K}})=
\mathrm{DGal}(L_0/{K})=e,$$
$$\dim_{\mathbb{C}}\mathcal E(\widehat{\mathbf{H}})=2,\quad \dim_{\mathbb{C}}\mathcal E(\widehat{H})=\dim_{\mathbb{C}} \mathcal E(H)=4.$$

 Now, for $\lambda\neq 0$ we see that conditions $\{c_2,\infty_1\}$ of case 1 leads us to

$$\begin{array}{l}
[\sqrt{r}]_{0}=[\sqrt{r}]_{-1}=[\sqrt{r}]_{1}=[\sqrt{r}]_{\infty}=0,
\quad\quad\alpha^\pm_{\infty}={1\pm\sqrt{1+\lambda}\over 2},\\ \\
 \alpha^+_{-1}=\alpha^+_{1}=\frac54,\quad
 \alpha^-_{-1}=\alpha^-_{1}=-\frac14,\quad \alpha^+_{0}=2,\quad\alpha^-_{0}=-1.
\end{array}$$ By step 2 we have the following possibilities for
$n\in\mathbb{Z}_+$ and for $\lambda\in\Lambda$:
$$
\begin{array}{lll}
\Lambda_{++++}) & n=\alpha^+_\infty -
\alpha^+_{-1}-\alpha^+_{1}-\alpha^+_0={\sqrt{\lambda+1}-8\over
2},& \lambda=4·n^2 + 32·n + 63, \\&&\\
\Lambda_{+++-}) &n=\alpha^+_\infty -
\alpha^+_{-1}-\alpha^+_{1}-\alpha^-_0={\sqrt{\lambda+1}-2\over
2},& \lambda=4·n^2 + 8·n + 3, \\&&\\
\Lambda_{++-+}) & n=\alpha^+_\infty -
\alpha^+_{-1}-\alpha^-_{1}-\alpha^+_0={\sqrt{\lambda+1}-5\over
2},& \lambda=4·n^2 + 20·n + 24, \\&&\\
\Lambda_{++--}) & n=\alpha^+_\infty -
\alpha^+_{-1}-\alpha^-_{1}-\alpha^-_0={\sqrt{\lambda+1}+1\over
2},& \lambda=4·n^2 -4n, \\&&\\
\Lambda_{+-++}) & n=\alpha^+_\infty -
\alpha^-_{-1}-\alpha^+_{1}-\alpha^+_0={\sqrt{\lambda+1}-5\over
2},& \lambda=4·n^2 + 20·n + 24, \\&&\\
\Lambda_{+-+-}) &n=\alpha^+_\infty -
\alpha^-_{-1}-\alpha^+_{1}-\alpha^-_0={\sqrt{\lambda+1}+1\over
2},& \lambda=4·n^2 -4n, \\&&\\
\Lambda_{+--+}) & n=\alpha^+_\infty -
\alpha^-_{-1}-\alpha^-_{1}-\alpha^+_0={\sqrt{\lambda+1}-2\over
2},& \lambda=4·n^2 + 8·n + 3, \\&&\\
\Lambda_{+---}) & n=\alpha^+_\infty -
\alpha^-_{-1}-\alpha^-_{1}-\alpha^-_0={\sqrt{\lambda+1}+4\over
2},& \lambda=4·n^2 -16n + 15,
\end{array}
$$
obtaining
$\Lambda_n\subseteq\Lambda_a\cup\Lambda_b\cup\Lambda_c\cup\Lambda_d\cup\Lambda_e,$
where
$$\begin{array}{ll}
\Lambda_a=\left\{4n^2 + 32·n +63:n\in\mathbb{Z}_+\right\},&
\Lambda_b=\left\{4n^2+8n+3:n\in\mathbb{Z}_+\right\}\\&
\\
\Lambda_c=\left\{4n^2 + 20·n +24:n\in\mathbb{Z}_+\right\},&
\Lambda_d=\left\{4n^2-4n:n\in\mathbb{Z}_+\right\},\\&
\\
\Lambda_e=\left\{4n^2 - 16·n +15:n\in\mathbb{Z}_+\right\}.&
\end{array}$$

Now, the rational function $\omega$ is given by:
$$\begin{array}{llll}
\Lambda_{++++}) &\omega={5/4\over  z+1}+{5/4\over  z-1}+{2\over
 z},& \Lambda_{+++-}) &\omega={5/4\over  z+1}+{5/4\over
 z-1}+{-1\over  z},\\ && & \\
\Lambda_{++-+}) &\omega={5/4\over  z+1}+{-1/4\over  z-1}+{2\over
 z}, & \Lambda_{++--}) & \omega={5/4\over  z+1}+{-1/4\over
 z-1}+{-1\over  z},\\& &\\ \Lambda_{+-++}) &\omega={-1/4\over
 z+1}+{5/4\over  z-1}+{2\over  z},&\Lambda_{+-+-})
&\omega={-1/4\over  z+1}+{5/4\over  z-1}+{-1\over  z},\\&&&\\
\Lambda_{+--+})&\omega={-1/4\over  z+1}+{-1/4\over  z-1}+{2\over
 z},& \Lambda_{+---}) &\omega={-1/4\over  z+1}+{-1/4\over
 z-1}+{-1\over  z}.
\end{array}
$$

By step 3, there exists a monic polynomial of degree $n$
satisfying the relation (\ref{recu1}),
$$
\begin{array}{ll}
\Lambda_{++++}) & \partial_z^2\widehat P_n+{(3· z + 2)·(3· z -
2)\over  z·( z + 1)·( z -
1)}\partial_z\widehat P_n+{n(n + 8)\over ( z + 1)·(1 -  z)}P_n=0, \\& \\
\Lambda_{+++-}) &\partial_z^2\widehat P_n+{3· z^2 + 2\over  z·( z + 1)·( z - 1)}\partial_z\widehat P_n+{n·(n + 2)\over( z + 1)·(1 -  z)}\widehat{P}_n=0, \\& \\
\Lambda_{++-+}) & \partial_z^2\widehat P_n+{6· z^2 - 3·x - 4\over  z·( z + 1)·( z - 1)}\partial_z\widehat P_n+{n· z·(n + 5) + 6\over  z·( z + 1)·(1 -  z)}\widehat{P}_n=0, \\& \\
\Lambda_{++--}) & \partial_z^2\widehat P_n+{3 z - 2\over  z·( z + 1)·(1 -  z)}\partial_z\widehat P_n+{n· z·(n - 1) - 12\over 4· z·( z + 1)·(1 -  z)}\widehat{P}_n=0, \\& \\
\Lambda_{+-++}) & \partial_z^2\widehat P_n+{6· z^2 + 3· z - 4\over  z·( z + 1)·( z - 1)}\partial_z\widehat P_n+{ z·(n^2 + 5) - 6\over  z·( z + 1)·(1 -  z)}\widehat{P}_n=0, \\& \\
\Lambda_{+-+-}) &\partial_z^2\widehat P_n+{3· z + 2\over  z·( z + 1)·( z - 1)}\partial_z\widehat P_n+{n· z·(n - 1) + 3 \over  z·( z + 1)·(1 -  z)}\widehat{P}_n=0, \\& \\
\Lambda_{+--+}) & \partial_z^2\widehat P_n+{3· z^2 - 4 \over  z·( z + 1)·( z - 1)}\partial_z\widehat P_n+{n·(n + 2)\over ( z + 1)·(1 -  z)}\widehat{P}_n=0, \\& \\
\Lambda_{+---}) & \partial_z^2\widehat P_n+{3· z^2 - 2\over  z·( z
+ 1)·(1 -
 z)}\partial_z\widehat P_n+{n·(4 - n)\over ( z + 1)·( z - 1)}P_n=0.
\end{array}
$$

The polynomial $P_n$ of degree $n$ exists for
$\lambda_n\in\Lambda_b$ with $n$ even, that is,
$\Lambda_n=\{n\in\mathbb{Z}:16n^2+16n+3\}$, for $\Lambda_{++-+})$
and $\Lambda_{+--+})$. Therefore
$E=E_n=\{n\in\mathbb{Z}:-4n^2-4n\}$.

\noindent The corresponding solutions for $\Lambda_n$ are
$$
\begin{array}{lllll}
\Lambda_{+++-})\quad&
\Phi_{1,n}=\widehat{P}_{2n}\widehat{f}_n\Phi_{1,0},&\Phi_{1,0}={\sqrt[4]{(
z^2-1)^5}\over  z}&\widehat{f}_n=1,&\widehat{\Psi}_{1,0}= z-\frac1{ z},\\&&&\\
\Lambda_{+--+}) &
\widehat{\Phi}_{2,n}=\widehat{Q}_{2n}\widehat{f}_n\widehat{\Phi}_{2,0},&\widehat{\Phi}_{2,0}={
z^2\over \sqrt[4]{ z^2 -1}}&\widehat{f}_n=1&\widehat{\Psi}_{2,0}={
z^2\over \sqrt{ z^2 -1}}.
\end{array}
$$
These two solutions are equivalent to the same solution of the
original Schr\"odinger equation and corresponds to the well known
supersymmetric quantum mechanics approach to this P\"oschl-Teller
potential, \cite{cokasu,cokasu2}. Furthermore, for all
$\lambda\in\Lambda_n$,
$$\mathrm{DGal}(\widehat{\mathbf{L}}_\lambda/{\mathbb{C}(x)})=\mathbb{G}^{[4]},\quad \mathrm{DGal}(\widehat{{L}}_\lambda/{\widehat{K}})=
\mathrm{DGal}(L_\lambda/{K})=e,$$
$$\dim_{\mathbb{C}}\mathcal E(\widehat{\mathbf{H}}-\lambda)=2,\quad \dim_{\mathbb{C}}\mathcal E(\widehat{H}-\lambda)=\dim_{\mathbb{C}} \mathcal E(H-\lambda)=4.$$
\medskip

\noindent\textbf{\large Searching Potentials From Parameterized
Differential Equations.}\\

The main object to search new potentials using
$\widehat{\partial}_z$ is the family of differential equations
presented by Darboux in \cite{da1}, see section
\ref{darbouxsection} and equation \eqref{orda7}, which can be
written in the form
\begin{equation}\label{paramet1}
\partial_z^2\widehat{y}+ \widehat{P}\partial_z\widehat{y}+(\widehat{Q} -\lambda \widehat{R})\widehat{y} = 0,\quad \widehat{P},\widehat{Q},\widehat{R}\in \widehat{K}.\end{equation} We recall that some Riemann's differential equations, presented in section
\ref{riemansection}, corresponds to this kind.\\

 When we have a
differential equation in the form \eqref{paramet1}, we reduce it
to put it in the form of the reduced algebrized Schr\"odinger
equation $\widehat{\mathbf{H}}\Phi=\lambda\Phi$, checking that
$\mathrm{Card}(\Lambda)>1$. Thus, starting with the potential
$\widehat{\mathbf{V}}$ and arriving to the potential $V$ we obtain
the Schr\"odinger equation $H\Psi=\lambda\Psi$. This methodology
(heuristic) is detailed below.\\

\begin{enumerate}
\item Reduce a differential equation of the form \eqref{paramet1} and put it in the form $\widehat{\mathbf{H}}\Phi=\lambda\Phi$, checking that
$\mathrm{Card}(\Lambda)>1$ and to avoid triviality, $\alpha$ must
be a non-constant function.

\item Write $\mathcal{W}={1\over 4}\partial_z(\ln\alpha)$ and
obtain
$\widehat{V}(z)=\alpha(\widehat{\mathbf{V}}-\partial_z\mathcal{W}-\mathcal{W}^2)$.

\item Solve the differential equation $(\partial_xz)^2=\alpha$, write
$z=z(x)$, $V(x)=\widehat{V}(z(x))$.
\end{enumerate}
\medskip

\noindent To illustrate this method, we present the following examples.\\

\textbf{Bessel Potentials}\\
\begin{itemize}\item (From Darboux transformations over $V=0$) In
the differential equation $$\partial_z^2\Phi=\left({n(n+1)\over
z^2}+\mu\right)\Phi,\quad \mu\in\mathbb{C},$$ we see that
$\lambda=-n(n+1)$ and $\alpha=z^2$. Applying the method, we obtain
$\widehat{\mathbf{V}}=\mu$ we obtain $\widehat{V}(z)=\mu
z^2+\frac14$ and $z=z(x)=e^{\pm x}$. Thus, we have obtained the
potentials $V(x)=\widehat{V}(z(x))=\mu e^{\pm2x}+\frac14$ (compare
with \cite[\S 6.9]{gapa}).

\item (From Bessel differential equation) The equation
$$\partial_z^2y+{1\over z}\partial_zy+{z^2-n^2\over
z^2}y=0,\quad n\in \frac12+\mathbb{Z},$$ is transformed to the
reduced equation
$$\partial_z^2\Phi=\left(\frac{n^2}{z^2} - {4z^2 + 1 \over 4z^2}\right)\Psi.$$ We can see that $\lambda=-n^2$, $\alpha=z^2$, obtaining
$\widehat{\mathbf{V}}=-z^2-\frac14$,
$\widehat{V}=-z^4-\frac14z^2+\frac14$ and $z=z(x)=e^{\pm x}$.
Thus, we have obtained the potential
$V(x)=\widehat{V}(z(x))=-e^{\pm4x}-\frac14e^{\pm2x}+\frac14$
(compare with \cite[\S 6.9]{gapa}).
\end{itemize}

We remark that the previous examples give us potentials related
with the Morse potential, due to their solutions are given in term
of Bessel functions.\\

We can apply this method to equations such as Whittaker,
Hypergeometric and in particular, differential equations involving
orthogonal polynomials (compare with \cite[\S 5]{cokasu}).\\

\chapter*{Final Remark}
\thispagestyle{empty}\markboth{Final Remark }{Final Remark}
\addcontentsline{toc}{chapter}{Final Remark}

The aim of this work is to give, in contemporary terms, a
formalization of original ideas and intuitions given by G.
Darboux, E. Witten and L. \'E. Gendenshte\"{\i}n in the context of
the Galois theory of linear differential equations. We found the
following facts.
\begin{itemize}

\item The superpotential is an algebraic
solution of the Riccati equation associated with a potential,
defined over a differential field.

\item Darboux transformation
was interpreted as an isogaloisian transformation, allowing to
obtain isomorphisms between their eigenrings.

\item  We introduced in a general way the Hamiltonian algebrization
method,which in particular allow to apply algorithmic tools such
as Kovacic's algorithm to obtain the solutions, differential
Galois groups and Eigenrings of second order linear differential
equations. We applied successfully this algebrization procedure to
solve problems in Supersymmetric quantum mechanics.
\item We can construct algebraically solvable and non-trivial algebraically quasi-solvable potentials  in the following ways.
\begin{enumerate}

\item Giving the potential in where for $\lambda=\lambda_0$ the
Schr\"odinger equation is integrable. After we put $\lambda\neq
\lambda_0$ checking that the Schr\"odinger equation is integrable
for more than one value of the parameter $\lambda$.

\item Giving a
superpotential to obtain the potential and after we check if the
Schr\"odinger equation is integrable for more than one value of
the parameter $\lambda$.

\item Since parameterized second order linear differential
equations applying an inverse process in the Hamiltonian
algebrization method. In particular, we can use algebraically
solvable and algebraically quasi-solvable potentials, special
functions with parameters (in particular with polynomial
solutions).
\end{enumerate}
\end{itemize}
\medskip

This thesis is a starting point to analyze quantum theories
through Galoisian theories. Therefore open questions and future work arise in a natural way: supersymmetric quantum mechanics with dimension greater than 2, relationship between algebraic and analytic spectrums, etc.\\

 As a conclusion, as happen in other
areas of the field of differential equations, in view of the many
families of examples studied along this thesis, we can conclude
that the \emph{differential Galois theory} is a natural framework
in which some aspects of \emph{supersymmetric quantum mechanics}
may appear more clearly.




\end{document}